\documentclass[aps,prd,floatfix,superscriptaddress,preprintnumbers,nofootinbib,tightenlines,preprint,longbibliography,groupedaddress]{revtex4-1}
\pdfoutput=1
\usepackage[english]{babel}
\usepackage{amsmath,amssymb,amsbsy,amstext,amsthm,booktabs,exscale,relsize,slashed,graphicx,amsfonts,upgreek,xcolor}
\usepackage{multirow,dcolumn,bm,enumerate,url,hyperref}
\usepackage{slashed}
\usepackage{bm}
\usepackage{latexsym}
\usepackage{graphicx}
\usepackage{epstopdf}
\usepackage{amsmath}
\usepackage{subfigure}
\usepackage{natbib}
\usepackage{hyperref}
\usepackage{pifont}
\usepackage{blindtext}
\usepackage{multirow}
\usepackage{tabularx}

\hypersetup{
  colorlinks   = true, 
  urlcolor     = magenta, 
  linkcolor    = magenta, 
  citecolor   = magenta 
}

\newcolumntype{P}[1]{>{\centering\arraybackslash}p{#1}}

\newcommand\Tstrut{\rule{0pt}{2.6ex}}         
\newcommand\Bstrut{\rule[-0.9ex]{0pt}{0pt}}   

\hypersetup{
     colorlinks   = true,
     citecolor    = violet,
     urlcolor     = violet,
     linkcolor    = violet
}

\newcommand{\nocontentsline}[3]{}
\newcommand{\tocless}[2]{\bgroup\let\addcontentsline=\nocontentsline#1{#2}\egroup}

\begin{document}

\title{The Enigmatic Galactic Center Excess: \\ Spurious Point Sources and Signal Mismodeling}
\preprint{MIT-CTP/5178}

\author{Rebecca K. Leane}
\thanks{{\scriptsize Email}: \href{mailto:rleane@mit.edu}{rleane@mit.edu}; {\scriptsize ORCID}: \href{http://orcid.org/0000-0002-1287-8780}{0000-0002-1287-8780}}
\affiliation{Center for Theoretical Physics \\ Massachusetts Institute of Technology \\ Cambridge, MA 02139, USA \\}

\author{Tracy R. Slatyer}
\thanks{{\scriptsize Email}: \href{mailto:tslatyer@mit.edu}{tslatyer@mit.edu}; {\scriptsize ORCID}:
\href{http://orcid.org/0000-0001-9699-9047}{0000-0001-9699-9047}}
\affiliation{Center for Theoretical Physics \\ Massachusetts Institute of Technology \\ Cambridge, MA 02139, USA \\}

\date{\today}

\begin{abstract} 
The Galactic Center GeV excess (GCE) has garnered great interest as a possible signal of either dark matter annihilation or some novel astrophysical phenomenon, such as a new population of gamma-ray emitting pulsars. In a companion paper, we showed that in a $10^\circ$ radius region of interest (ROI) surrounding the Galactic Center, apparent evidence for GCE point sources (PSs) from non-Poissonian template fitting (NPTF) is actually an artifact of unmodeled north-south asymmetry of the GCE. In this work, we develop a simplified analytic description of how signal mismodeling can drive an apparent preference for a PS population, and demonstrate how the behavior pointed out in the companion paper also appears in simpler simulated datasets that contain no PS signals at all. We explore the generality of this behavior in the real gamma-ray data, and discuss the implications for past and future studies using NPTF techniques. While the drop in PS preference once north-south asymmetry is included is not ubiquitous in larger ROIs, we show that any overly-rigid signal model is expected to yield a spurious PS signal that can appear very convincing: as well as apparent significance comparable to what one would expect from a true PS population, the signal can exhibit stability against a range of variations in the analysis, and a source count function  that is very consistent with previous apparent NPTF-based detections of a GCE PS population. This contrasts with previously-studied forms of systematic mismodeling which are unlikely to mimic a PS population in the same way. In the light of this observation, and its explicit realization in the region where the GCE is brightest, we argue that a dominantly smooth origin for the GCE is not in tension with existing NPTF analyses.
\end{abstract}

\maketitle

\tableofcontents

\section{Introduction}

The \textit{Fermi} Gamma-Ray Space Telescope has detected a robust gamma-ray excess surrounding the Galactic Center (GC)~\cite{Goodenough:2009gk, Hooper:2010mq, Hooper:2011ti, TheFermi-LAT:2015kwa}. This Galactic Center Excess (GCE) is peaked at energies $\sim 1-3$ GeV, and extends out to 10$^\circ$ from the Galactic Center~\cite{Hooper:2013rwa}. A number of studies have found that its morphology is broadly consistent with a signal of dark matter (DM) annihilation~\cite{Daylan:2014rsa,Calore:2014xka,Karwin:2016tsw}, being approximately spherically symmetric around the GC (with axis ratios within $20\%$ of unity), and scaling as $r^{-2\gamma}$ --- consistent with a generalized NFW profile with index $\gamma\sim1.1-1.4$~\cite{Gordon:2013vta,  Zhou:2014lva}. More recent analyses have shown that, depending on the diffuse gamma-ray background model and region over which the fit is performed, the GCE morphology can be more consistent with tracing stellar mass in the Galactic bulge and nuclear stellar cluster~\cite{Macias:2016nev, Bartels:2017vsx, Macias:2019omb}. In either case, the GCE could potentially be explained by a new population of millisecond pulsars
(MSPs)~\cite{Abazajian:2014fta, Abazajian:2012pn, Hooper:2013nhl, Mirabal:2013rba, Calore:2014oga, Cholis:2014lta, Yuan:2014yda, OLeary:2015qpx,Ploeg:2017vai,Bartels:2018xom}.

The GCE was first discovered using template fitting methods~\cite{Goodenough:2009gk}, which build up a model for the gamma-ray sky as a linear combination of spatial ``templates''  for distinct physical contributions to the gamma-ray emission. Gamma-ray source templates can have Poissonian statistics (i.e. a DM signal, or backgrounds from cosmic-ray interactions with the gas and starlight, or contributions from known point sources (PSs); in all cases the template is fully characterized by the expected emission in each pixel, and only its overall normalization is floated), or non-Poissonian (i.e., characterizing populations of astrophysical point-sources (PSs), such as pulsars, when their individual positions are unknown). We will loosely refer to these two cases as ``smooth'' and ``point-like'' / ``PS'' contributions, although templates with Poissonian statistics can still have sharp variations in the expected flux from pixel to pixel. To distinguish smooth and PS contributions with identical spatial morphologies, the template fitting method has been extended to include non-Poissonian template fitting (NPTF)~\cite{Lee:2015fea}, which exploits the differences in photon statistics between the two cases~\cite{Malyshev:2011zi,Lee:2014mza}.

In 2015, two papers claimed strong statistical evidence for the presence of unresolved gamma-ray PSs (with angular extension below the resolution of the detector) in the Inner Galaxy, using NPTF methods~\cite{Lee:2015fea} and a complementary technique employing wavelets~\cite{Bartels:2015aea}, and inferred that the GCE was dominantly comprised of PSs. More recently, two independent studies have demanded reconsideration of both the wavelet~\cite{Zhong:2019ycb} and NPTF~\cite{Leane:2019xiy} results. This article will focus on the NPTF-based evidence; let us briefly discuss the status of the wavelet argument. 

The evidence for GCE PSs from wavelet methods, as presented in Ref.~\cite{Bartels:2015aea}, consisted of a detection of numerous wavelet peaks which did not appear to be associated with the thick disk of the Milky Way, and whose rate appeared consistent with explaining the bulk of the GCE under the assumption of a rather typical luminosity function for the PSs ($dN/dL \propto L^{-1.5}$ up to a cutoff, where $L$ indicates the source luminosity and $N$ the number of sources). However, a recent re-analysis~\cite{Zhong:2019ycb} has demonstrated that these wavelet peaks appear to almost entirely correspond to PSs in the recently released 4FGL \textit{Fermi}-LAT source catalog \cite{Fermi-LAT:2019yla}, and that a significant fraction of these sources cannot be part of the GCE. The remaining sources cannot contribute more than a small fraction of the GCE flux, and masking out all the 4FGL sources does not appear to measurably affect the intensity of the GCE~\cite{Zhong:2019ycb}. Thus while the methods of Ref.~\cite{Bartels:2015aea} appear to have successfully detected a number of PSs or candidate PSs several years earlier than the official catalog, Ref.~\cite{Zhong:2019ycb} highlights that the presence of additional PSs is expected in the region of the sky where the GCE is found, and their existence does not guarantee they are contributing to the GCE. Indeed, in a similar vein to Ref.~\cite{Zhong:2019ycb}, an earlier study~\cite{Bartels:2017xba} using a previous PS catalog (2FIG)~\cite{Fermi-LAT:2017yoi} found that a number of newly detected pulsar-like sources in the inner Galaxy did not contribute significant flux to the GCE, in that masking them out did not affect the GCE signal.

The NPTF approach explicitly seeks to separate PSs in the inner Galaxy into different physical contributions, from the Galactic disk, extragalactic isotropic PSs, and the GCE itself. In this sense it is less prone than the wavelet analysis to misinterpretation based on the detection of PSs that are real but not associated with the GCE; however, possible systematics from errors in the spatial templates have always been a concern \cite{Lee:2015fea}.  Our companion paper~\cite{PhysRevLett.125.121105} argues that in a $10^\circ$ radius region of interest (ROI) around the Galactic Center, there is apparent strong evidence for GCE PSs, but this evidence is spurious, with the data being better explained by a smooth GCE with a pronounced north-south asymmetry.

In this article, we aim to place this result in the context of previous studies of possible systematic effects in NPTF analyses, and then to build up an understanding of how an unmodeled asymmetry can drive a convincing-looking but spurious PS detection, using analytic estimates and simplified simulations. We will then explore the degree to which the mechanism we have identified persists under variations of the analysis, and discuss the implications for previous and future NPTF studies. We begin in Section~\ref{sec:methods} by summarizing our data selection, priors, template choices, and regions of interest, and then discuss key aspects of previous NPTF analyses in Section~\ref{sec:previous}. In Section~\ref{sec:analytic} we provide a simple analytic description of the NPTF, in the limit where the likelihoods can be approximated as Gaussian, and use this to understand how an overly-rigid signal template for a smooth population can drive a preference for PSs. In Section~\ref{sec:smooth} we validate the analytic description by considering simulations of scenarios where a north-south asymmetric smooth component is incorrectly modeled as symmetric, and no PSs are simulated at all; this is a simpler version of the realistic NPTF analysis where the fit contains templates for Galactic and extragalactic PS populations. Having understood how asymmetry can drive a spurious PS preference, in Section~\ref{sec:variations} we explore how much our results change under variations to the Galactic diffuse emission model, the cuts on the photon dataset, the prior on the SCF, the ROI, and the particular asymmetry under study. In Section~\ref{sec:discussion} we compare the SCFs preferred by the various analyses and discuss the implications of our results for current and future NPTF analyses, before presenting our conclusions in Section~\ref{sec:conclusion}.

\section{Methodology and Data Selection}
\label{sec:methods}

\tocless\subsection{NPTF Implementation}

To employ the NPTF method, we use the NPTF package \texttt{NPTFit} \cite{Mishra-Sharma:2016gis}, interfaced with the Bayesian interference tool \texttt{MultiNest}~\cite{Feroz:2008xx}. The total number of live points for all \texttt{MultiNest} runs is \mbox{\texttt{nlive = 500}} unless specified otherwise.

\tocless\subsection{Fermi Data Selection}
We use the \texttt{Pass 8} \textit{Fermi} data, in the energy range $2-20$ GeV and collected over 573 weeks, from August 4th 2008 to June 19th 2019. We employ only events from the \textsc{UltracleanVeto} (1024) class, which has the most stringent cosmic-ray rejection cuts. This is further restricted to the top three quartiles of events graded by angular reconstruction (PSF1$-$PSF3), with quality cuts \texttt{DATA\_QUAL==1 \&\& LAT\_CONFIG==1}. The maximum zenith angle is $90^\circ$.

\tocless\subsection{Simulated Data}
Simulated contributions from Poissonian templates are generated by taking a random Poisson draw in \texttt{Python} of the product of the template and a predetermined normalization. Mock data for PS populations are generated using \texttt{NPTFit-Sim}~\cite{NPTFSim}. 
When creating simulated data, we first perform a fit on the real data using templates as defined in the relevant section. We then calculate the medians of the posterior distributions for the various model parameters, and simulate data based on those parameter values.

Table~\ref{tab:sim_values_one} describes the parameters used in simulations throughout this work, taken from the posterior medians in the fit to the real data when the smooth GCE template is broken into independently-floated northern and southern components, and no template for GCE PSs is included. 

To facilitate direct comparison with Ref.~\cite{PhysRevLett.125.121105}, we define a ``baseline mock dataset'', which is based on the simulated realization shown in Fig.~3 of Ref.~\cite{PhysRevLett.125.121105}.

\begin{table*}[h]
\centering
\renewcommand{\arraystretch}{1.5}
\setlength{\tabcolsep}{5.2pt}
\begin{tabular}{ccc}
\hline
\multicolumn{3}{c}{\textsc{Simulation Parameters}}\Tstrut\Bstrut \\
\hline
Parameter & Simulation Value (using \texttt{p6v11}) & Simulation Value (using \texttt{Model A}) \Tstrut\Bstrut \\
\hline 
\hline
$\log_{10}A_\text{iso}$   & ${-1.49}$ & $0.21$  \Tstrut\Bstrut \\ 
$\log_{10}A_\text{dif}$   & ${1.15}$& $-$  \Tstrut\Bstrut \\
$\log_{10}A_\text{ics}$   & $-$  & $0.56$ \Tstrut\Bstrut \\
$\log_{10}A_\text{pibrem}$   & $-$  & $0.98$ \Tstrut\Bstrut \\
$\log_{10}A_\text{bub}$  & ${-1.20}$ & $-0.73$ \Tstrut\Bstrut \\ 
$\log_{10}A_\text{GCE}^\text{north}$  & ${0.67}$& $0.60$  \Tstrut\Bstrut \\ 
$\log_{10}A_\text{GCE}^\text{south}$  & ${0.34}$& $0.44$  \Tstrut\Bstrut \\ 
$\log_{10}A_\text{PS}^\text{disk}$  & $-1.53$ & $-0.92$  \Tstrut\\
 $S_b^\text{disk}$    & $12.95$ & $6.83$    \\
 $n_1^\text{disk}$      & $2.55$ & $2.50$    \\
 $n_2^\text{disk}$      & $-1.18$ & $-0.37$  \\
 $\log_{10}A_\text{PS}^\text{iso}$ & $-4.63$ & $-4.66$  \Tstrut\\
 $S_{b,1}^\text{iso}$     & $31.06$ & $30.18$    \\
 $n_1^\text{iso}$    & $3.60$  & $3.67$   \\
 $n_2^\text{iso}$      & $-0.52$ & $-0.60$  \\
\hline
\hline
\end{tabular}
\caption{Parameter values used to generate the simulated data from Poissonian and non-Poissonian templates for the case where the smooth GCE template is broken into northern and southern components which are floated independently (normalizations controlled by $A_\text{GCE}^\text{north,south}$), and no GCE PS template is included in the simulation. These values are taken from the posterior medians in the corresponding fit to real data.}
\label{tab:sim_values_one}
\end{table*}

\begin{table*}[h]
\renewcommand{\arraystretch}{1.5}
\setlength{\tabcolsep}{5.2pt}
\begin{center}
\begin{tabular}{ c  c c  }
\hline
\multicolumn{3}{c}{\textsc{Prior Ranges}}\Tstrut\Bstrut		\\  
\hline Parameter 	 & \textit{Fermi} \texttt{p6v11} diffuse model & Models A, F \Tstrut\Bstrut \\
\hline 
\hline
$\log_{10}A_\text{iso}$   & $[-3,1]$ & $[-3,1]$  \Tstrut\Bstrut \\ 
$\log_{10}A_\text{dif}$   & $[0,2]$  & $-$ \Tstrut\Bstrut \\
$\log_{10}A_\text{GCE}^\text{north}$   & $[-3,1]$ & $[-3,1]$  \Tstrut\Bstrut \\ 
$\log_{10}A_\text{GCE}^\text{south}$   & $[-3,1]$ & $[-3,1]$  \Tstrut\Bstrut \\ 
$\log_{10}A_\text{GCE}$ & $[-3, 1]$  & $[-3, 1]$  \Tstrut\Bstrut	\\
$\log_{10}A_\text{ics}$   & $-$  & $[-2,2]$ \Tstrut\Bstrut \\
$\log_{10}A_\text{pibrem}$   & $-$  & $[-2,2]$ \Tstrut\Bstrut \\
$\log_{10}A_\text{bub}$  & $[-3, 1]$ & $[-3, 1]$\Tstrut\Bstrut \\ 
$\log_{10}A_\text{PS}$  & $[-6, 1]$ & $[-6, 1]$ \Tstrut\\
$S_b^\text{PS}$      & $[0.05 ,80]$  & $[0.05 ,80]$  \Tstrut\Bstrut \\
$n_1^\text{PS}$     & $[2.05, 5]$  & $[2.05, 5]$  \Tstrut\Bstrut  \\
$n_2^\text{PS}$      & $[ -3 ,1.95]$  & $[ -3 ,1.95]$   \Bstrut\\
\hline
\hline
\end{tabular}
\end{center}
\caption{Parameters and associated prior ranges used in all analyses unless explicity stated otherwise in the text. If the GCE Smooth normalization is allowed to go negative, the prior range used is $A_{\rm GCE}=[-9,9$].}
\label{tab:priors}
\end{table*}

\tocless\subsection{Template Modeling}

Our model of the gamma-ray sky includes Poissonian templates for the Galactic diffuse emission, isotropic emission (``Iso''), emission in the \textit{Fermi} Bubbles (``Bub''), and the GCE (``GCE Smooth''), and non-Poissonian templates for PSs tracing the Galactic disk (``Disk PS''), isotropic emission (``Iso PS''), and the GCE (``GCE PS''). These are the same templates defined in Ref.~\cite{Leane:2019xiy} (up to a factor of the difference in exposure between the datasets), although the templates we label as ``GCE Smooth'' and ``GCE PS'' are labeled respectively as ``NFW DM'' and ``NFW PS'' in that work. ``NFW'' stands for Navarro-Frenk-White~\cite{Navarro:1995iw}, a commonly-employed prescription for the DM density profile; we use a generalized NFW profile with inner slope 1.25, which has previously been found to be a good description of the GCE. For GCE PSs, the assumed PS distribution is technically NFW$^2$, to match the morphology of an annihilating DM signal.

For the diffuse model, we use the \textit{Fermi} \texttt{p6v11} diffuse model as a baseline (to facilitate comparison with earlier NPTF analyses that found evidence for GCE-correlated PSs), and check results with the GALPROP-based~\cite{Strong:1998pw} Galactic diffuse emission models denoted \texttt{Model A} and \texttt{Model F} in Ref.~\cite{Calore:2014xka}. We do not employ the more recent \textit{Fermi} \texttt{Pass 7} and \texttt{Pass 8} diffuse models because choices were made in their construction that render them unsuited for studies of extended diffuse emission (see \cite{Leane:2019xiy} for further discussion). In the figures
that follow, we use the \texttt{p6v11} diffuse model unless
noted otherwise.  When testing the spatial morphology of various components, templates may be broken into sub-regions, as described below. For all non-Poissonian templates, we assume a broken power-law form for the SCF.  

Table~\ref{tab:priors} details the priors used in our analyses. Unless specified otherwise, all components are restricted to have non-negative fluxes. The SCF for non-Poissonian templates is parameterized as $dN/dF = A (F/F_b)^{-n_2}$ for $F < F_b$, $dN/dF = A (F/F_b)^{-n_1}$ for $F \ge F_b$. By convention we state our prior on $F_b$ in terms of the average counts at the break; the conversion factor from $F_b$ to the average counts $S_b$ is the average exposure in the ROI ($2.79 \times10^{11}$ cm$^2$ s in our default dataset).  $A$ is the normalization of the relevant template.

\tocless\subsection{Bayes Factors (BFs)}

When testing for the presence of PSs, we compare the best fits with and without the GCE PS template; the BF between two fits with and without GCE PS thus describes the strength of evidence in favor of NFW-distributed (GCE) PSs. A BF is the ratio of Bayesian evidences for two models, and hence can be understood as the ratio of probabilities for the models, given the data. For example, a BF of 100 can be broadly compared to a $99\%$ confidence limit; assuming a Gaussian distribution, this would correspond to $\sim2.6\sigma$. A BF of 1 means no model is favored over the other, and a BF of less than one means the preference has swapped to the other model under study.\\

\tocless\subsection{Regions of Interest}

We will consider three different regions of interest (ROI):

\begin{enumerate}
 \item A 10$^\circ$ radius circle of the Galactic Center, excluding the band with $|b|< 2^\circ$ along the Galactic plane. This is the same as our companion paper Ref.~\cite{PhysRevLett.125.121105}.
 \item A 30$^\circ$ radius circle of the Galactic Center, excluding the band with $|b|< 2^\circ$ along the Galactic plane. This is the same as previous NPTF studies~\cite{Lee:2015fea,Leane:2019xiy,Chang:2019ars}.
 \item A 40 by 40 square, with $|b|< 2^\circ$ masked along the Galactic plane. This facilitates comparison with Ref.~\cite{Calore:2014xka}.
 
\end{enumerate}

 As previous studies have found the GCE extends out to at least 10$^\circ$ from the Galactic Center~\cite{Daylan:2014rsa}, the 10$^\circ$ radius is a well-motivated choice.  At the same time, there is reason to think that systematic effects from the mis-modeling of the Galactic diffuse emission are likely to be less severe in smaller ROIs~\cite{bennick,Daylan:2014rsa,Chang:2018bpt}. Nonetheless, we will investigate the effects of mismodeling in larger ROIs.

 \section{Summary of Selected NPTF Analyses of the GCE}
 \label{sec:previous}
 
 To place the rest of this paper in context, we now briefly summarize relevant analyses performed in previous NPTF studies of the GCE, and the inferences drawn from those studies regarding the likely effect of systematic uncertainties on a GCE PS detection.
 
 \subsection{Summary of Relevant Results from Lee et al (2016) (Ref.~\cite{Lee:2015fea})}
 
 The NPTF-based evidence for a GCE-correlated PS population, presented by Ref.~\cite{Lee:2015fea}, used \textit{Fermi} \texttt{Pass 7} data and the top half of events according to a cut designed to select events with good angular resolution, in a $30^\circ$ radius region of interest (ROI), and employed the  \textit{Fermi} \texttt{p6v11} model for the Galactic diffuse gamma-ray emission. The authors of that work noted the possibility for systematic errors in the templates to affect their results, and consequently they also tested a range of other diffuse models, including \texttt{Model A} and \texttt{Model F}. They found a strong and generic preference for a GCE PS population, with a Bayes factor (BF) in favor of PSs ranging from $10^6-10^9$ across the various diffuse models; in the same analysis the use of the \texttt{p6v11} diffuse model led to a BF of $10^7$. In all analyses, the posterior probability distribution for the flux associated with the different templates consistently assigned the flux of the GCE to the PS population.

Another potentially important output of the analysis of Ref.~\cite{Lee:2015fea} was the posterior probability distribution for the source count function (SCF), the number of inferred PSs as a function of their flux (denoted $dN/dF$, where $F$ is the photon flux in the energy band of interest). The SCF was modeled as a broken power law, and was consistently found to have a rather peaked shape, with the photon flux being dominated by sources near the break in $dN/dF$. However, the position of the break could differ by a factor of $\sim 2-3$ between analyses using different diffuse models. (It is perhaps worth noting that in the analysis of alternate diffuse models, while the 3FGL sources were masked, no models were included for the faint counterparts of known source populations, and so the GCE might have picked up unrelated unresolved sources; it is possible that including other PS populations would produce wider variation in the GCE population inferred with different diffuse models.)

Furthermore, Ref.~\cite{Lee:2015fea} tested the effect of simulating the data with an alternate diffuse model, and then fitting with \texttt{p6v11}; they found that within the range of diffuse models tested, the GCE PS population was always successfully recovered with a BF comparable to real data, and that when a smooth GCE was simulated, the BF in preference of PSs was not more than $\mathcal{O}(10)$. When the GCE was part smooth and part PSs, the simulations could not reliably reconstruct the SCF, but the BF in preference of PSs was substantially lower than when the GCE was entirely PSs. 

That work also tested a range of other systematics which are less relevant to our present analysis and which we will not detail here. Putting all this together, tentative conclusions from Ref.~\cite{Lee:2015fea} included:
\begin{itemize}
 \item Within the range of diffuse models tested, the differences between models were not sufficient to produce a spurious high-significance detection of GCE PSs.
 \item The SCF should not be trusted well below the break in the power law, especially in cases where it was suspected there might be both a smooth and PS-like component.
 \item The BF in favor of PSs was relatively stable under variations of the ``true'' background model, and could potentially be used to accurately exclude scenarios where the GCE had a significant smooth component.
\end{itemize}

 \subsection{Summary of Relevant Results from Leane et al (2019) (Ref.~\cite{Leane:2019xiy})}

More recently~\cite{Leane:2019xiy}, we demonstrated that large DM signals (several times greater than the GCE) injected into the real data were misattributed to PSs using the NPTF pipeline described in Ref.~\cite{Lee:2015fea} and implemented as publicly-available code in Ref.~\cite{Mishra-Sharma:2016gis}. The pipeline worked well on simulated data, implying that the origin of the misattribution was a systematic mismatch between one or more of the spatial templates and the real gamma-ray data they were meant to describe (and not, for example, a bug in the code or breakdown of the statistical method). The misattribution we observed can also be understood as the DM template preferring a large unphysical negative coefficient in the original gamma-ray data; until a large enough DM signal is added to overcome this negative coefficient, the best-fit coefficient for the DM signal will not rise above zero. 

Whatever systematic effect was responsible for the misattribution of injected signals, it seemed plausible that if a DM signal were in fact powering the GCE (in whole or in part), it could potentially have been misidentified as originating from PSs due to the same systematic. Furthermore, changing the prior on the smooth GCE component (to allow it to run negative) or injecting a smooth GCE-like signal modified the bright end of the SCF, not only the faint end, suggesting that correcting this mismodeling could potentially affect the SCF beyond just the faint sources. We also showed in real data that the severity of the problem -- e.g. parameterized by the degree to which the DM template coefficient preferred to run negative -- varied between different choices of modeling for the diffuse gamma-ray background, suggesting that errors in the diffuse model could contribute significantly to this behavior~\cite{Leane:2019xiy}. 

Using simulated data, we provided a proof-of-principle example of one way that this behavior could arise, as a consequence of additional unresolved PS populations that are not well-modeled by the chosen spatial templates. We showed that such an unmodeled source population could cause a high-significance spurious detection of GCE PSs, although the proof-of-principle example did not correspond to a real detection of a new PS population in the data, and it was not clear if a similar example could be found that would match all aspects of the apparent GCE PS detection. This proof-of-principle example also involved positing a new population of PSs with a SCF comparable to the SCF extracted for GCE PSs in the real data; thus it was still unclear if the bright end of the SCF observed in real data could be reproduced without actually positing a new PS population with fluxes in the relevant range.

 \subsection{Summary of Relevant Results from Chang et al (2019) (Ref.~\cite{Chang:2019ars})}

Even more recently, the degree to which biases in the NPTF method could result in a misattribution of the GCE to PSs was explored~\cite{Chang:2019ars} using simulated data. That work argued that the bright end of the SCF was likely to be more robust to errors than the faint end, and that inherent biases in the NPTF pipeline and errors in the diffuse background models -- while present -- were unlikely to provide strong evidence for a population of bright GCE PSs where none exist. However, the simulations in question did not quantitatively reproduce the behavior observed in real data in Ref.~\cite{Leane:2019xiy}, and left open the question of whether systematics capable of reproducing the results of Ref.~\cite{Leane:2019xiy} would have the same inability to generate a preference for a spurious source population with the observed SCF (see Appendix~\ref{sec:chang} for further discussion).

 \subsection{Summary of Relevant Results from Leane et al (2020) (Ref.~\cite{PhysRevLett.125.121105})}

In the light of these previous results, the new contribution of our companion paper~\cite{PhysRevLett.125.121105} is to present for the first time an explicit example, realized in the \textit{Fermi} data, of a form of mismodeling that gives rise to a spurious high-significance detection of GCE PSs with a SCF consistent with fits to the real data. Specifically, the companion paper demonstrates that in a $10^\circ$ radius ROI surrounding the Galactic Center, allowing for north-south asymmetry in the GCE removes a previously strong apparent preference for a GCE PS population, and furthermore that this behavior can be understood in detail using simulated data containing no GCE PSs. In contrast to previously studied forms of mismodeling/bias (which may also plausibly exist in real analyses), we have shown that:
\begin{itemize}
 \item The entire SCF is affected, not only the faint end, even though no GCE PSs are simulated.
 \item The BF for spurious GCE PSs can be very large, greater than or equal to that expected if the spurious PS population were genuine.
\end{itemize}
 To our knowledge, this is the first demonstration of an explicit mechanism for producing such a convincing but spurious signal, where accounting for the systematic removes the preference for GCE PSs in an analysis of real data. We also found for the first time a marked north-south asymmetry for the GCE in real data. However, in this work we will show that it is unclear at this stage if the asymmetry is also an artifact of mismodeling, as it does not appear consistently across all ROIs and Galactic diffuse emission models.

\section{Point Sources as a Proxy for Higher Variance}
\label{sec:analytic}

Let us first develop an understanding of how a mismodeling of a signal (or background) template can lead to spurious evidence for PSs. For this purpose, it is helpful to work in the limit of a large number of counts and sources, where we can approximate all Poisson distributions as Gaussian distributions with mean equal to their variance. This approximation will not be universally valid for the signals we are interested in, but it suffices to provide a simple and analytically tractable picture of the problems that can arise due to mismodeling.

\subsection{Probability Distribution for Counts from an Unresolved Source Population} 

Consider the probability to obtain $N$ photons from a source population, in the case where the expected number of sources is $n_0$, and the expected number of counts per source is $s$. Suppose first that the number of sources $n$ is fixed, while the number of counts per source is drawn from a Poisson distribution with mean $s$. Since the sum of variables drawn from Poisson distributions also follows a Poisson distribution, which can be approximated by a Gaussian for $n s \gg 1$, the total number of photons from the source population approximately follows the distribution:
\begin{equation} P(N|\{n, s\}) = \frac{1}{\sqrt{2 \pi n s}} e^{-(n s-N)^2/(2 n s)}.\end{equation}

Now to obtain the probability of obtaining $N$ photons given that the number of sources is Poisson-distributed with mean $n_0$, this distribution needs to be convolved with $P(n|n_0)$, i.e. the probability of drawing exactly $n$ sources when we expect $n_0$. We can approximate this probability distribution by a Gaussian with expectation value and variance equal to $n_0$. Thus we obtain:
\begin{align}
 P(N|\{n_0, s\}) &= \int dn P(N|\{n, s\})  P(n|n_0)\nonumber  \\
 &\approx \int dn  \frac{1}{\sqrt{2 \pi N}} e^{-(n s-N)^2/(2 N)}  \frac{1}{\sqrt{2 \pi n_0}} e^{-(n_0-n)^2/(2 n_0)}.
\end{align}
This expression can be simplified by the further approximation that $n s \approx N$ in regions where the integrand is large, and so we can approximately replace $n s \rightarrow N$ in the variance and normalization prefactor for $P(N|\{n, s\})$. With this replacement the integral can be done analytically, yielding:
\begin{align}
  P(N|\{n_0, s\}) &\approx \frac{1}{\sqrt{2 \pi (N + n_0 s^2)}} e^{-(N - n_0 s)^2/(2 (N + n_0 s^2))}\nonumber\\
  &\approx  \frac{1}{\sqrt{2 \pi n_0 s (1+s)}} e^{-(N - n_0 s)^2/(2 n_0 s (1+s))}
\end{align}
 where the second approximation holds for $N \approx n_0 s$, i.e. where the total number of photons is not too far from its expected value, so that the variance in the model can be approximated by the variance expected from the data. We see that the resulting distribution is Gaussian, with mean $\langle N \rangle = n_0 s$ as expected, and variance $n_0 s (1+s)$ -- that is, compared to a Poisson distribution which has equal mean and variance, the variance is inflated by a factor of $1+s$. We see that in the limit of $s \ll 1$, so that the sources are all very faint, we recover the usual (Gaussian approximation to the) Poisson distribution. Within these approximations, the characteristic feature of a source population is an inflated pixel-to-pixel variance, with the factor of inflation relative to the expectation value being $1+s$, where $s$ is the expected number of photons per source. Of course, when the number of sources/pixel or photons/pixel is small, the large-number approximation will break down, and the probability distribution can have a more complicated non-Gaussian form.
 
 We have so far assumed that the SCF is a delta function, with all sources having the same expected flux $s$. We can treat the case with a broader SCF as the sum of several source populations with different values of $s$. Summing variables drawn from Gaussian distributions, both the means and variances simply add; thus if a given population with counts/source $s_i$ contributes an expected $n_{0,i}$ sources, its contribution to the mean will be $n_{0,i} s_i$, and to the variance $n_{0,i} s_i (1 + s_i)$. Taking the continuum limit, and expressing the SCF as $f(s) = dn_0/ds$, we can write the expected number of photons as $\int ds s f(s)$, and the variance as $\int ds s f(s) (1+s)$. If the signal model also contains a smooth Poissonian component with an expected number of counts $s_0$, the contribution to both the mean and variance will be $s_0$.
 
\subsection{Effects of a Mismatch in Total Count Number on Inferred Variance}

Now suppose we have a signal that is drawn (in each pixel) from a Gaussian probability distribution with expectation value $X$ and variance $\sigma$. Suppose we try to fit an ensemble of such pixels with a model governed by a Gaussian distribution with mean $Y$ and variance $\tau$. For a single pixel, where the draw from the true distribution has yielded a value of $x$ for the number of counts, the likelihood (probability of drawing $x$ given the model) is then 
\begin{equation}
 \mathcal{L}=P(x|\{Y,\tau\}) = e^{-(x-Y)^2/(2\tau^2)} / \sqrt{2 \pi \tau^2}. 
\end{equation}
When we combine the likelihoods from many pixels, we will sum their log likelihoods; as the number of pixels becomes large, this sum should converge on the number of pixels multiplied by the expected log likelihood. We can compute the expectation value of the pixel log likelihood by integrating over the distribution of $x$:
\begin{align} \langle \ln \mathcal{L} \rangle & = \int dx e^{-(x-X)^2/(2 \sigma^2)}/\sqrt{2 \pi \sigma^2}  \left[-(x-Y)^2/2\tau^2  - (1/2) \ln (2\pi\tau^2)\right] \nonumber \\
& = -[(X-Y)^2 + \sigma^2]/2\tau^2 -(1/2) \ln (2\pi\tau^2), \end{align}
which can be exponentiated to give an effective ``average'' likelihood:
\begin{align} \mathcal{L} = \frac{1}{\sqrt{2\pi\tau^2}} e^{-[(X-Y)^2 + \sigma^2]/(2\tau^2)}.\end{align}

This probability is maximized for $Y=X$, $\tau=\sigma$, as should be the case. However, if for some reason it is not possible to achieve $X=Y$ -- for example, this can occur when the model requires the same value of $Y$ in two pixels, but their $X$ values are very different -- then $\mathcal{L}$ is maximized for 
\begin{equation}
 \tau^2 \rightarrow \sigma^2 + (X-Y)^2.
\end{equation}
If the true distribution is the large-number limit of a Poisson distribution, we will have $\sigma^2=X$ (mean=variance), and thus
\begin{equation}
 \tau^2 \rightarrow X + (X-Y)^2.
\end{equation}
Thus the effect of $X\ne Y$ is to inflate the reconstructed variance of the model, relative to its expectation value. As discussed above, this is exactly the signature of a PS population with multiple photons per source, where the variance is a factor of $1+s$ larger than the expectation value. Thus we expect that an asymmetry which forces $Y \ne X$ in some pixels will drive a preference for $s > 0$, and we see the preferred value of $s$ will be $s \rightarrow (X-Y)^2/X$.

\subsection{The Effect of Multiple Signal/Background Components} 

Now suppose in addition to the mis-modeled signal component, there is a well-modeled Poissonian background with expectation value $B$ counts (within the same spatial region where the signal expectation is $Y$ counts). As the sum of variables drawn from Gaussian distributions also follows a Gaussian distribution, the model expectation value and variance in this case will be $(B+Y, B+\tau^2)$ respectively. If the true signal normalization (in this region) is $X$ and the background normalization is $B$, the best-fit variance will be given by 
\begin{equation}
 B + \tau^2 \rightarrow B + X + (B + X - (B+Y))^2
\end{equation}
and thus as previously we expect 
\begin{equation}
 \tau^2 \rightarrow X + (X-Y)^2.
\end{equation}
This suggests that the presence of a well-modeled smooth background should not affect the spurious preference for PSs, or their preferred SCF.

In the case of a broader SCF, we expect that the non-Poissonian enhancement to the variance -- played by the factor $1+s$ above -- will generalize to a factor of
\begin{equation}
 1 + \frac{\int ds s^2 f(s) }{ \int ds s f(s)},
\end{equation}
 the preferred value for $s$ discussed above will translate to a preferred value for the integrated quantity 
 \begin{equation}
  \bar{s} \equiv  \frac{ \int ds s^2 f(s)}{\int ds s f(s)}.
 \end{equation}
Suppose the SCF is a power law, 
\begin{equation}
 f(s)=A (s/s_b)^{-\alpha},
\end{equation}
for $s < s_b$, with a sharp cutoff at $s=s_b$; then we would obtain 
\begin{equation}
 \bar{s} = \frac{2-\alpha}{3-\alpha} s_b
\end{equation}
assuming $\alpha < 2$. Thus $\bar{s}$ is mostly controlled by $s_b$ in this case, but there is some degeneracy between $s_b$ and $\alpha$, especially when $\alpha \rightarrow 2$. 

If the signal model also contains a smooth component contributing $s_0$ counts, then the mean will be 
\begin{equation}
 s_0 + \int ds s f(s)
\end{equation}
and the variance 
\begin{equation}
 s_0 + \int ds s f(s) (1+s) \equiv (1+\bar{s}) (s_0 + \int ds s f(s)),
\end{equation}
and thus we expect to find a preference for an effective average number of counts/source of 
\begin{equation}
 \bar{s} = \int s^2 f(s) ds/(s_0 + \int ds s f(s)).
\end{equation}
Under the same sharply-cutoff power-law model for the SCF, and writing the total counts produced by the PS population as $s_\text{PS}$ for convenience, we find  
\begin{equation}
 \bar{s} =   \frac{2-\alpha}{3-\alpha} s_b s_\text{PS}/(s_0 + s_\text{PS}).
\end{equation}
Thus at this level of approximation, there is a degeneracy between the properties of the SCF and the fraction of flux associated with PSs vs smooth emission; a preference for a particular value of $\bar{s}$ can be accommodated by a wide range of values of $s_b$ and $\alpha$, as the fraction of counts in sources vs smooth emission,
\begin{equation}
 s_\text{PS}/(s_0 + s_\text{PS}),
\end{equation}
is varied. Breaking this degeneracy requires going beyond the approximations used in this section, most likely by taking into account the non-Gaussianity of the probability distributions. (In a real analysis, the choice of priors may also play a role in how this degeneracy is broken.) However, if there is a preference for a non-zero $\bar{s}$, there will always be a preference for some PS population; as the fraction of the flux attributed to this source population is decreased, the individual brightness of the required sources will increase.

\section{Spurious Point Source Signals from Only Smooth Templates} 
\label{sec:smooth}

Given these results, we expect that it should be possible to see a preference for spurious PSs even in simpler simulations that include no simulated PSs at all, in contrast to the simulations of Ref.~\cite{PhysRevLett.125.121105} which included the full set of templates used to analyze the real data. For this section, we first perform a fit to the real data using all smooth/Poissonian templates, including separate north and south templates for the GCE, and templates for the isotropic and disk PS populations; the resulting posterior median values for the parameters are provided in Table~\ref{tab:sim_values_one}. However, we then generate realizations that include only the asymmetric smooth GCE and (optionally) the Galactic diffuse emission model; the normalizations of all other components are set to zero.

\begin{figure*}[t]
\leavevmode
\centering
\subfigure{\includegraphics[width=0.42\textwidth]{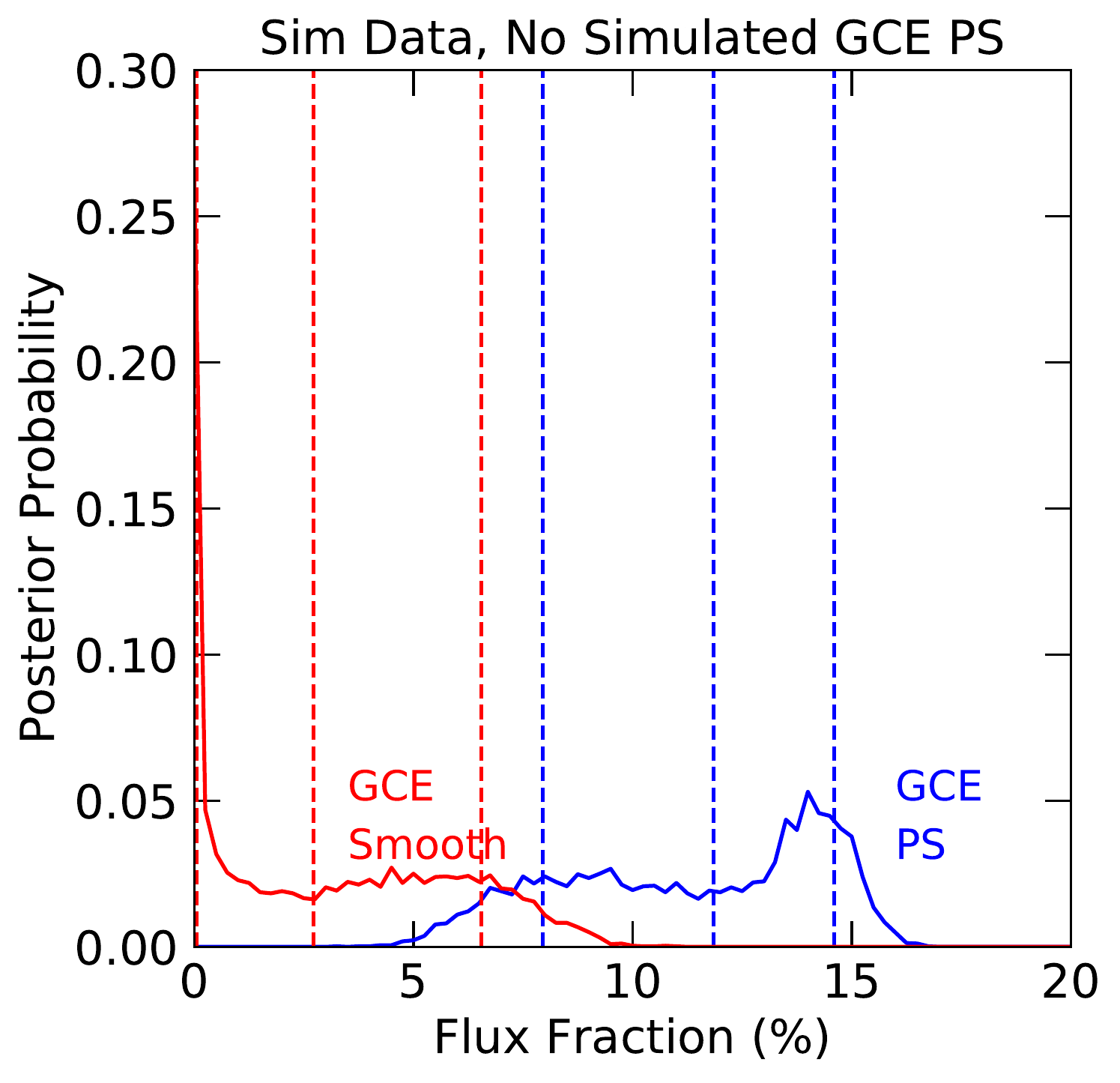}}
\hspace{1mm}
\subfigure{\includegraphics[width=0.4\textwidth]{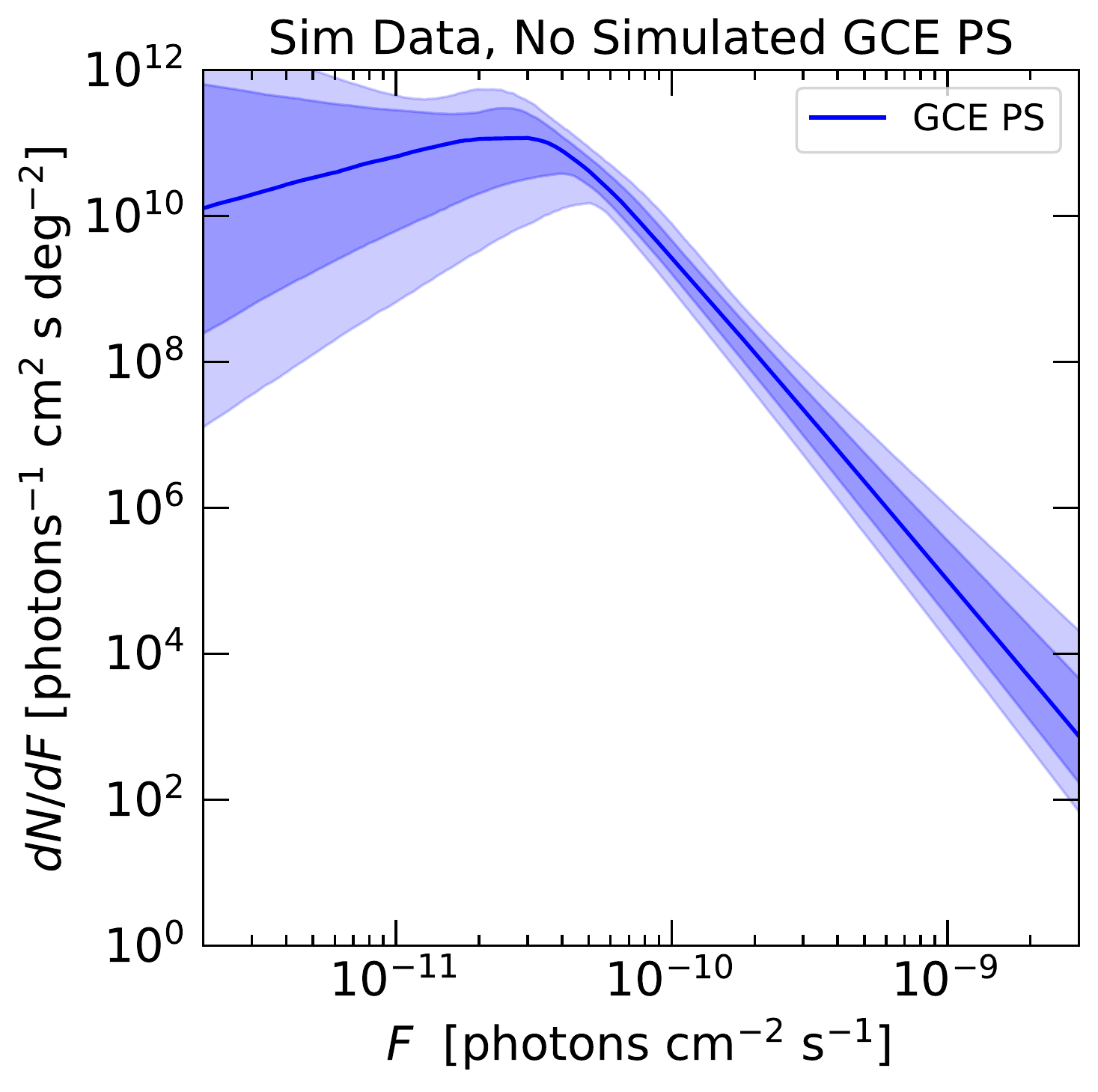}}
\caption{Results of an analysis using single GCE templates (smooth and PS) over the whole ROI, with a simulated dataset containing only smooth GCE and Galactic diffuse emission models (no PS templates are included in the simulation). The smooth GCE template is separated into northern and southern components which have different normalizations. \textbf{Left panel:} flux posteriors; the GCE Smooth component is constrained to have positive coefficient. \textbf{Right panel:} SCF for the (spurious) GCE PS component. Spurious PSs are still found, albeit the peak of the SCF is at lower flux than when the isotropic and disk PSs are simulated.}
\label{fig:10deg_dmdiff}
\end{figure*}

\begin{figure*}[t]
\leavevmode
\centering
\subfigure{\includegraphics[width=0.42\textwidth]{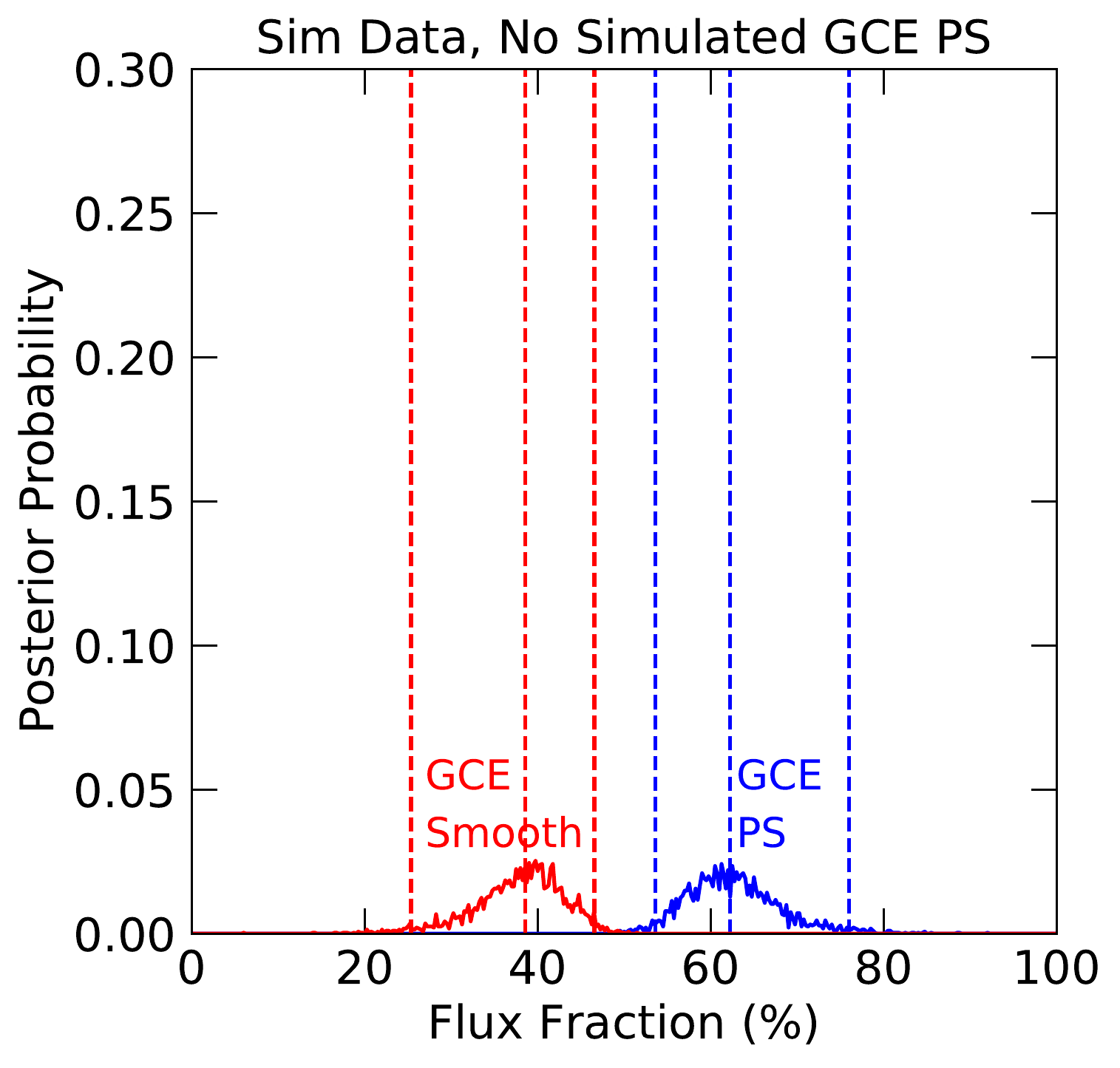}}
\hspace{1mm}
\subfigure{\includegraphics[width=0.4\textwidth]{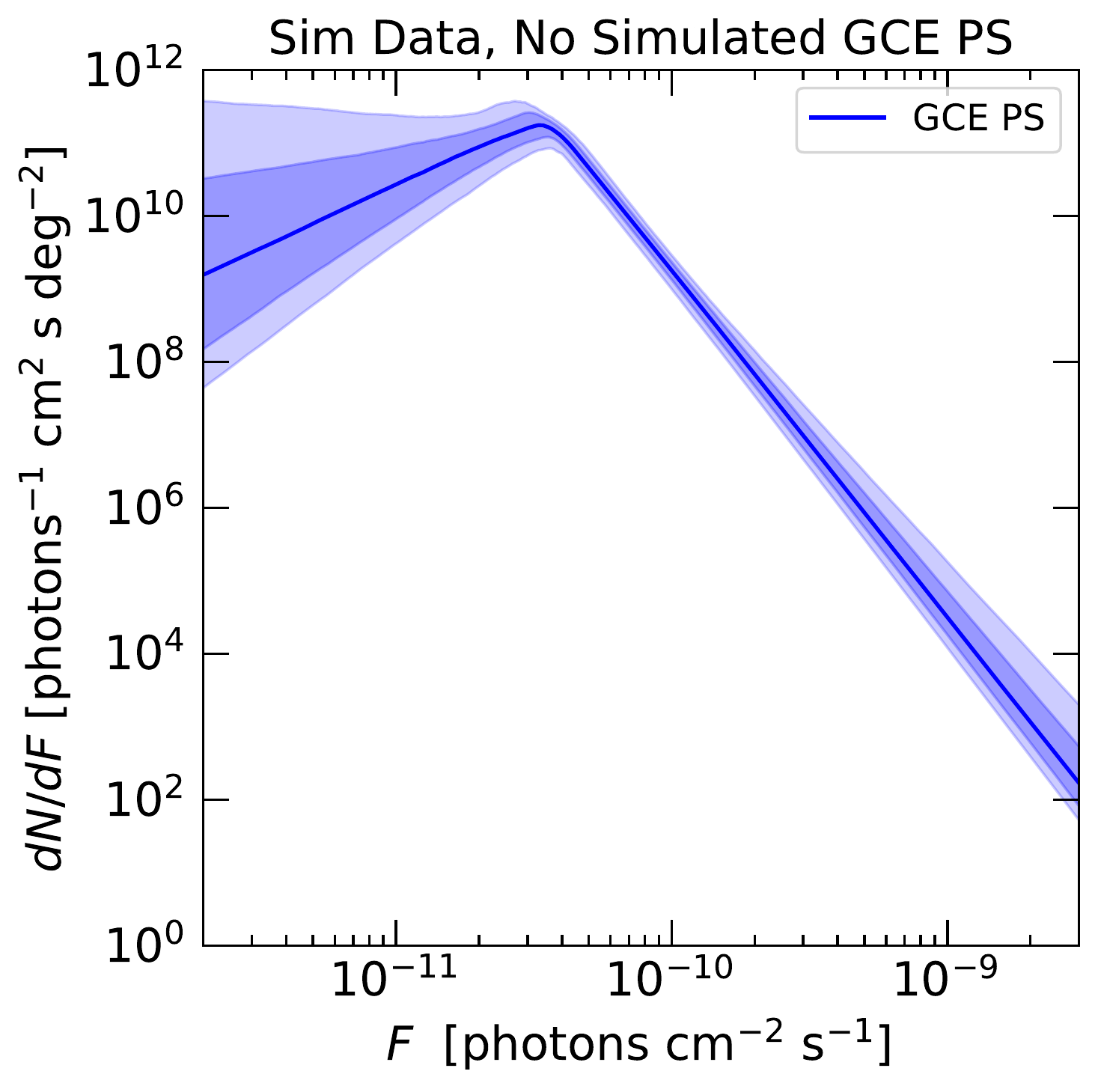}}
\caption{Same as Fig.~\ref{fig:10deg_dmdiff}, but only the smooth north and south GCE templates are included in the simulation; the Galactic diffuse emission is not simulated.}
\label{fig:10deg_dm}
\end{figure*}

\begin{figure*}
\leavevmode
\centering
\subfigure{\includegraphics[width=0.4\textwidth]{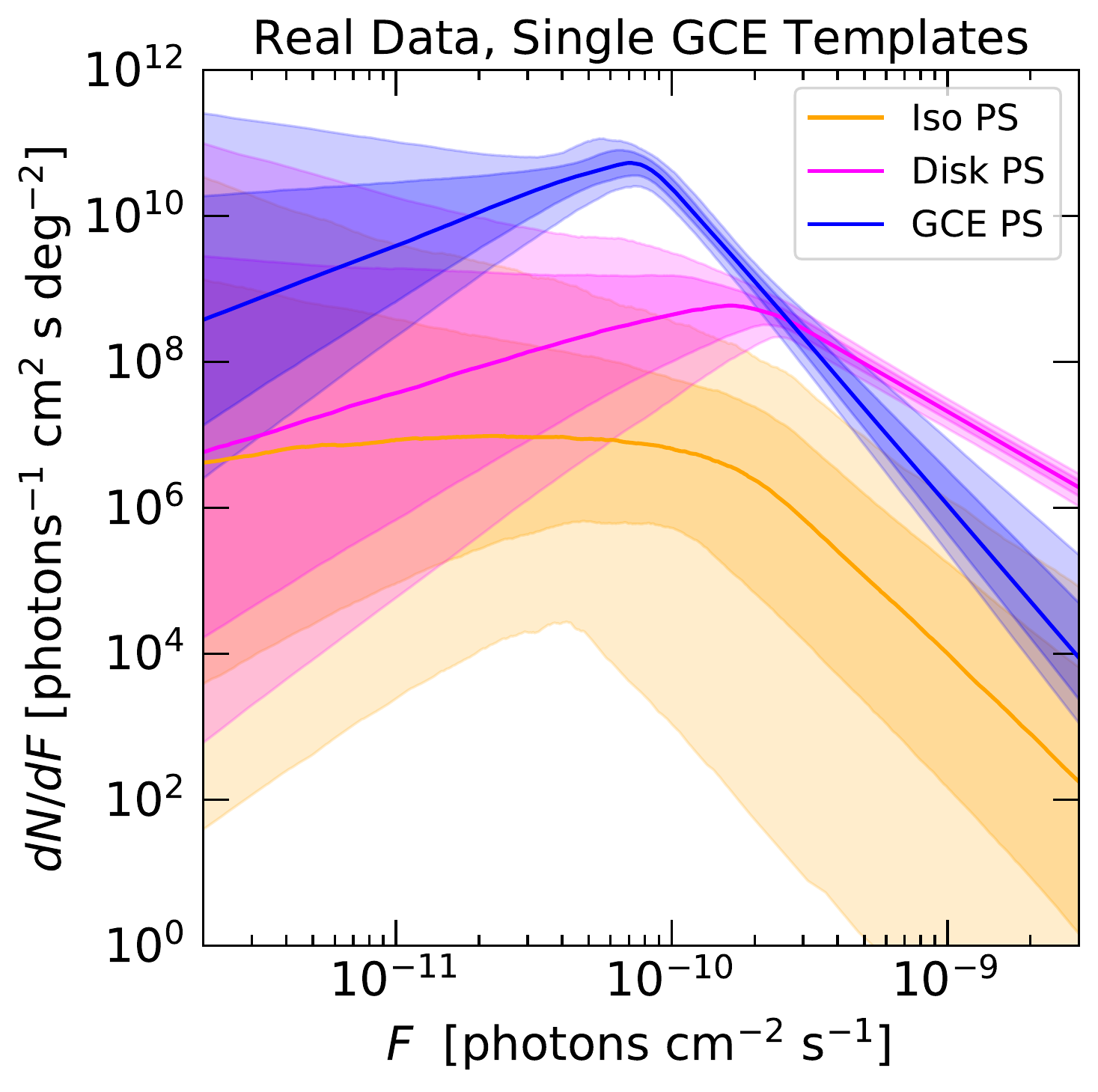}}
\hspace{1mm}
\subfigure{\includegraphics[width=0.4\textwidth]{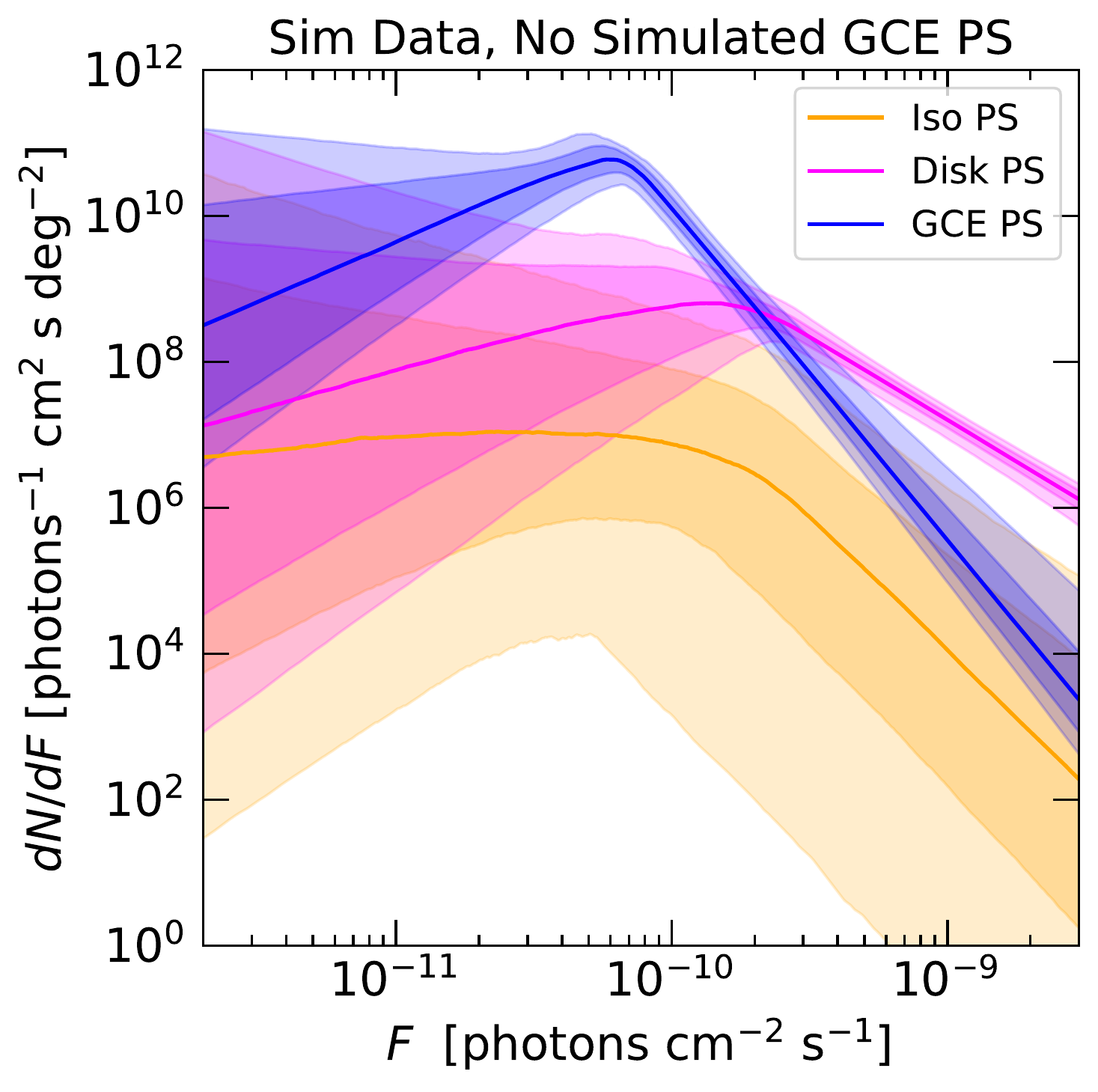}}\\
\subfigure{\includegraphics[width=0.4\textwidth]{Plots/Run_6_onlyDMdif_scf}}
\hspace{1mm}
\subfigure{\includegraphics[width=0.4\textwidth]{Plots/Run_6_onlyDM_scf}}
\caption{SCF comparison across scenarios, for the illustrative realizations previously presented in Figs.~\ref{fig:10deg_dmdiff} and \ref{fig:10deg_dm}, as well as the real data and the baseline mock dataset. All template normalizations in simulated data are based on the posterior median for the fit to the real data with no GCE PS template, but separately floated north and south components for the GCE Smooth template. \textbf{Top-Left:} Real data, analyzed with all background templates (see text) and single symmetric GCE PS and GCE Smooth templates. \textbf{Top-Right:} Simulated data including all background models and north-south-asymmetric smooth GCE (and no GCE PSs). Analyzed with same templates as top-left. \textbf{Bottom-Left:} Simulated Galactic diffuse emission and north-south-asymmetric smooth GCE.  Analyzed with single symmetric GCE PS and GCE Smooth templates, plus the Galactic diffuse emission template. \textbf{Bottom-Right:}  Simulated north-south-asymmetric smooth GCE only. Analyzed with single symmetric GCE PS and GCE Smooth templates.}
\label{fig:scf_compare}
\end{figure*}

\subsection{Simulating a Signal + Diffuse Background}

We first simulate data containing an asymmetric smooth GCE and the \texttt{p6v11} diffuse background model, and analyze 25 resulting realizations with a symmetric GCE Smooth template, a symmetric GCE PS template, and the \texttt{p6v11} template. 

Figure~\ref{fig:10deg_dmdiff} shows the results for an illustrative realization of this scenario. Even though there are zero PSs (GCE-correlated or otherwise) in the simulated data, spurious GCE PS populations are reconstructed for all of the 25 realizations. The BFs for GCE PSs range from $\sim10^{16}-10^{33}$. The realization shown in Fig.~\ref{fig:10deg_dmdiff} is fairly typical and has a BF $\sim10^{26}$ in favor of GCE PSs. The break in the reconstructed SCF corresponds to $11.4_{-3.5}^{+5.6}$ photons/source (or a flux of $\sim 3-4\times 10^{-11}$ photons/cm$^2$/s). Across 25 realizations, the break varies between $\sim2-7\times 10^{-11}$ photons/cm$^2$/s. This cutoff value is mildly lower than observed in simulations including the full range of templates, described in Ref.~\cite{PhysRevLett.125.121105}; this could indicate that more complicated fits drive the peak of the SCF for the spurious PSs to higher values, or alternatively that when simulated disk or isotropic PSs are present, they may be mis-allocated to the spurious source population. The latter mechanism does appear to occur in the baseline mock dataset of Ref.~\cite{PhysRevLett.125.121105}, as the reconstructed flux assigned to the disk PSs is considerably lower than its simulated value; likewise, in the real data, the reconstructed flux in disk PSs is higher when the GCE is allowed to be asymmetric.  This behavior contrasts with the ``Bubbles PSs'' scenario explored in Ref.~\cite{Leane:2019xiy} in that the driver of the mis-allocation would be errors in the smooth emission templates, not a mismodeled novel PS distribution.

\subsection{Signal Only Simulations (Background Free)}

We can go beyond the case with only smooth templates and consider the extreme scenario where we simulate no backgrounds at all, only a north-south-asymmetric smooth GCE, which is analyzed with symmetric GCE Smooth and GCE PS templates. 

Figure~\ref{fig:10deg_dm} shows the results for an illustrative realization; again a spurious point source population is reconstructed, and the SCF in this case has a cutoff corresponding to $10.1_{-1.3}^{+1.4}$ photons/source. In this case, over 25 simulations, the BF in favor of including a PS template is always huge ($\sim10^{224}-10^{306}$). We will discuss why the BF is so large in the zero-background case later in this section, via a quantitative comparison to the results of Section~\ref{sec:analytic}.

\subsection{Source Count Function Comparisons}

Figure~\ref{fig:scf_compare} summarizes the reconstructed SCFs for the realizations we have previously plotted and the baseline mock dataset taken from Ref.~\cite{PhysRevLett.125.121105}, as well as the real data.
In all cases the (simulated or real) data are analyzed with a pipeline containing a single set of north-south-symmetric GCE templates; the other templates match those used for the simulation, or in the case of the real data, they match the templates used to simulate the baseline mock dataset. We see that failing to allow enough freedom to model the asymmetry leads to spurious PS populations, containing sources with photons/source well above the degeneracy limit, mimicking the apparent detection of PSs in the real data. The flux corresponding to the peak of the SCF depends on which sources of background are simulated, but is $\sim 10$ photons/source even when no background templates are simulated at all.

\subsection{Comparison with Analytic Results from Sec.~\ref{sec:analytic}}

Suppose there is a 2:1 asymmetry in expectation value for number of counts/pixel, between the northern and southern hemispheres of a Poissonian signal (i.e. $X_\text{north} = 2 X_\text{south}$); this is approximately the level of asymmetry found in the real data by Ref.~\cite{PhysRevLett.125.121105}. Suppose further that the model we are fitting to the data assumes north/south symmetry (expectation value $Y$ independent of whether the pixel is in the north or south). The overall best-fit choice of $Y$ is expected to be approximately the mean of the northern and southern values, $Y \approx 1.5 X_\text{south}$. Thus the fit will prefer $\tau^2 \approx \sigma^2 + 0.25 X_\text{south}^2$; since the true distribution is Poissonian, $\sigma^2 = X$, and so we expect on average $\tau^2 \rightarrow 1.5 X_\text{south} + 0.25 X_\text{south}^2 \approx 1.5 X_\text{south}  (1 + 0.17 X_\text{south})$, thus suggesting a preferred value for the averaged number of counts/source of $\bar{s} \approx 0.17 X_\text{south}$.

Within the $10^\circ$ region of interest and with the plane masked for $|b|<2^\circ$, the average number of counts per pixel in the southern half of the ROI in the simulation of Fig.~\ref{fig:10deg_dm} is roughly 10. Thus we would expect a preferred number of counts/source of $\bar{s} \approx 2$. A more careful analysis, computing the estimated likelihood in each pixel as a function of $s$ within the approximations of this section (assuming a delta-function SCF) and taking the product of pixel likelihoods, finds that the best fit is obtained for $\bar{s}=3$, and that this corresponds to an improvement in the likelihood relative to $\bar{s}=0$ of $\Delta \ln \mathcal{L} \approx 700$, or a likelihood ratio of $\sim 10^{300}$.

In the analysis depicted in Fig.~\ref{fig:10deg_dm}, the BF in favor of a PS population is always very large, $\sim10^{224}-10^{306}$; we see from the analytic estimate that this is to be expected in the zero-background case. (Note that the faintest pixels in the region of interest have an expected number of photon counts of $\sim 4$, so we do not expect the Gaussian approximation to be especially accurate for these pixels.) The break in the preferred SCF corresponds to $s_b=10.1_{-1.3}^{+1.4}$, and the preferred slope is $\alpha = -1.75_{-0.9}^{+1.2}$.
The peak of the posterior attributes roughly 40\% of the counts to a smooth component and 60\% to PSs; by our reasoning above, this corresponds to a preference for $\bar{s} \approx 0.8 \times 0.6 \times 10 \approx 5$ in the Gaussian approximation, which is similar to the preference for $\bar{s} \approx 3$ we predicted from our analytic estimates. We note that a similar SCF is observed in Fig.~\ref{fig:10deg_dmdiff}, where a template for the diffuse background is included in both the simulations and the fit; within our approximations, we would not expect a large difference in the preferred value of $\bar{s}$ between the cases with and without a (well-known) diffuse model. Of course, mismodeling in the diffuse background could contribute to the discrepancy that sources the preference for non-zero $\bar{s}$, and increase the preferred value of $\bar{s}$. 

It is beyond the scope of this simplified analysis to explore the impact of adding additional PS populations with a different morphology to the signal, although from the simulations it seems that the presence of such PS populations has the potential to increase the value of $s$ corresponding to the peak in the SCF.

\section{Point Sources vs Asymmetry under Variations of the Analysis of Ref.~\cite{PhysRevLett.125.121105}}
\label{sec:variations}

Now that we have demonstrated both analytically and via simulations that an unaccounted-for asymmetry can generate a spurious preference for GCE PSs, we will explore the circumstances under which this situation is realized in the \textit{Fermi} gamma-ray data. In the companion paper \cite{PhysRevLett.125.121105} we demonstrated that in the $10^\circ$ radius ROI and using the \text{p6v11} model with the top three quartiles of data by angular resolution, there is an apparent strong preference for GCE PSs (BF $\sim 10^{15}$) that becomes insignificant (BF $<10$) when the GCE is allowed to be north-south asymmetric. In this section we will consider the impact of alternate signal and background models, different photon selection, variations to the SCF priors, and larger ROIs.

\subsection{Analyses in the 10$^\circ$ ROI with Alternate Diffuse Models} 

\begin{figure}[t]
\leavevmode
\centering
\includegraphics[width=0.41\textwidth]{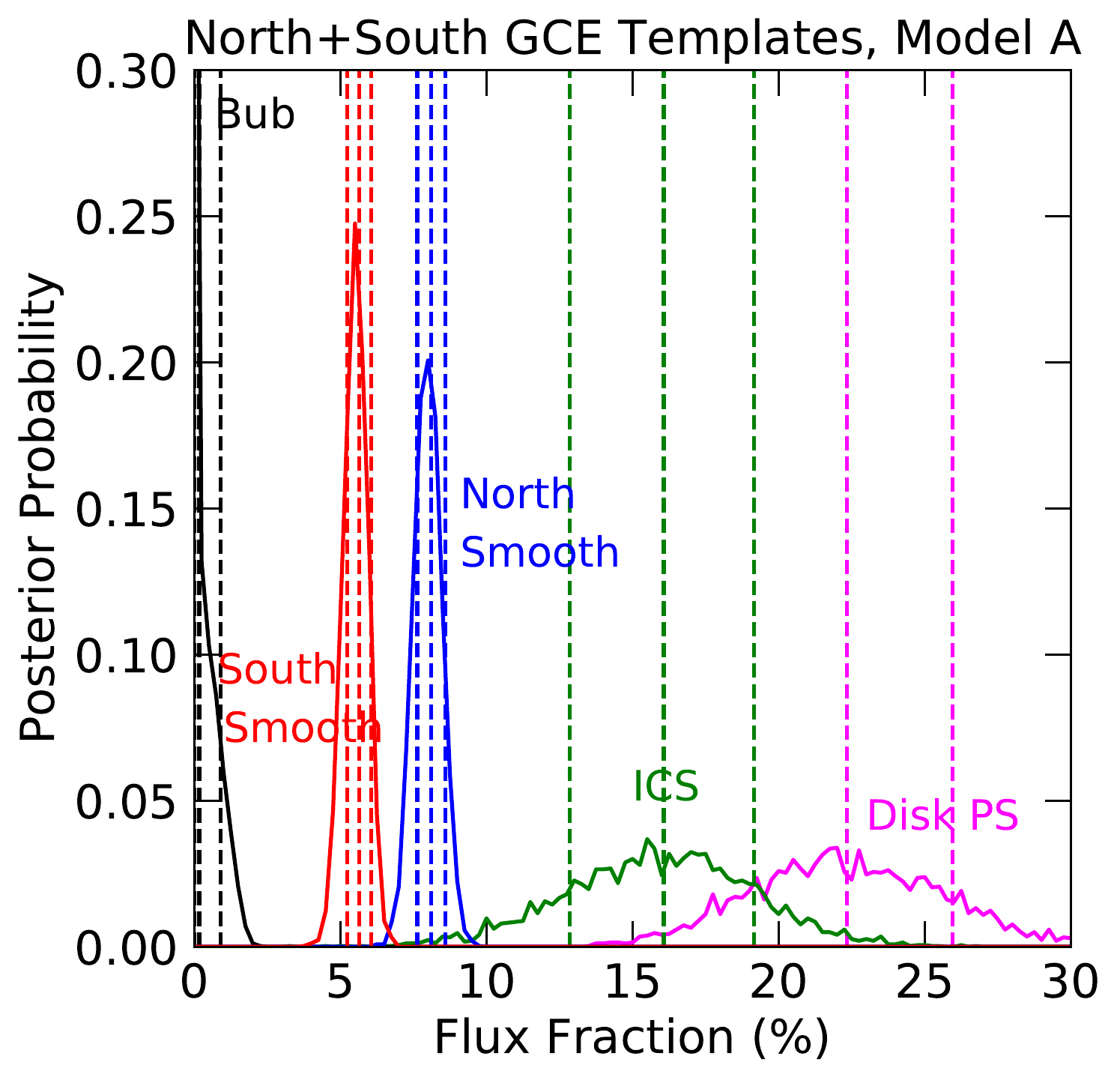} \hspace{3mm}
\includegraphics[width=0.41\textwidth]{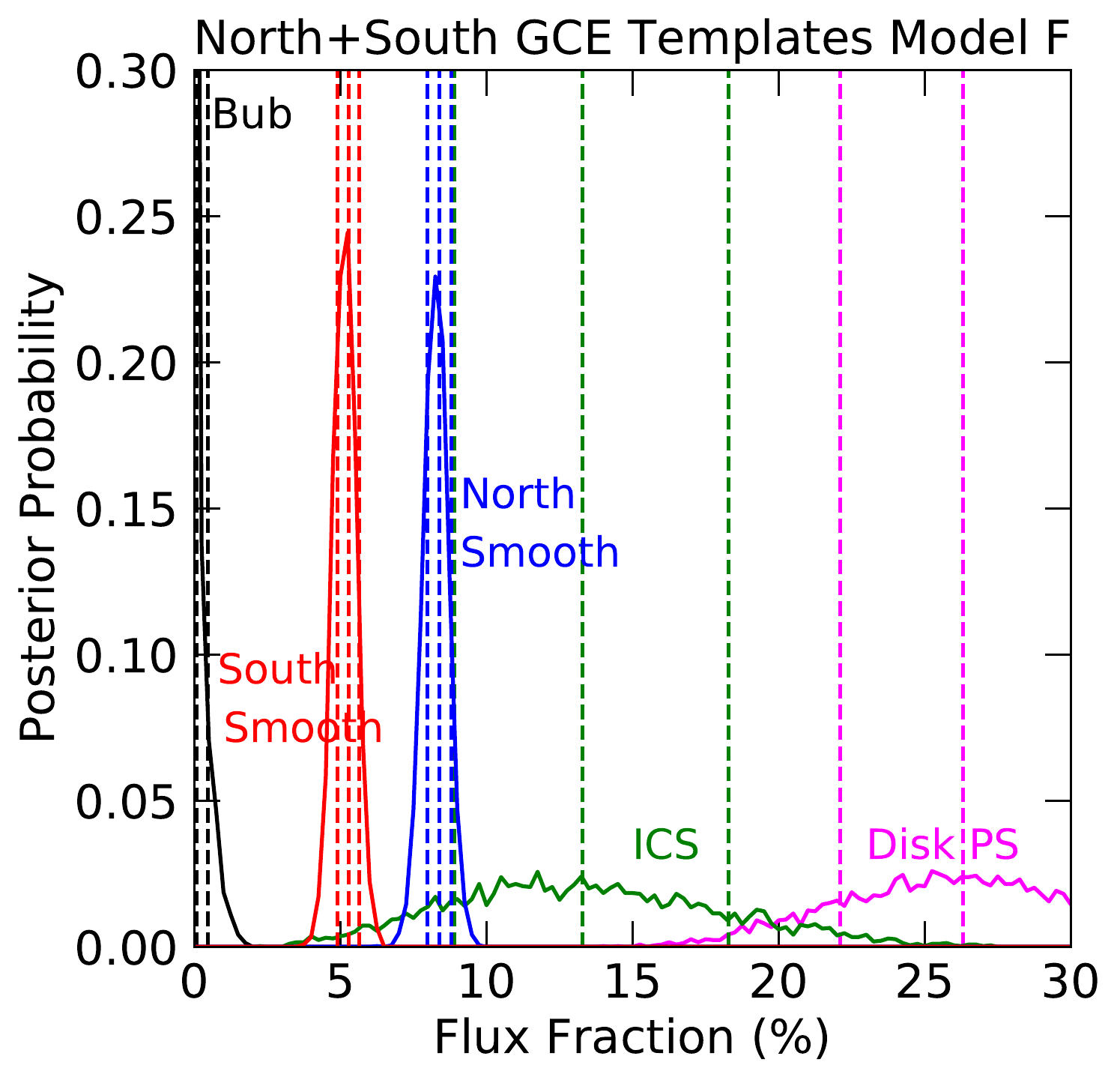}
\caption{Selected flux posteriors demonstrating the impact of providing a smooth GCE template with additional freedom, in the real data, with diffuse model \texttt{Model A} (\textit{left}) and \texttt{Model F} (\textit{right}). No GCE PSs are included in the fit; the smooth GCE template is divided into independent north and south regions. The GCE in the northern hemisphere is found to be nearly twice as bright as the south with diffuse models \texttt{Model A} and \texttt{Model F}.}
\label{fig:10degasymAF}
\end{figure}

\begin{figure*}[t]
\leavevmode
\centering
\subfigure{\includegraphics[width=0.33\textwidth]{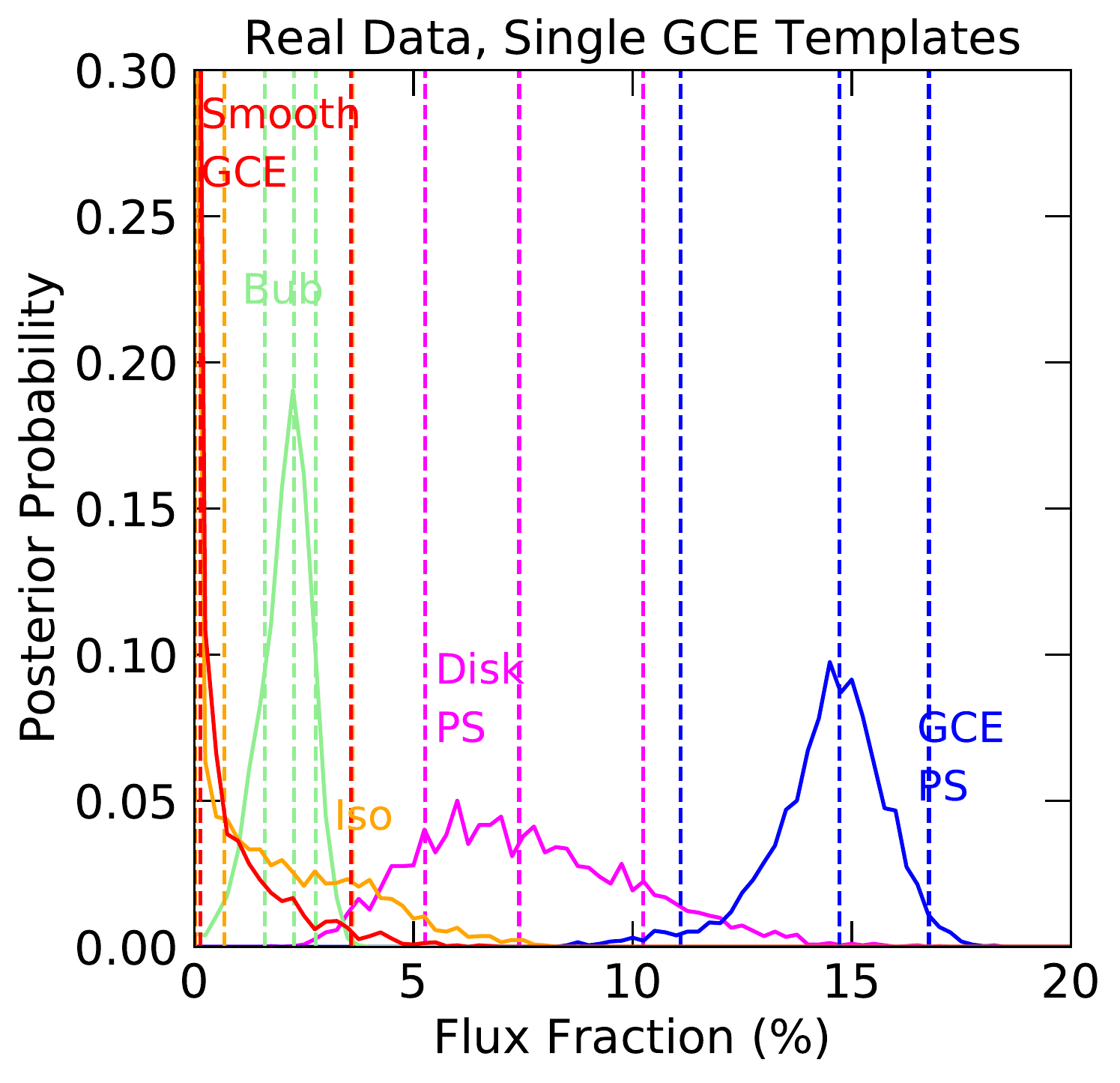}}
\subfigure{\includegraphics[width=0.33\textwidth]{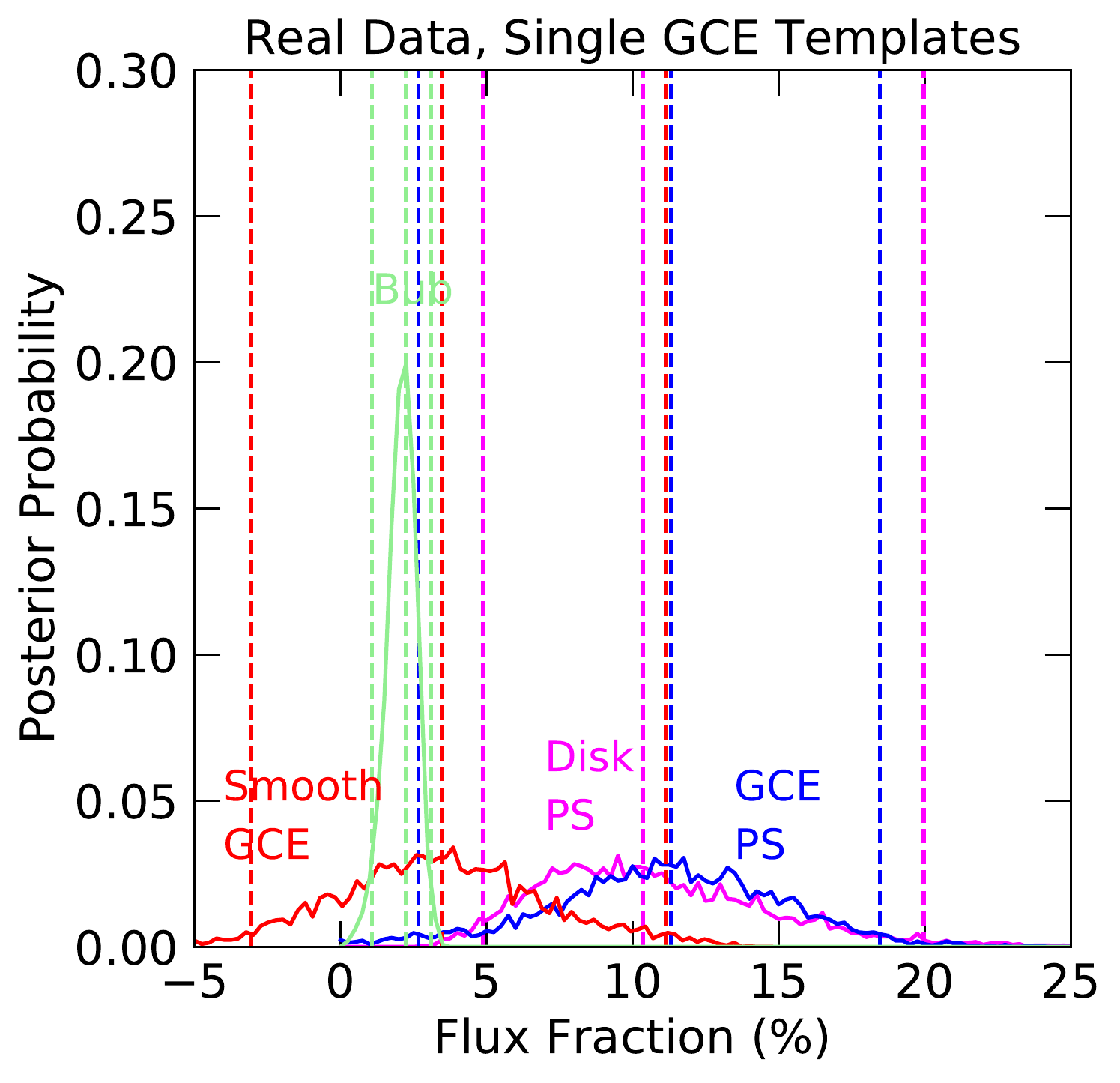}}
\subfigure{\includegraphics[width=0.32\textwidth]{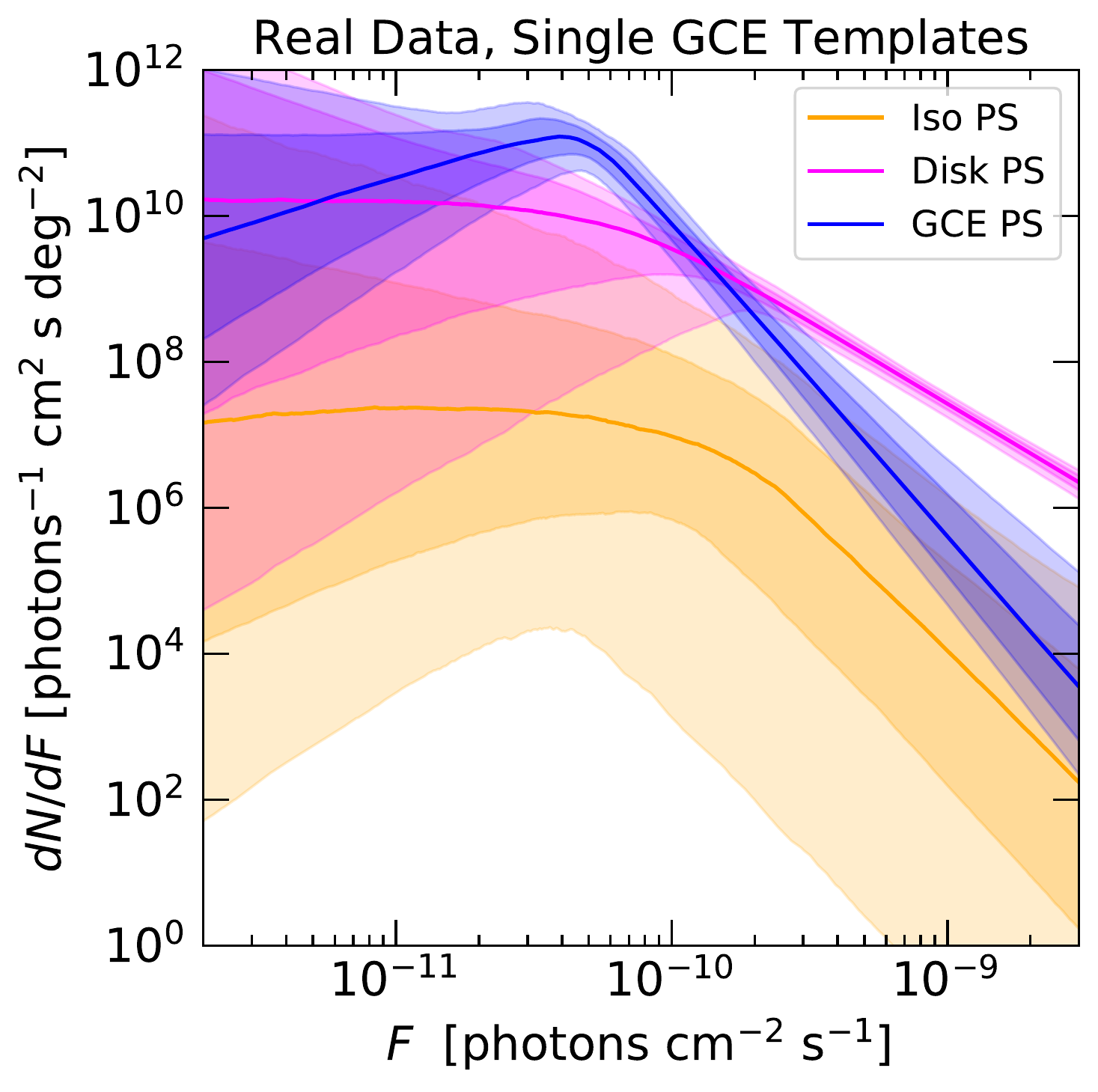}}\\
\subfigure{\includegraphics[width=0.33\textwidth]{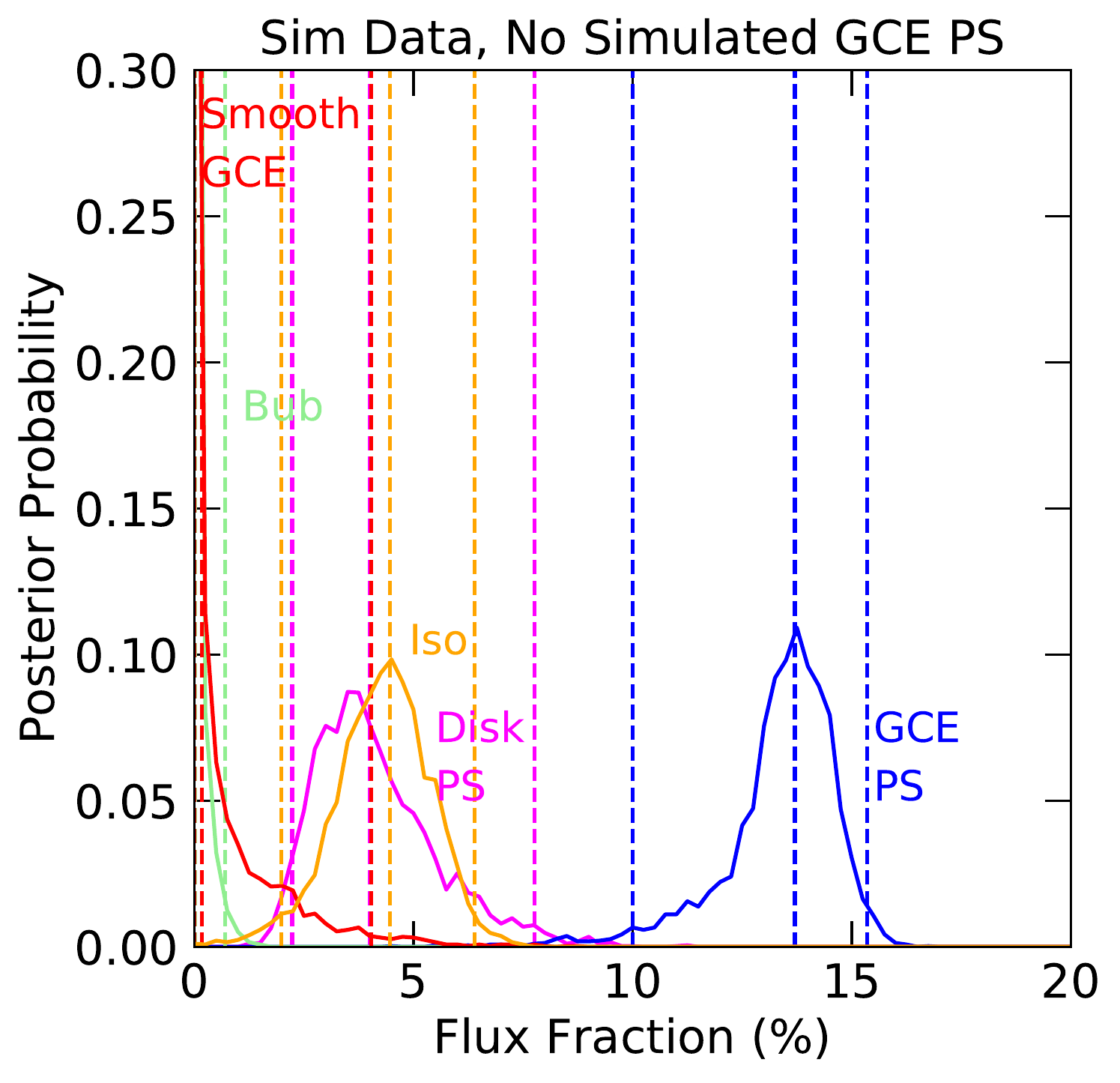}}
\subfigure{\includegraphics[width=0.33\textwidth]{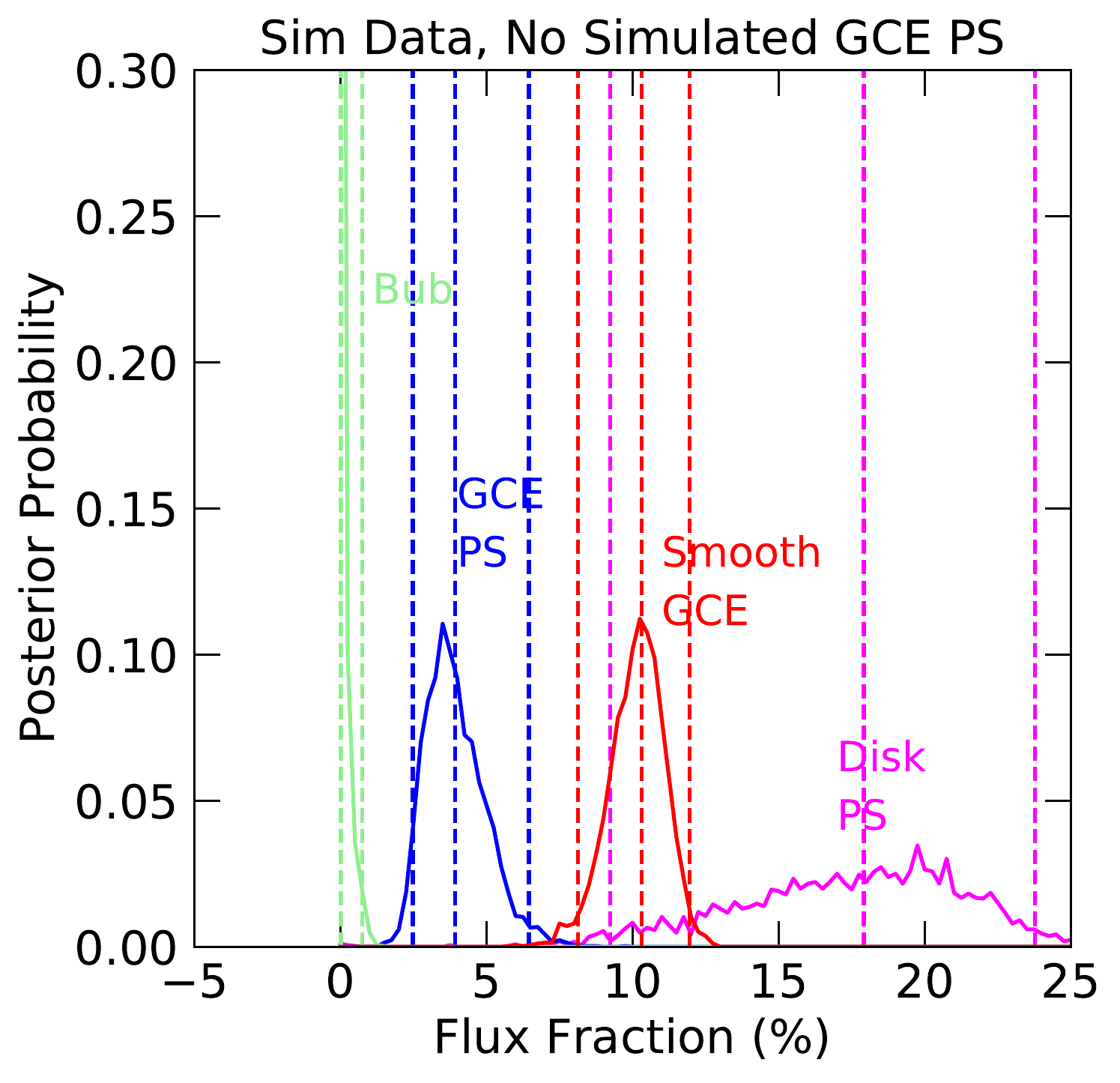}}
\subfigure{\includegraphics[width=0.32\textwidth]{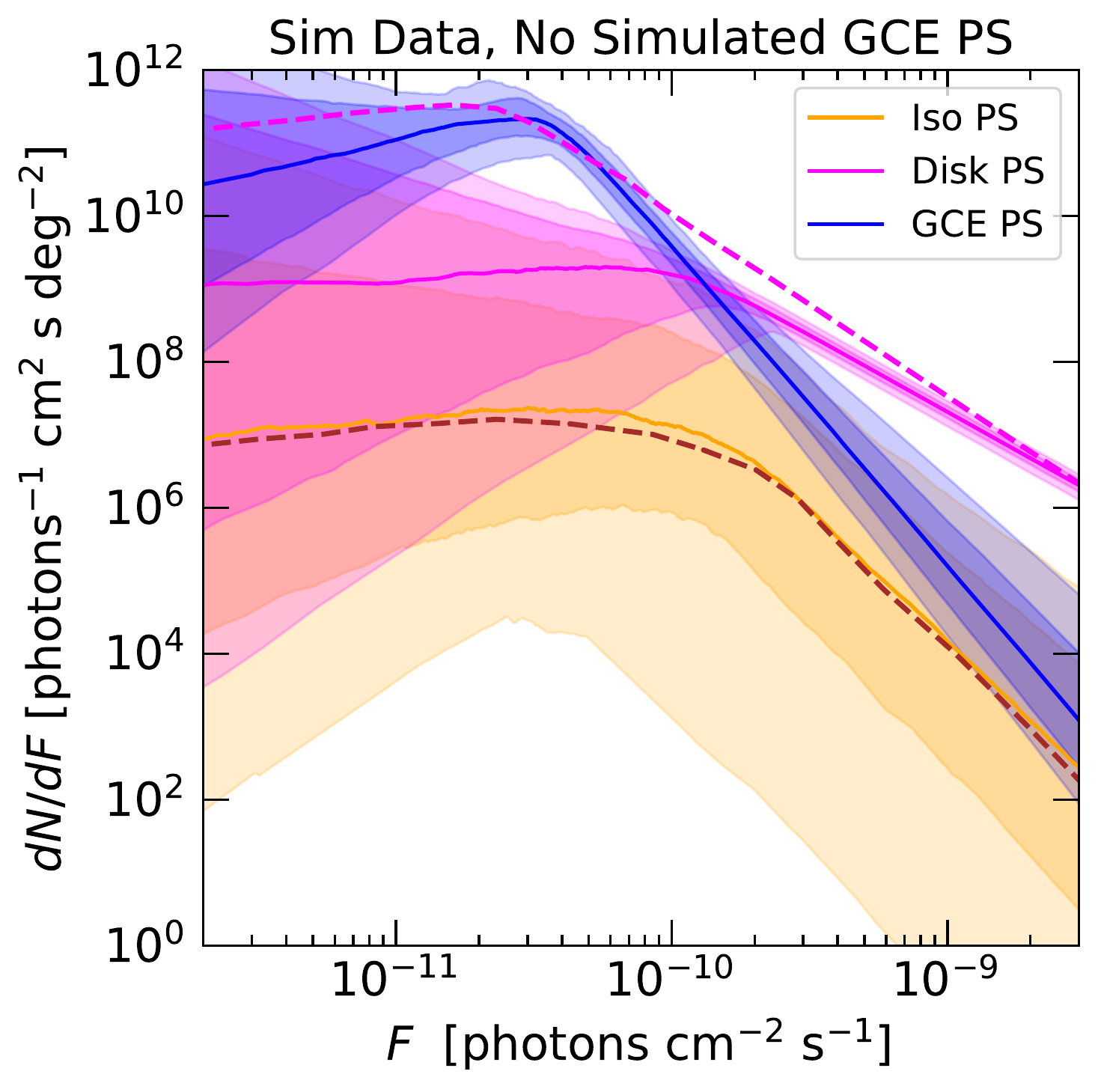}}\\
\caption{Results using diffuse model \texttt{Model A}. Comparison of real (\textit{top row}) and simulated (\textit{bottom row}) data; in all cases the analyses used symmetric GCE templates (smooth and PS). The simulated dataset is based on the posterior medians from the fit to the real data (fluxes shown in the left panel of Fig.~\ref{fig:10degasymAF}) using separate Poissonian templates for the northern and southern GCE; no GCE PSs were simulated. The Bayes factors in favor of PSs in the real and simulated data are both $\sim 4\times10^{2}$.
\textbf{Left column:} Flux posteriors for various templates in the fit where the GCE Smooth component is constrained to have positive coefficient. \textbf{Middle column:} Flux posteriors for various templates in the fit where the GCE Smooth component is allowed to float to negative values. \textbf{Right column:} SCF corresponding to the left column. The dashed lines show the simulated disk SCF (pink) and simulated Iso PS SCF (brown).}
\label{fig:10degA}
\end{figure*}

\begin{figure*}[t]
\leavevmode
\centering
\subfigure{\includegraphics[width=0.35\textwidth]{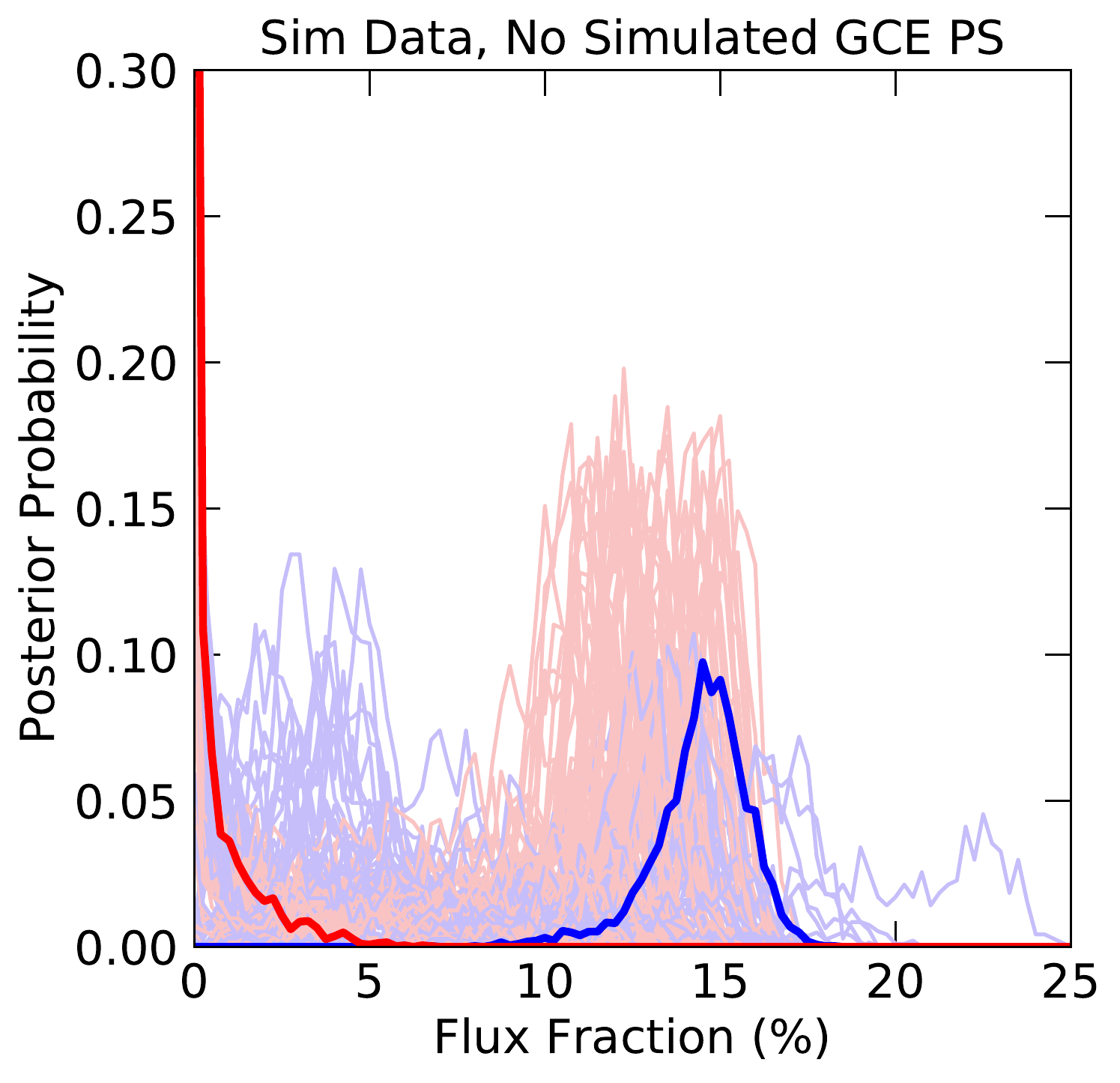}}
\subfigure{\includegraphics[width=0.34\textwidth]{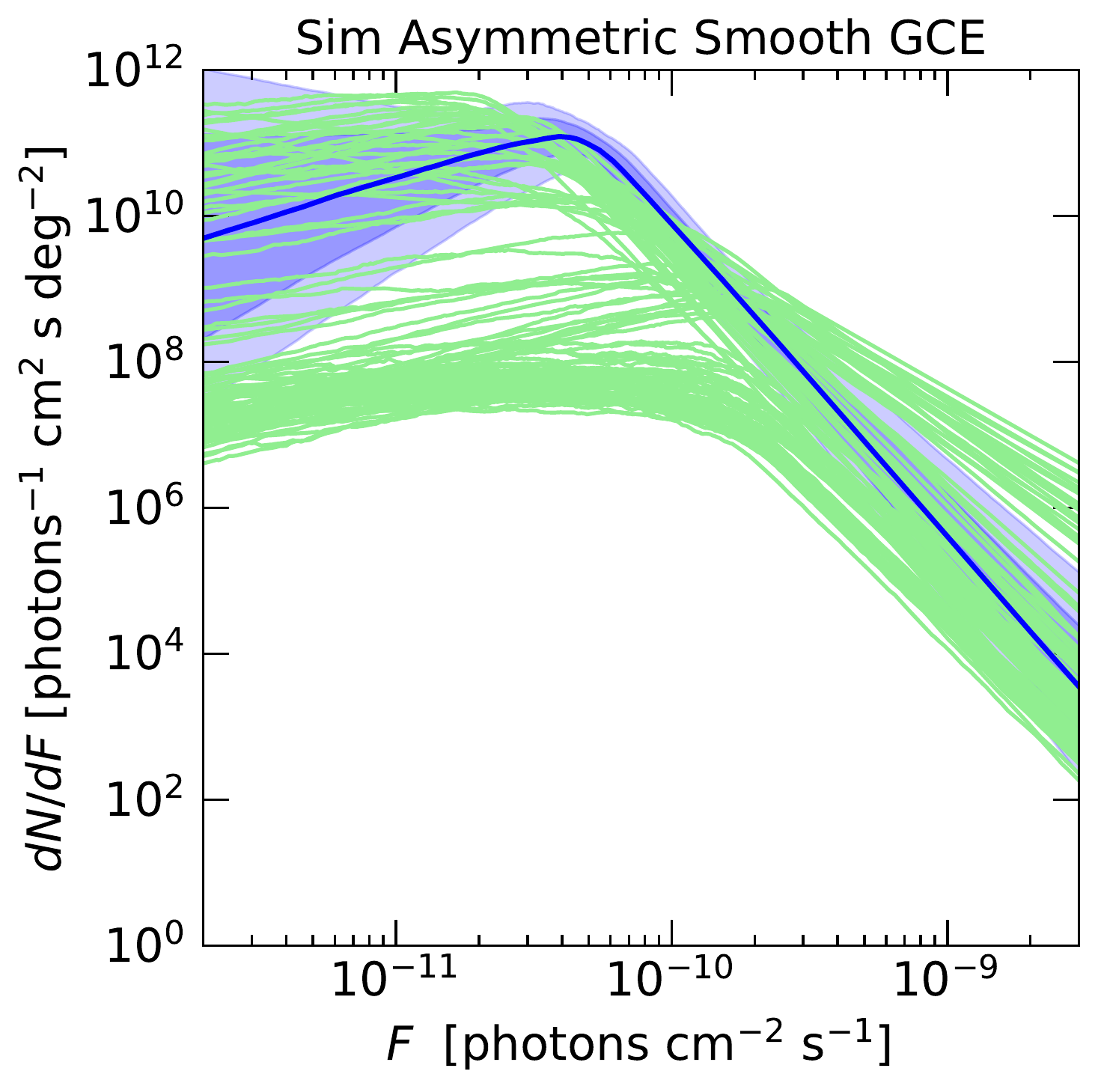}}
\caption{Spread of analysis results over 100 simulated-data realizations (for diffuse model \texttt{Model A}), extending Fig.~\ref{fig:10degA}. In all cases, the analyses use symmetric GCE templates (PS and Smooth). The simulated dataset is based on the posterior medians (fluxes shown in left panel of Fig.~\ref{fig:10degasymAF}) from the fit to real data using separate GCE Smooth templates for the northern and southern GCE; no GCE PSs were simulated. \textbf{Left:} Flux fraction posteriors.  Fainter blue (red) lines correspond to the GCE PS (GCE Smooth) posteriors for simulated realizations, bold darker lines are the real data. \textbf{Right:} the SCF obtained in the real data using one symmetric GCE PS template is shown in blue, the posterior median values of the reconstructed SCFs for GCE PSs, in the simulations, are shown in green.}
\label{fig:10degAspread}
\end{figure*}

\begin{figure*}[t]
\leavevmode
\centering
\subfigure{\includegraphics[width=0.39\textwidth]{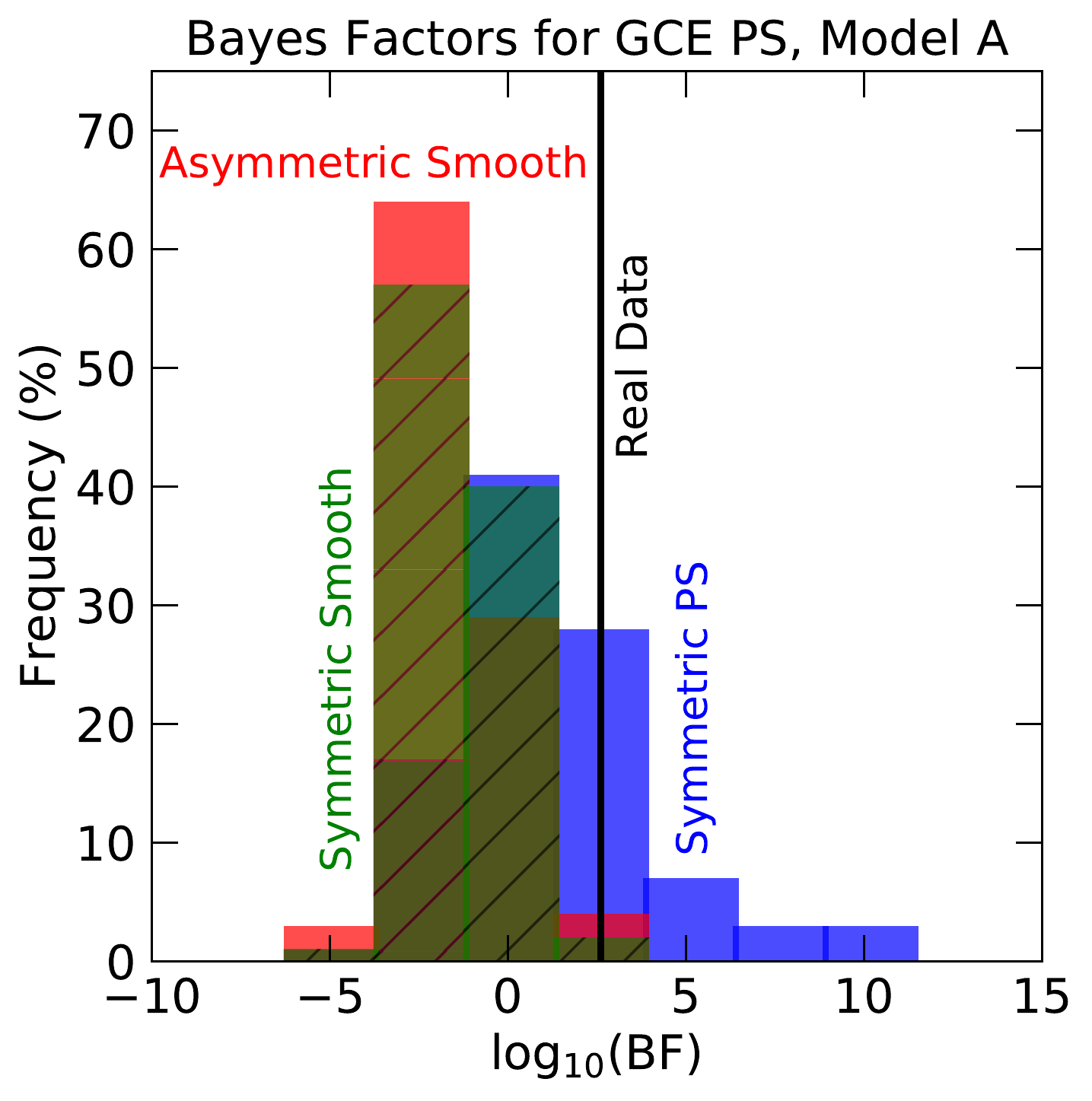}}\\
\subfigure{\includegraphics[width=0.32\textwidth]{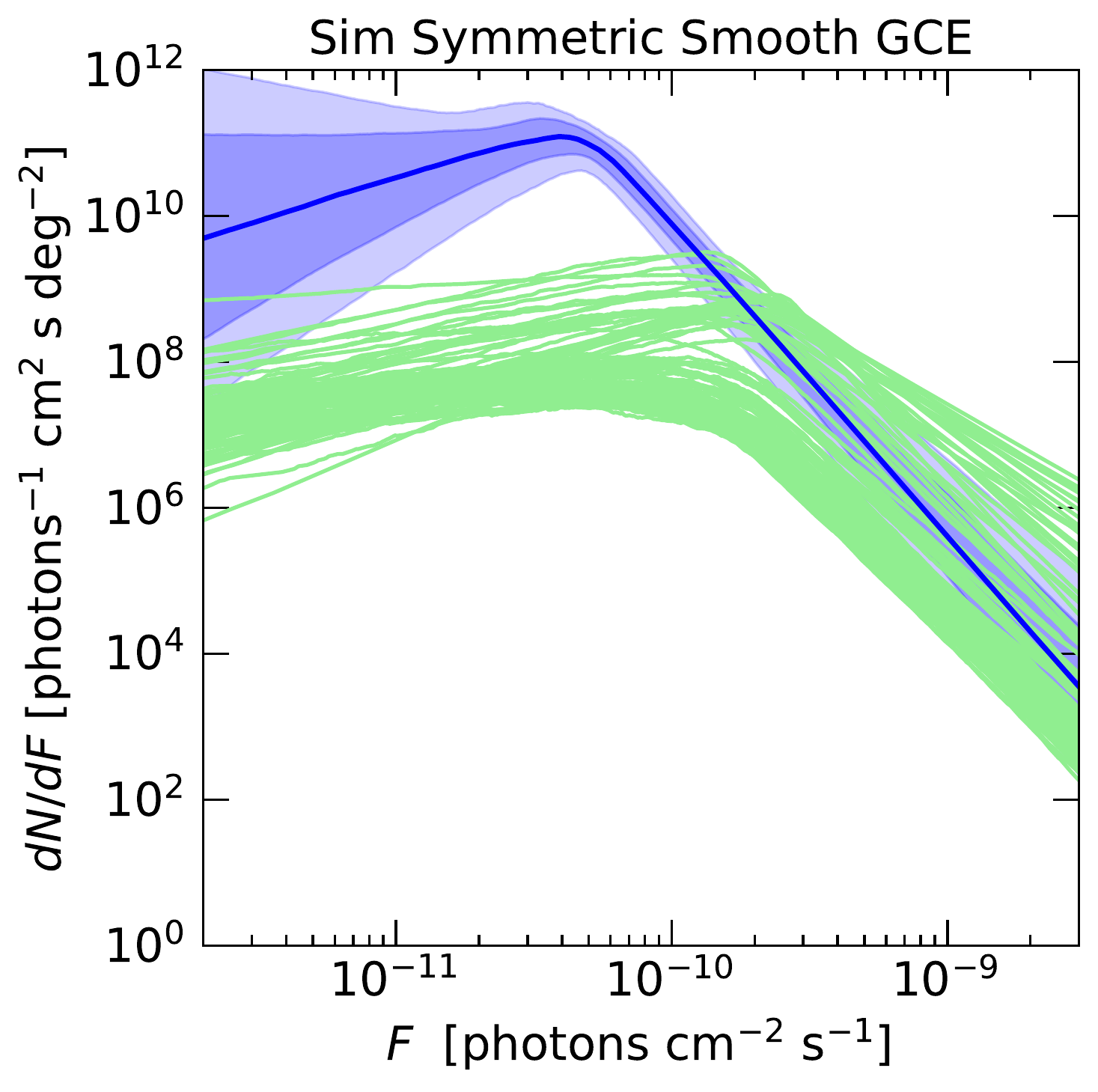}}
\subfigure{\includegraphics[width=0.32\textwidth]{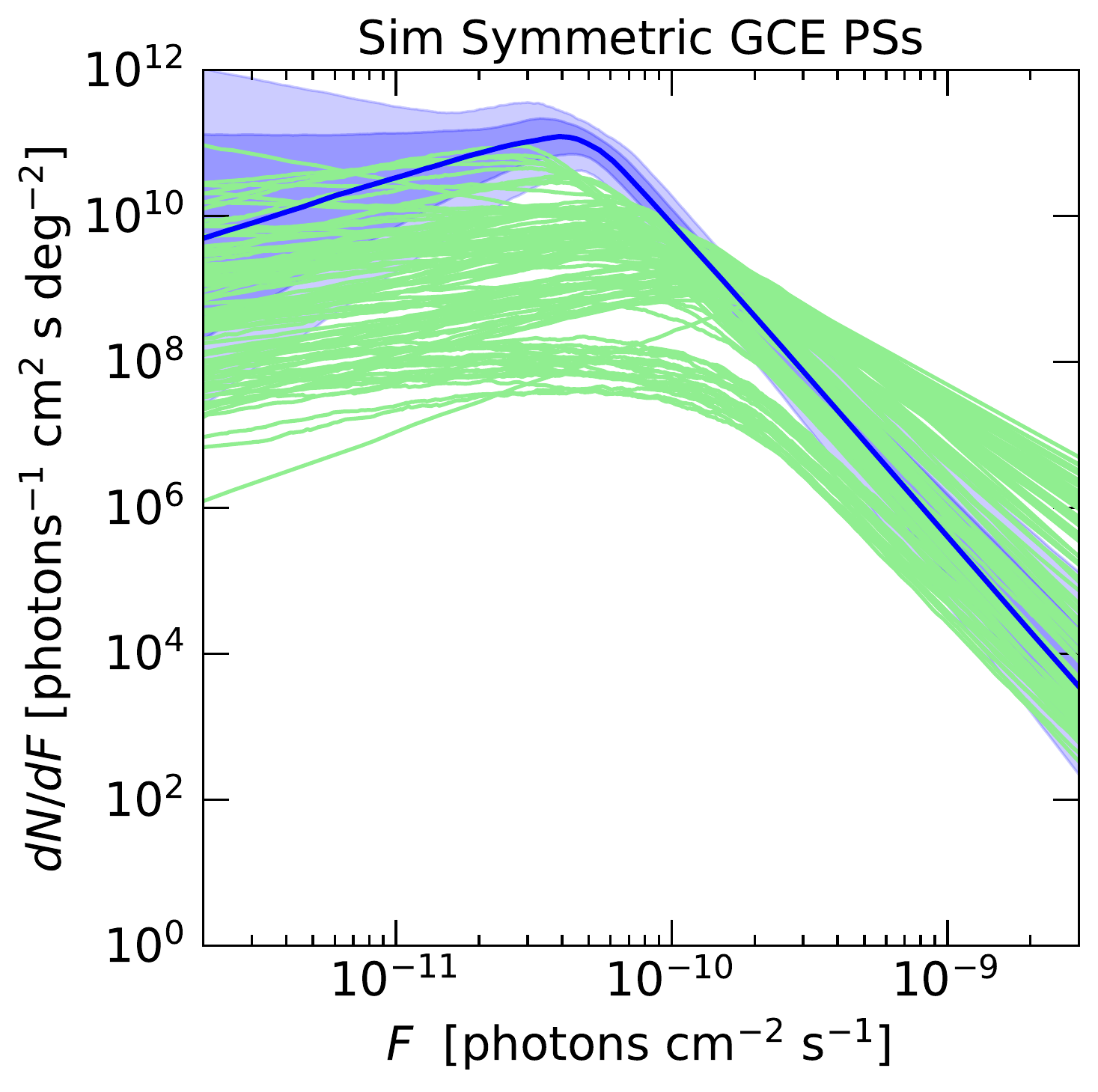}}
\subfigure{\includegraphics[width=0.32\textwidth]{Plots/Run_6a_scf_spread}}
\caption{Spread of analysis results over 100 simulated-data realizations (for diffuse model \texttt{Model A}). \textbf{Top:} Histogram of $\log_{10}$(BF) for each of three simulated scenarios: (1) where the GCE is $100\%$ smooth and symmetric (parameters based on analysis of real data with no GCE PSs) (2) where the GCE is $100\%$ PSs (parameters based on analysis of real data including both GCE Smooth and GCE PS templates), and (3) where the GCE is smooth and asymmetric (parameters based on analysis of real data, with only a GCE Smooth template, subdivided into independent north and south components). In all cases, the analyses use symmetric GCE templates (PS and Smooth). \textbf{Bottom:} The SCF obtained based on the three simulated scenarios shown in the BF plot. The SCF obtained in the real data using one symmetric GCE PS template is shown in blue, the posterior median values of the reconstructed SCFs for GCE PSs, in the simulations, are shown in green.}
\label{fig:10degAspread2}
\end{figure*}

In this subsection we show results for the 10$^\circ$ radius ROI used in Ref.~\cite{PhysRevLett.125.121105}, but using the diffuse models \texttt{Model A} and \texttt{Model F} instead of \texttt{p6v11}. These models provide a better fit to the data at high energies than \texttt{p6v11}, and have additional freedom, as they have separate templates for gas-correlated emission and gamma rays from inverse Compton scattering (ICS).

Figure~\ref{fig:10degasymAF} shows the preference for flux asymmetry that appears when the Galactic diffuse emission is modeled with \texttt{Model A} and \texttt{Model F}, and the GCE is assumed to be smooth but allowed to be asymmetric. In this figure, no GCE PSs have been simulated -- this is the analogue of Fig.~3 in Ref.~\cite{PhysRevLett.125.121105}.

We found in the companion paper \cite{PhysRevLett.125.121105} that \texttt{Model F} does not prefer any GCE PSs in the real data in this ROI. Consequently, despite the preference for asymmetry, we would not expect detectable spurious PSs to be generated in the scenario with an asymmetric smooth component matching the best fit to real data. We test this explicitly by generating 30 realizations based on the posterior median parameter values from a fit to real data, where the GCE is assumed to be smooth but allowed to be asymmetric (fluxes shown in the right panel of Fig.~\ref{fig:10degasymAF}), and fitting the simulated data with symmetric GCE PS and GCE Smooth templates, along with background templates. As expected, we find in simulations that the realizations generally do not find PSs, with the BF in favor of GCE PSs varying from $10^{-6}$ to $10^{2}$, and the BF being $\sim1$ or less in the majority of realizations. 

In contrast, the analysis with \texttt{Model A} in Ref.~\cite{PhysRevLett.125.121105} found modest evidence in favor of GCE PSs in the real data (BF $\sim 400$). We simulate 100 realizations based on the posterior median parameter values from a fit to the real data, in an analysis where the GCE is assumed to be smooth but allowed to be asymmetric (fluxes shown in the left panel of Fig.~\ref{fig:10degasymAF}). We then fit these simulated datasets using symmetric GCE PS and GCE Smooth templates, along with background templates. 

Figure~\ref{fig:10degA} shows the results for a selected realization with BF $\sim 400$ in favor of GCE PSs, demonstrating that spurious GCE PSs can be created in simulations with \texttt{Model A} as well, with BF, SCF and flux posteriors that are very similar to the real data. Note that as the BF for GCE PSs in this scenario (and in the real data) is not nearly as large as when the \texttt{p6v11} model is employed, the posterior fluxes are more sensitive to prior choices; in particular, choosing a linear or log prior for the GCE normalization parameter changes the size of the uncertainty bands on the posterior fluxes. 

Figure~\ref{fig:10degAspread} shows the spread of results in 100 simulated datasets for \texttt{Model A}, extending the one realization shown in Fig.~\ref{fig:10degA}. Compared to the \texttt{p6v11} results, there is more spread in both the recovered flux posteriors and SCFs, although behavior similar to that observed in the real data remains quite frequent. Over the set of 100 realizations, the BF in favor of GCE PSs in this case ranges from $10^{-4}$ to $10^{3}$, consistent with the flux posteriors ranging from assigning very little flux to PSs to assigning the entire GCE to PSs. The SCF for GCE PSs commonly has a shape very similar to that observed in the real data, but (in other realizations) can also have a shape more similar to the SCFs for the disk and isotropic PSs.

Figure~\ref{fig:10degAspread2} compares the distribution of BFs and SCFs in three simulated scenarios for \texttt{Model A}: (1) where the GCE is smooth and symmetric (parameters based on analysis of real data with no GCE PSs) (2) where the GCE is 100\% PSs (parameters based on analysis of real data including both GCE Smooth and GCE PS templates), and (3) where the GCE is smooth and asymmetric (parameters based on analysis of real data, with only a GCE Smooth template, subdivided into independent north and south components). In all cases, the analyses use symmetric GCE templates (PS and Smooth). From the BF plot, all scenarios (symmetric GCE PSs or smooth, or asymmetric smooth GCE) appear to overlap with the real data. However, for the symmetric smooth simulations, while some BFs may be as high as observed in the real data, the inferred GCE PS SCFs in these cases do not (over 100 realizations) ever resemble the GCE PS SCF in the real data (as shown in the lower left panel); instead, it appears the GCE PS template may be picking up bright PSs that were simulated as part of the disk. As such, the SCF can serve as a diagnostic that the data (with this background model) do not appear to be well-described by simulations with only a symmetric smooth template. The expected BF distribution for symmetric PSs overlaps with the result in real data and extends to significantly higher BF values -- although the higher-BF realizations often recover incorrect SCFs for all three PS templates, e.g. with the GCE picking up bright disk or isotropic PSs. The results for an asymmetric smooth GCE overlap with the real data (although the BF is on the high end), and while there is a range of SCFs as discussed above, the SCFs matching the results on real data also often coincide with high BFs similar to what is observed in the data.

We see that the BFs in favor of spurious PSs are expected to be smaller with \texttt{Model A} and \texttt{Model F} than with \texttt{p6v11}, even though there is manifestly no difference between the true GCE PS populations in these scenarios (in all cases, no GCE PSs are simulated). One might ask if this is because of a difference in the preferred asymmetry or the posterior values for other model parameters, when the diffuse model is changed, but instead it seems to just reflect a difference in sensitivity between the pipeline using \texttt{p6v11} and \texttt{Models A/F}. 

To demonstrate this point, we use the baseline mock dataset (obtained by fitting the real data using the \texttt{p6v11} diffuse model, allowing for a smooth asymmetric GCE but no GCE PSs), and fit this mock dataset assuming symmetric GCE PS and GCE Smooth templates, with \texttt{Model A} or \texttt{Model F} as the diffuse model.

\begin{figure}[t]
\leavevmode
\centering
\subfigure{\includegraphics[width=0.32\textwidth]{Plots/Run_2_scf}}
\subfigure{\includegraphics[width=0.32\textwidth]{Plots/Run_2a_scf}}
\subfigure{\includegraphics[width=0.32\textwidth]{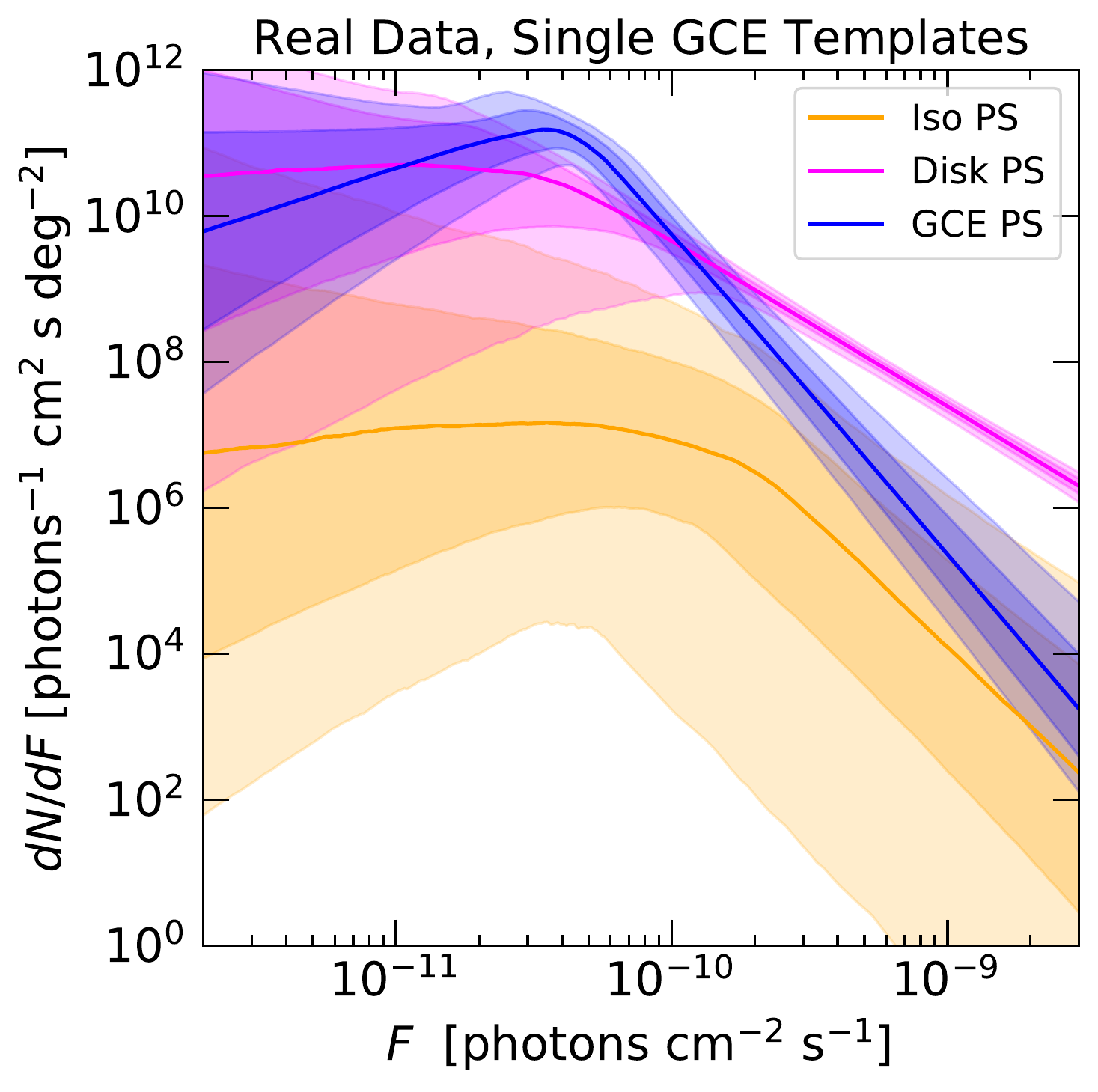}}\\
\subfigure{\includegraphics[width=0.32\textwidth]{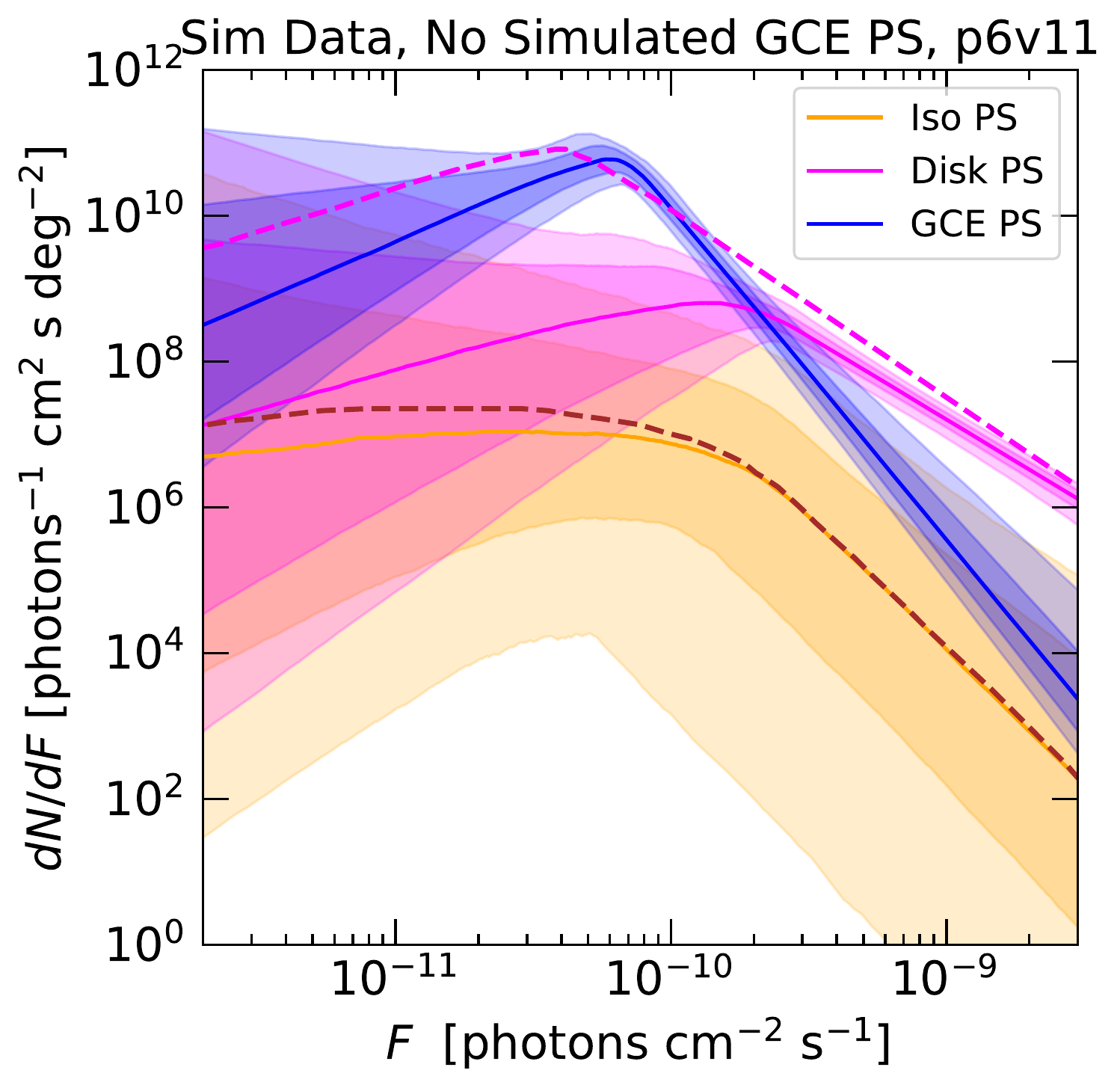}}
\subfigure{\includegraphics[width=0.32\textwidth]{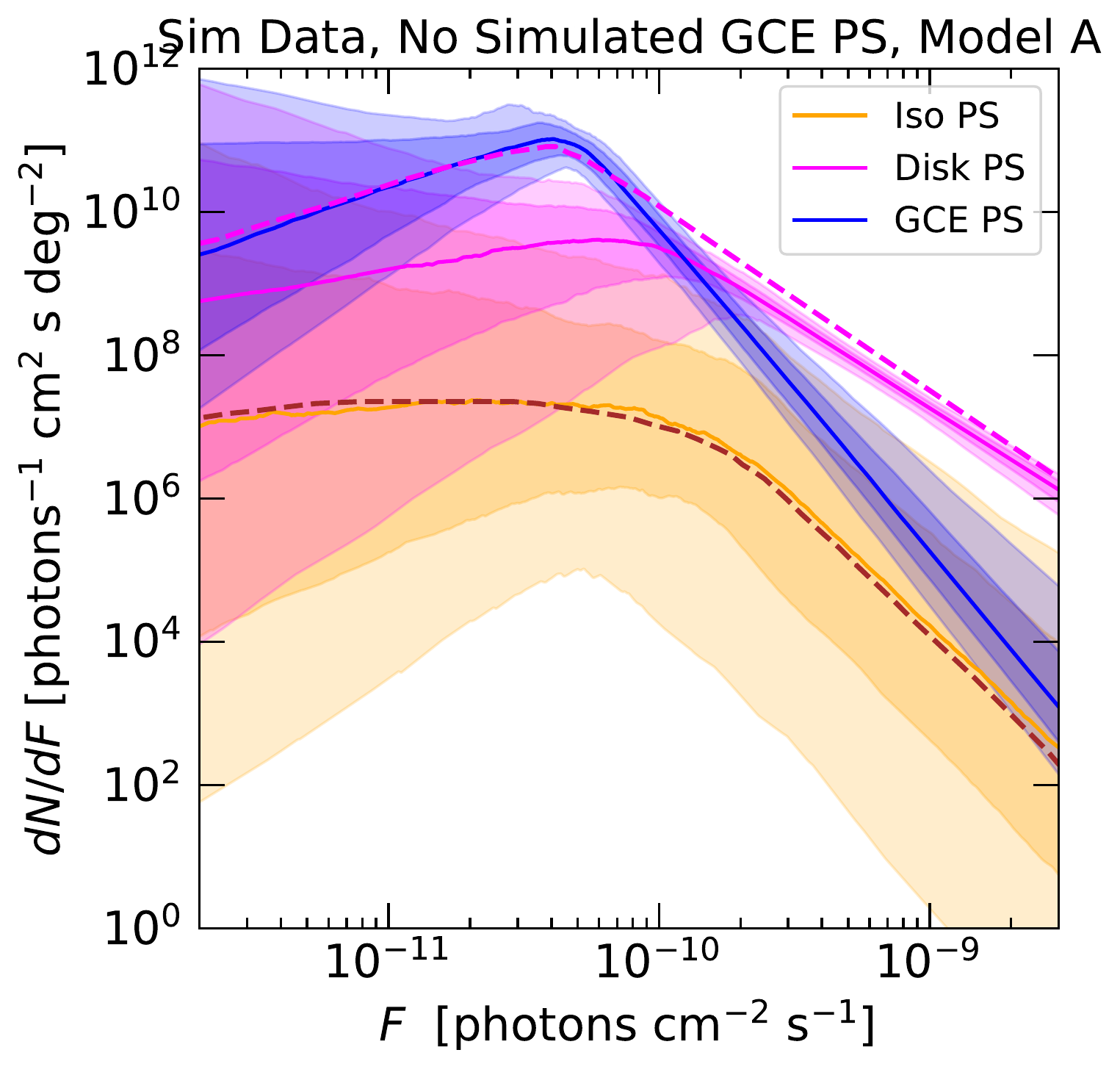}}
\subfigure{\includegraphics[width=0.32\textwidth]{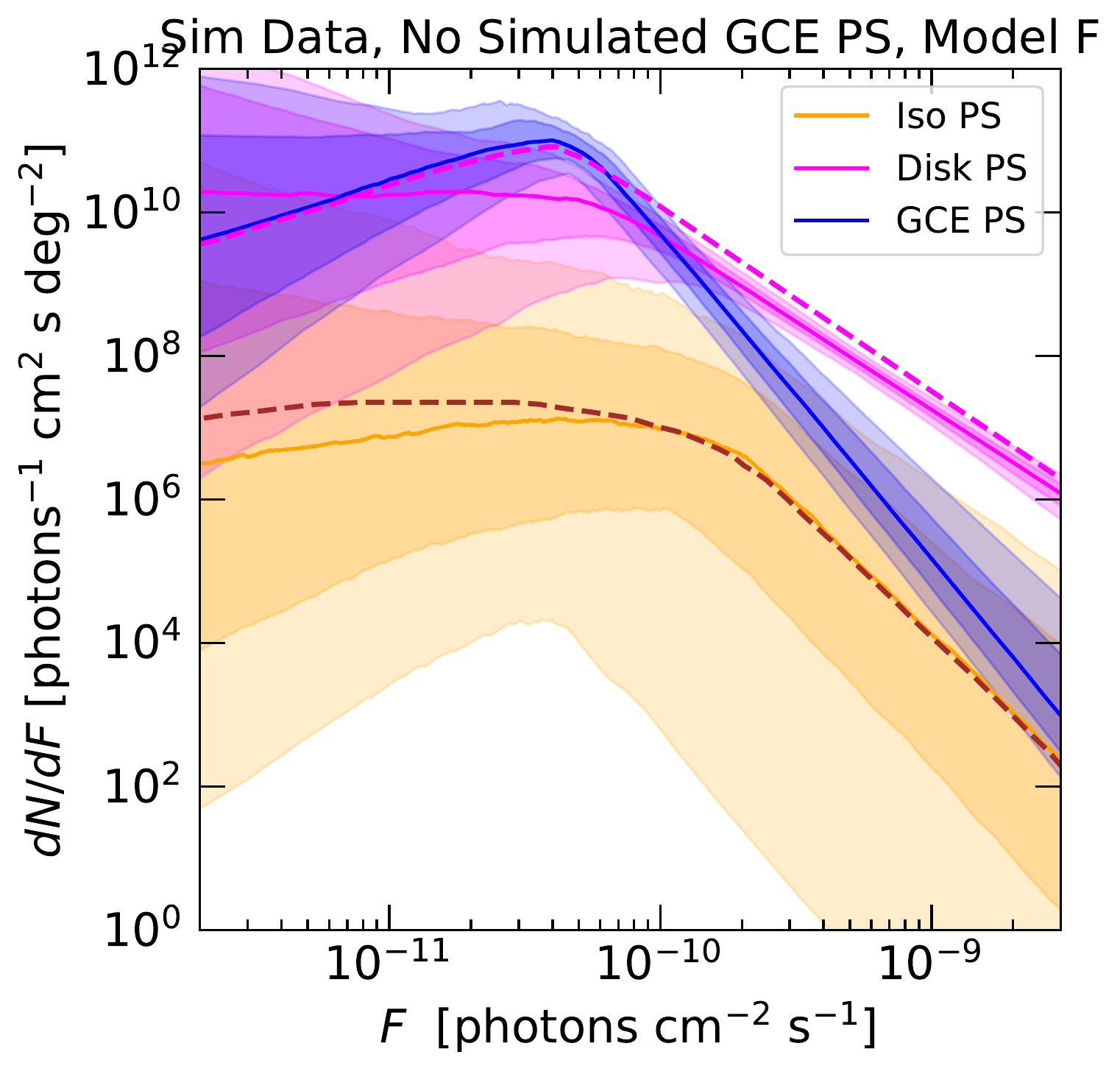}}\\
\caption{\textbf{Top row:} Comparison of SCFs found in the real data in analyses with symmetric GCE PS and GCE Smooth templates, for diffuse models \texttt{p6v11} (\textit{left}), \texttt{Model A} (\textit{middle}), and \texttt{Model F} (\textit{right}). \textbf{Bottom row:} SCFs recovered in the identical analysis on the baseline mock dataset, again using diffuse models \texttt{p6v11} (\textit{left}), \texttt{Model A} (\textit{middle}) and \texttt{Model F} (\textit{right}). Recall the baseline mock dataset does not contain any GCE PSs. The dashed lines show the simulated disk SCF (pink) and simulated Iso PS SCF (brown).}
\label{fig:modelsSCF}
\end{figure}

Figure~\ref{fig:modelsSCF} shows the SCF recovered in these cases, when the \texttt{p6v11} asymmetric GCE simulated data is analyzed using diffuse models \texttt{Model A} and \texttt{Model F}, and the comparison with the fit using \texttt{p6v11}; we also show the comparison to the results of identical analyses performed on the real data. The take-away point when comparing these SCF is that our baseline simulation, which was
created using \texttt{p6v11}, appears to reproduce the behavior of the real data in detail. The upper panel and its corresponding lower panel appear to produce comparable SCF. The BFs in favor of (spurious) GCE PSs are $\sim10^2$ (\texttt{Model A}) and $7$ (\texttt{Model F}), compared to $\sim 4\times10^{12}$ with \texttt{p6v11}. This highlights the point that these diffuse models are expected to give lower BFs in favor of spurious PSs even in identical datasets (note this is clearly not a matter of these models being ``better'' matches to the data and hence less likely to be fooled by spurious PSs, as the lower BFs are seen in simulations where they are actually the \textit{wrong} diffuse models relative to the simulated data). Interestingly, these BFs are comparable to the BFs found for GCE PSs when analyzing the real data with these diffuse models. Furthermore, the SCFs recovered when analyzing this simulated realization with different diffuse models, with symmetric GCE PS and Smooth templates, are very similar to the SCFs obtained with identical analyses on the real data. This suggests that the baseline mock dataset provides a good match to the real data in many respects.

We caution that the smaller preference for PSs when using \texttt{Model A} or \texttt{Model F} means that prior choices can have a larger impact, and it is not as clear-cut that the loss of significance for PSs when asymmetry is allowed is well beyond what one would expect just from the extra degree of freedom in the model. However, the results obtained with \texttt{Model A} or \texttt{Model F} in the real data are completely consistent with what we would expect for a smooth asymmetric GCE capable of mimicking a high-significance spurious GCE PS population in the analysis with the \texttt{p6v11} diffuse model.

\begin{figure}[t]
\leavevmode
\centering
\subfigure{\includegraphics[width=0.41\textwidth]{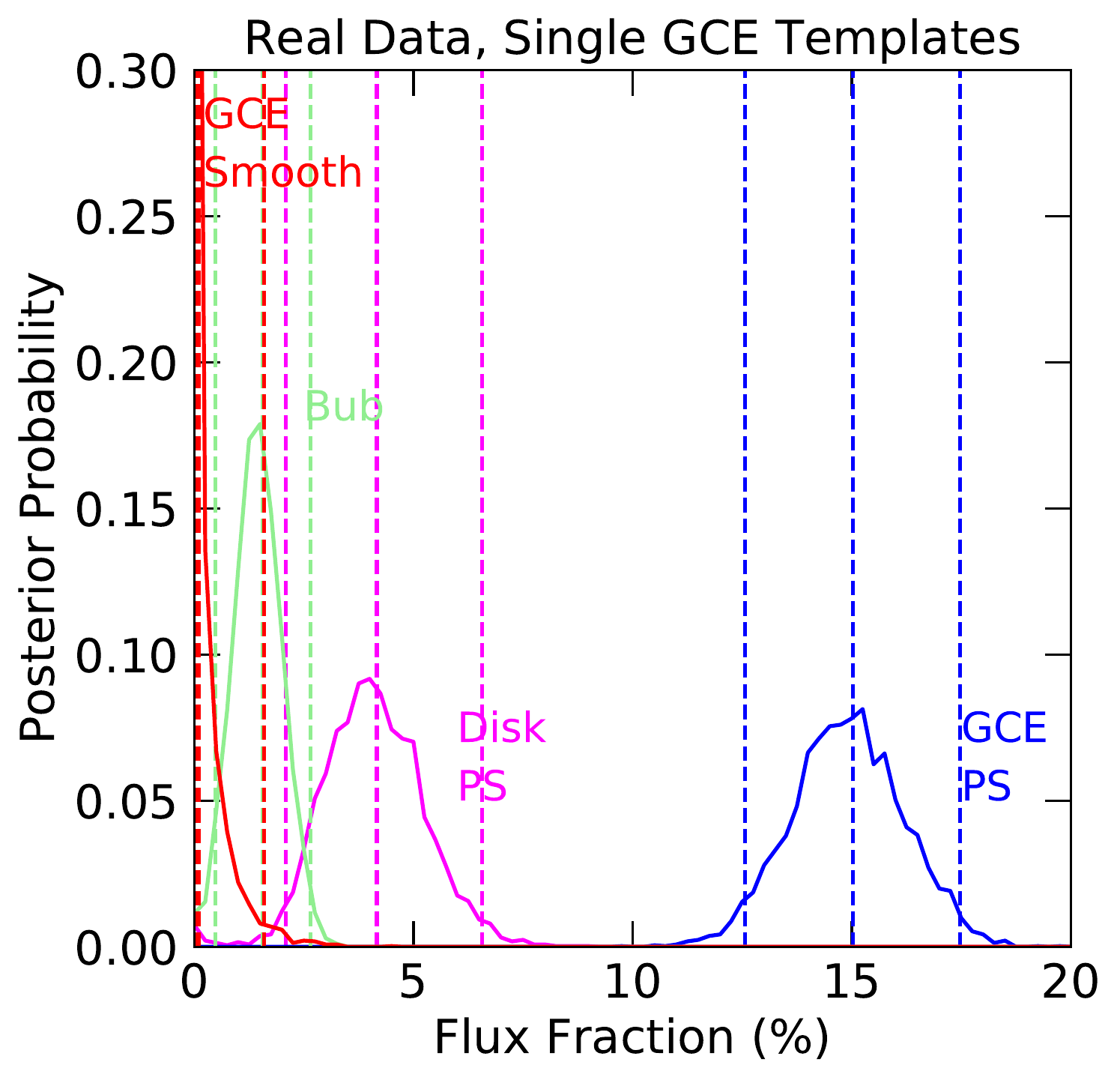}}
\subfigure{\includegraphics[width=0.40\textwidth]{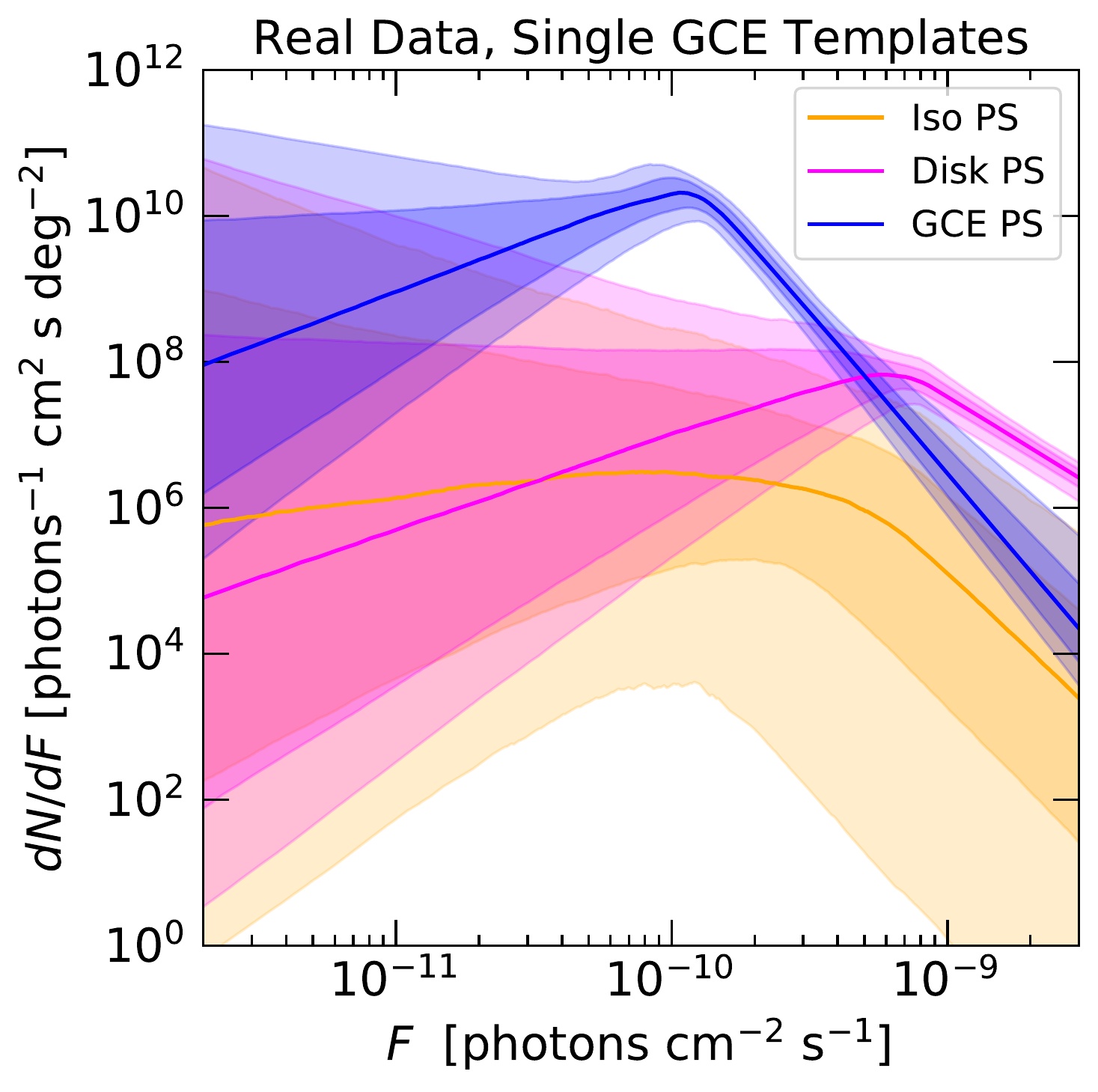}}\\
\subfigure{\includegraphics[width=0.41\textwidth]{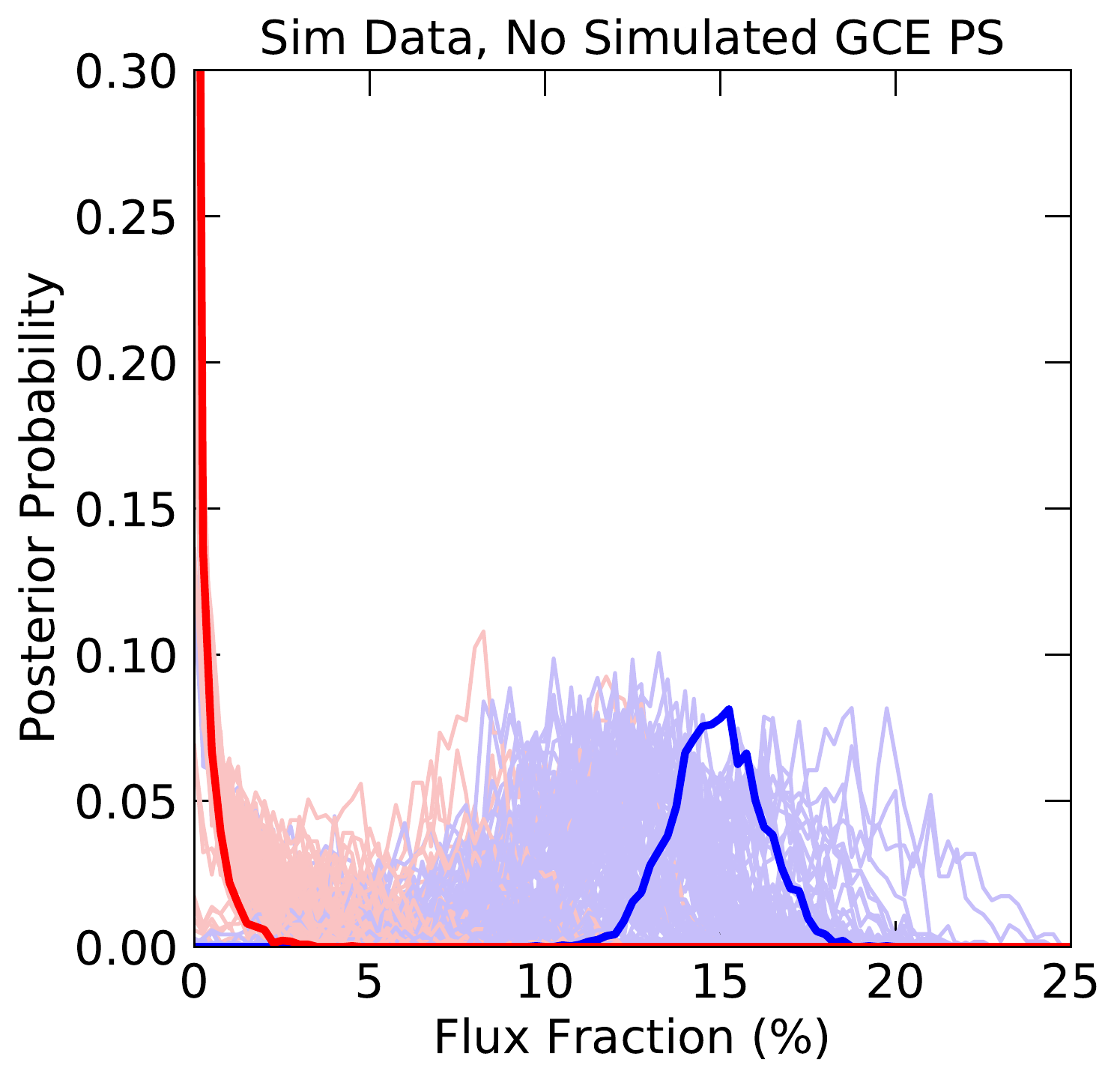}}
\subfigure{\includegraphics[width=0.39\textwidth]{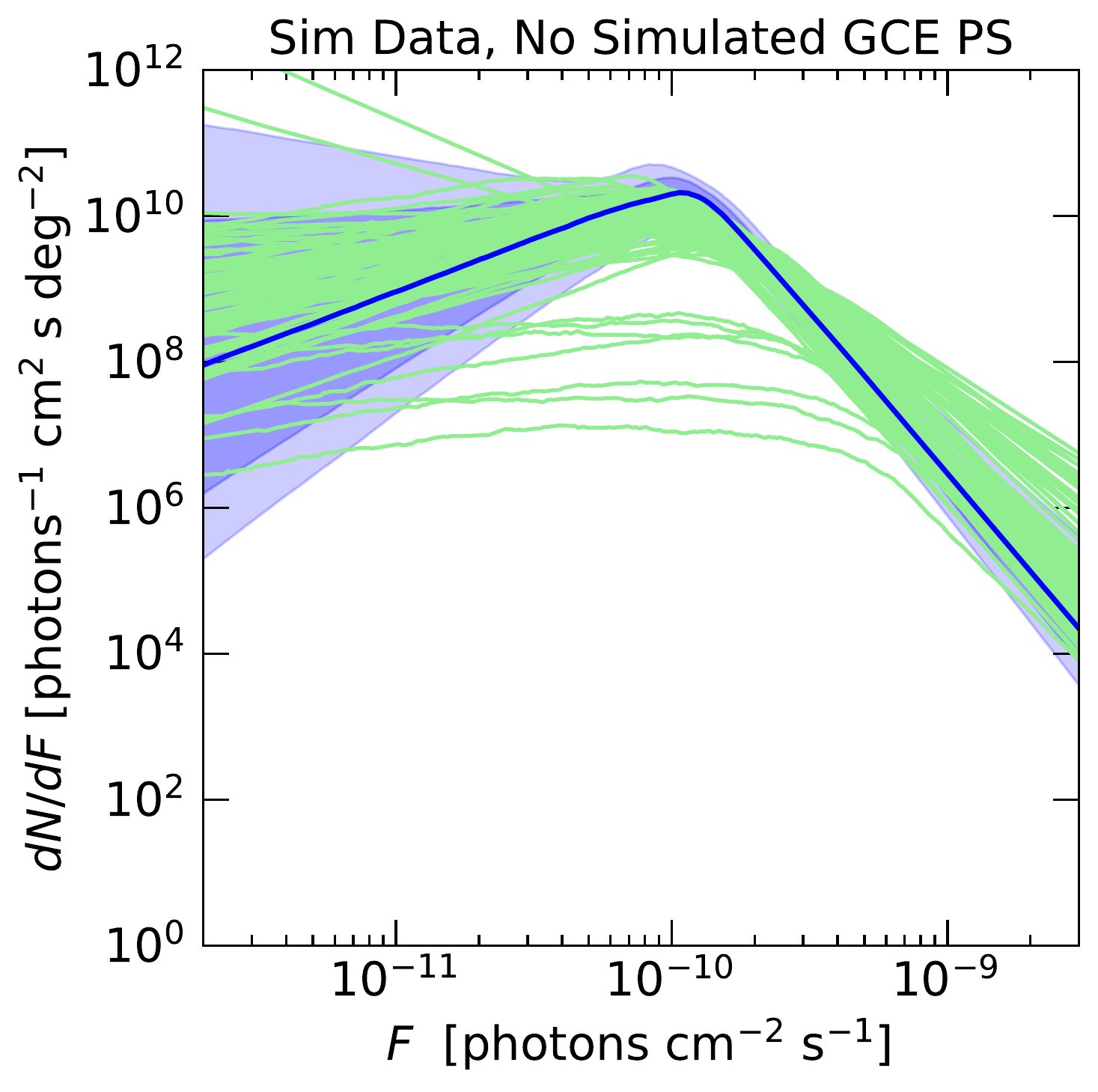}}
\caption{Summary of results using only the top quartile of data by angular resolution. All simulations include a smooth north-south asymmetric GCE; they are analyzed with symmetric GCE Smooth and GCE PS templates, plus backgrounds. {\bf Top-Left:} flux posteriors for analysis of the real data with symmetric GCE Smooth and GCE PS templates. {\bf Top-Right:} SCFs for PS populations from analysis of the real data with symmetric GCE Smooth and GCE PS templates. {\bf Bottom-Left:} flux posteriors from 100 realizations of the simulated data (fainter lines) and for the real data (solid lines), for GCE PSs (blue) and GCE Smooth (red). {\bf Bottom-Right:} posterior median SCFs from 100 realizations of the simulated data (green lines) and the posterior SCF for the real data (blue band). }
\label{fig:bestpsf}
\end{figure}

\subsection{Analyses with Highest Angular Resolution Quartile Only}

Previous NPTF results have generally used only the top quartile of data by angular resolution (with the exception of Ref.~\cite{Linden:2016rcf}), sacrificing statistics in favor of improved discrimination between point sources and diffuse emission. While by default we have used the top three quartiles, in this subsection, we explore the robustness of the main results of Ref.~\cite{PhysRevLett.125.121105} to using only the highest-resolution quartile.

Figure~\ref{fig:bestpsf} summarizes our results. We find comparable results to Ref~\cite{PhysRevLett.125.121105}; an unmodeled asymmetry also generates a spurious GCE PS population with features that are generically comparable to the real data, in the flux fraction allocated to GCE PSs and in the SCF of the inferred PSs. The BF preference in favor of PSs decreases modestly (compared to the three-quartile result) when using the \texttt{p6v11} diffuse model, both in the real data and in simulations built from the posterior medians of that fit.  Specifically, the BF in favor of (symmetric) GCE PSs in the real data is $\sim 10^{10}$ (vs $10^{15}$ with three quartiles), whereas across 100 realizations of simulated data, we find a range of BFs from $\sim 1-10^{10}$ (compared to $\sim 1-10^{20}$ with three quartiles). 

We also tested the preference for GCE PSs when the two GCE templates are broken into northern and southern pieces; in this case we find the preference for GCE PSs is $\sim1$, consistent with the three quartile analysis.

Lastly, we repeated the analysis on real data using the top angular resolution quartile and \texttt{Model A} rather than \texttt{p6v11}. We find that the BF preference for (symmetric) GCE PSs is $\sim10^3$ when no asymmetry is allowed (i.e. in this case we do not see a degradation in BF from restricting to the top quartile), but decreases to $\sim1$ when asymmetry is permitted, consistent with our other results.

\subsection{Effects of Increasing the GCE Source Count Function Slope Prior}

\begin{figure}[t]
\leavevmode
\centering
\subfigure{\includegraphics[width=0.4\textwidth]{Plots/Run_2_scf}}
\subfigure{\includegraphics[width=0.4\textwidth]{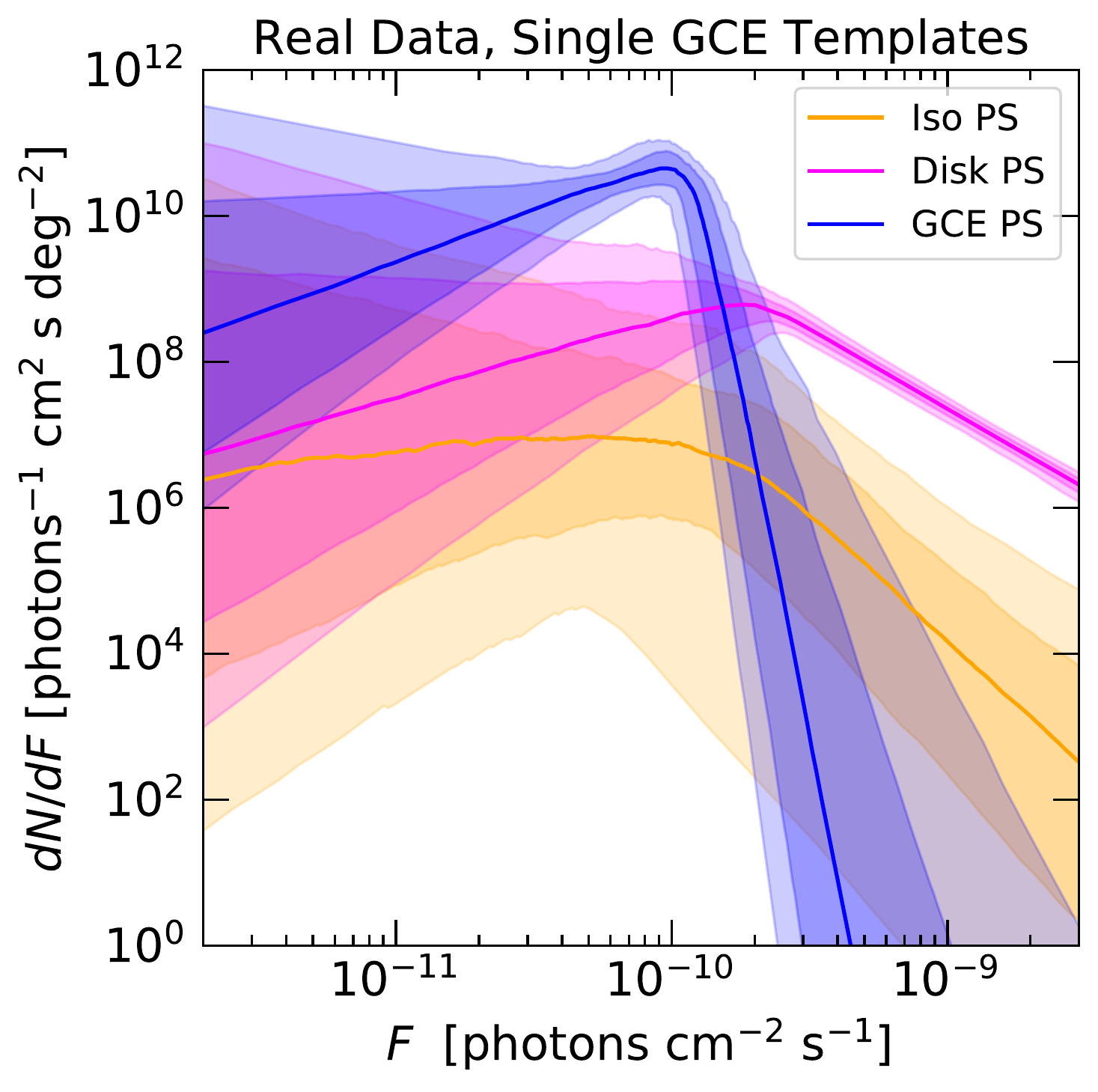}}\\
\subfigure{\includegraphics[width=0.4\textwidth]{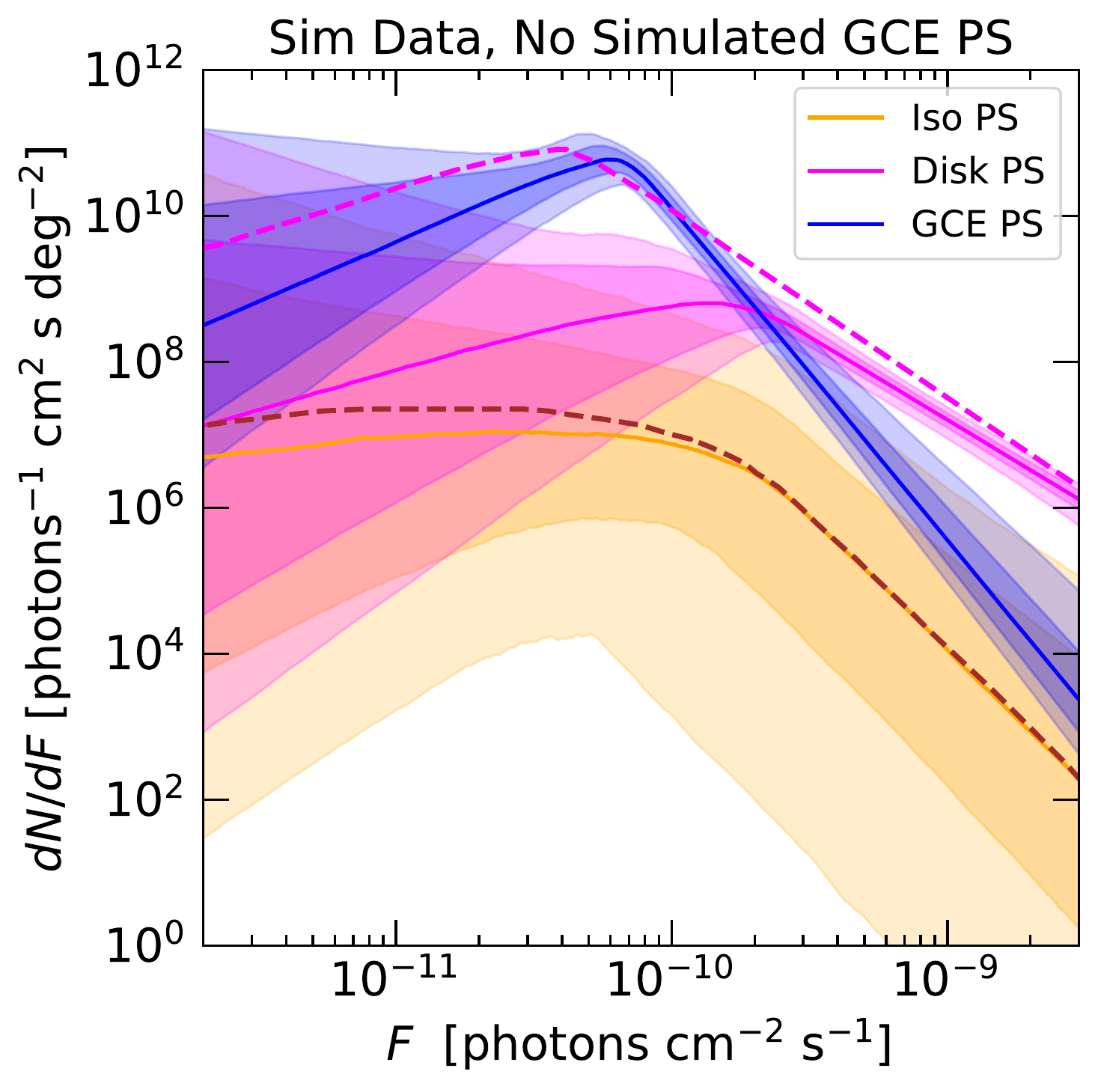}}
\subfigure{\includegraphics[width=0.4\textwidth]{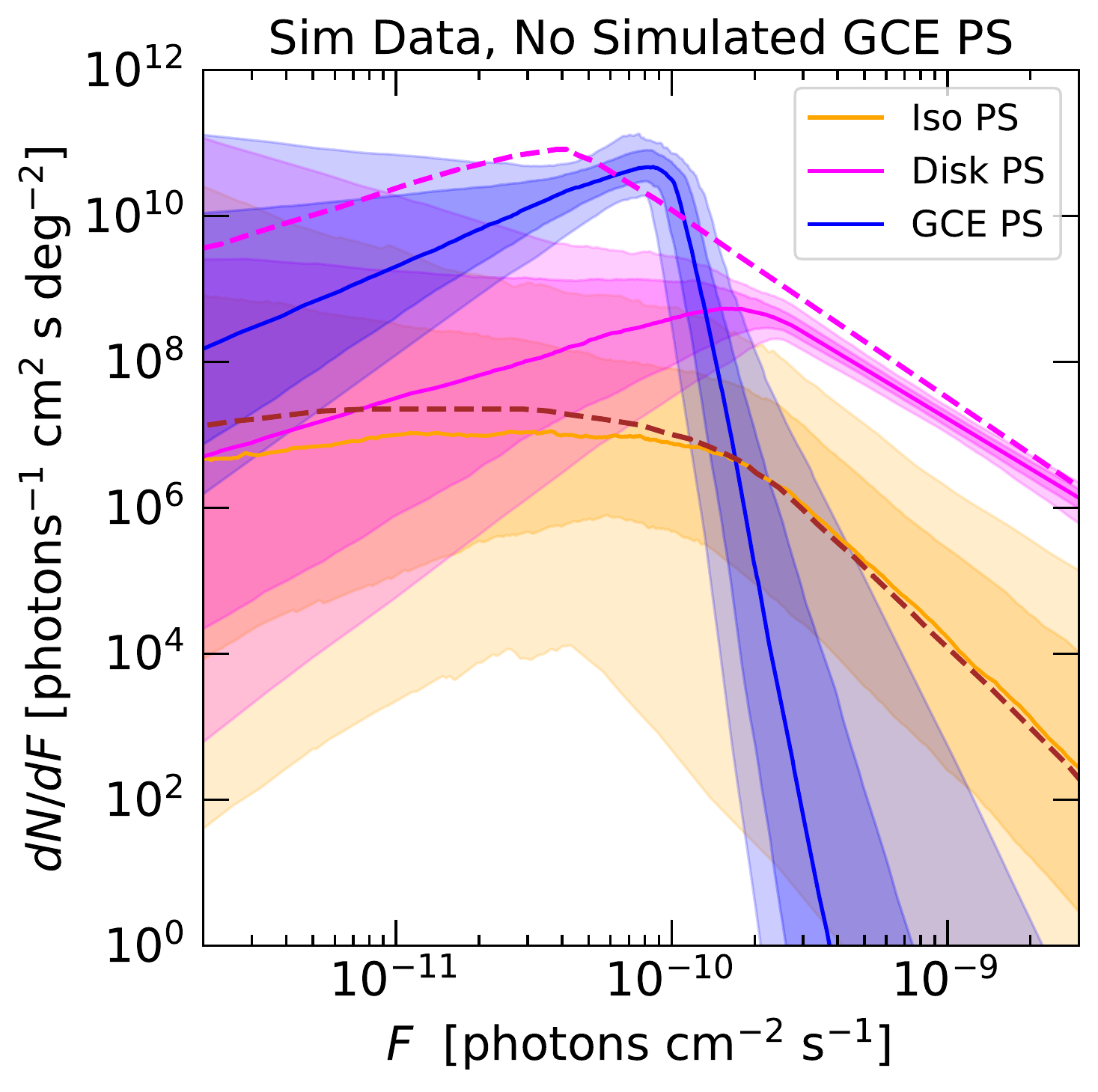}}
\caption{SCFs obtained for the real data (\textit{top row}) and baseline mock dataset (\textit{bottom row}) in our default analysis (\textit{left column}) and when the SCF slope above the break for GCE PSs is allowed to be steeper (\textit{right column}, see text for details). The fits employ symmetric GCE PS and GCE Smooth templates, plus other background templates. Recall that the simulated data includes no GCE PSs. In the bottom row, the dashed lines show the simulated disk SCF (pink) and simulated Iso PS SCF (brown).}
\label{fig:slopes}
\end{figure}

By default in Ref.~\cite{PhysRevLett.125.121105} and this work, we impose a prior on the high-flux slope of the SCF for GCE PSs of $n_1 < 5$, where $dN/dF \propto F^{-n_1}$ above the break in the SCF. We make this choice in order to reduce computation time, and because this is already an extremely steep slope compared to known astrophysical luminosity functions. However, previous NPTF studies allowed larger values of $n_1$, and indeed these larger values are nominally preferred by the data. To ensure that our results are not being sculpted by this choice of prior, and to more easily compare the SCFs we extract to results from previous studies, in this section we test the effects from increasing the prior range on the slope of the GCE PS SCF above the break.

Figure~\ref{fig:slopes} shows the different SCFs recovered when the prior range on $n_1$ for GCE PSs is increased from [2.05,5] to [2.05,30]. We show how the SCF changes in the fit on real data under this change in priors, as well as how the reconstructed SCF changes when considering the baseline mock dataset. We see that in both cases the SCF break moves up slightly; this is to be expected, as increasing $n_1$ means the SCF cuts off more quickly and there are fewer events directly above the break. Again we see that the baseline mock dataset behaves very similarly to the real data under this change of prior; while the SCFs for the reconstructed PS population change noticeably with the variation in prior in both cases, they change in much the same way. It is interesting that the simulation with an asymmetric smooth GCE predicts that the SCF will prefer a very steep slope above the cutoff, as observed in real data.

For the real data, increasing the prior range for the slope induces a small decrease in the BF in favor of PSs, from $\sim10^{15}$ to $\sim10^{12}$. In the baseline mock dataset, the BF preference for GCE PSs slightly increases with this change, from $\sim10^{12}$ to $\sim10^{13}$. Both of these changes are much smaller than the scatter in BF between different realizations, as discussed earlier. Thus the similarity of the SCFs and BFs between the real data and the baseline mock dataset does not appear to be prior-dependent.

\subsection{Analyses with a 30$^\circ$ circle, 2$^\circ$ plane mask ROI}

Previous NPTF analyses of the GCE have employed a larger ROI, extending to 30$^\circ$ from the Galactic Center \cite{Lee:2015fea,Leane:2019xiy}. We have previously demonstrated that in this large region, the smooth component of the GCE will converge to a very negative normalization if such is allowed \cite{Leane:2019xiy}, with the GCE PS contribution picking up a very large positive normalization to compensate. We suggested that mis-modeling of the PS populations in this large ROI could be one contributor to this behavior \cite{Leane:2019xiy}. It could also reflect the diffuse emission model being too bright in some relevant subregions because its normalization is fitted over the whole large ROI, leading to oversubtraction (\cite{bennick}; this effect has been discussed in the context of Poissonian template fitting in Refs.~\cite{Daylan:2014rsa,Chang:2018bpt}). Our choice of a 10$^\circ$ radius ROI in Ref.~\cite{PhysRevLett.125.121105} and the previous sections, is partly because this choice should mitigate the effects of fitting the diffuse model over too large a region, and does markedly reduce the degree to which the smooth GCE component prefers a negative coefficient. Consequently, it should not be surprising if results in the 30$^\circ$ radius ROI are not consistent with those in the 10$^\circ$ ROI. Nonetheless, we here present the results of allowing the GCE to possess north/south asymmetry in this larger ROI. We still mask the Galactic plane for $|b|<2^\circ$.

\begin{figure*}[t]
{\includegraphics[width=0.42\textwidth]{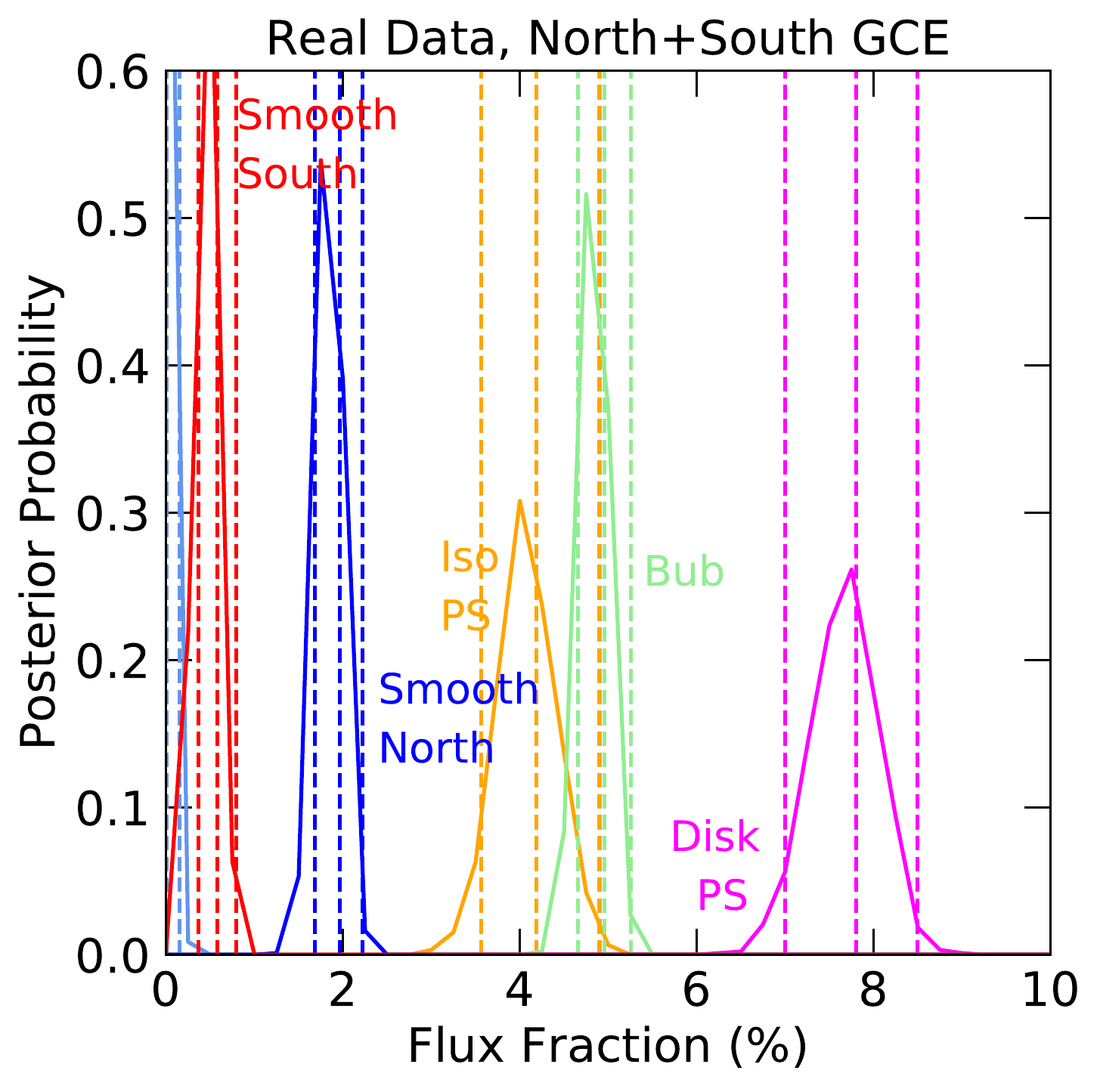}}
\caption{Flux posteriors when floating north and south smooth GCE templates separately in the $30^\circ$ radius ROI.}
\label{fig:asym30}
\end{figure*}

Figure~\ref{fig:asym30} shows the flux posteriors for a selection of templates in the 30$^\circ$ ROI; as in the 10$^\circ$ radius ROI, the GCE is brighter in the north than in the south. Using the \texttt{p6v11} Galactic diffuse emission model, we still find a significant preference for north-south asymmetry of the smooth GCE, with a BF of $\sim10^{23}$ in favor of asymmetry.

\begin{figure*}[t]
\leavevmode
\centering
\subfigure{\includegraphics[width=0.32\textwidth]{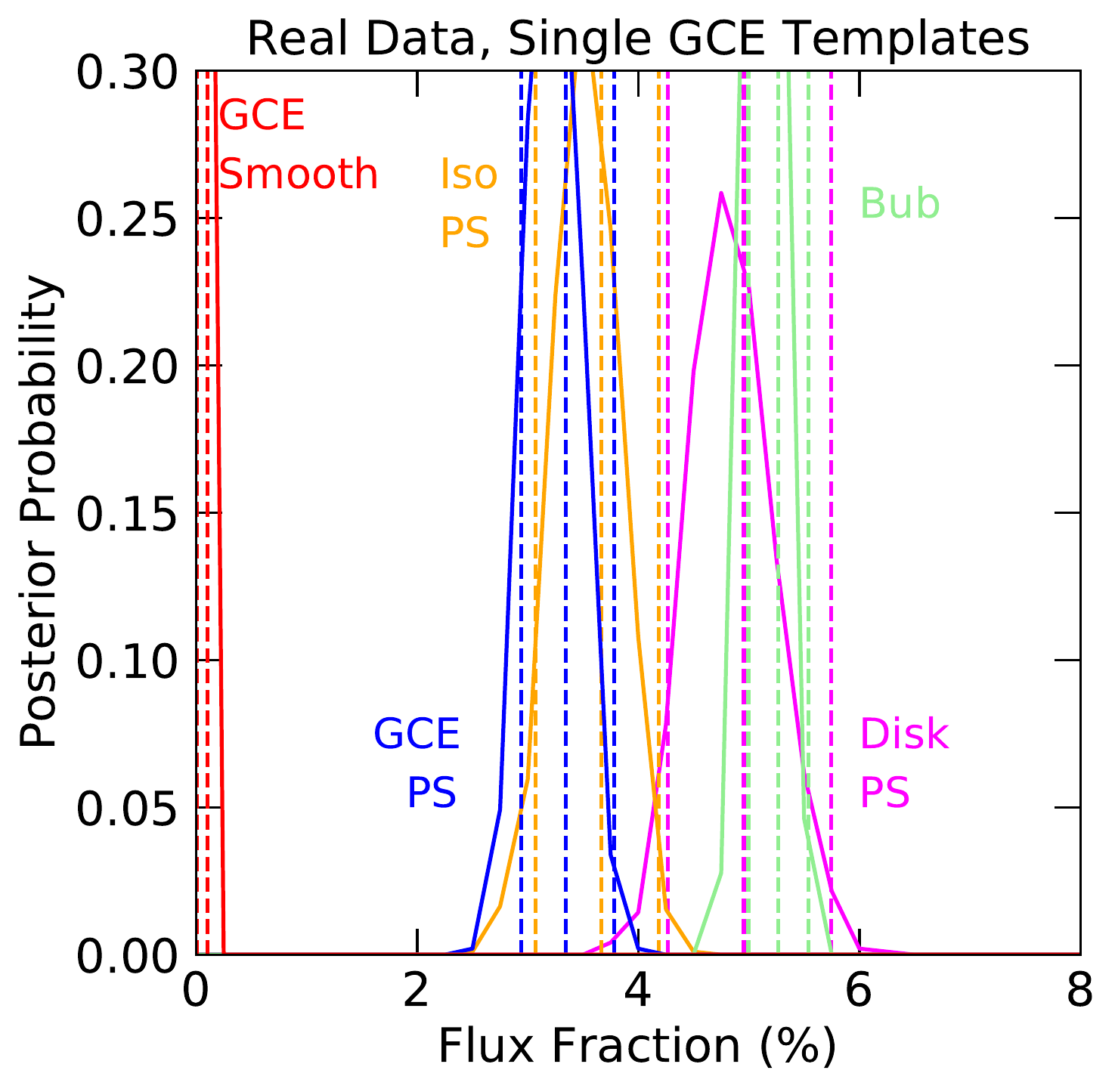}}
\subfigure{\includegraphics[width=0.32\textwidth]{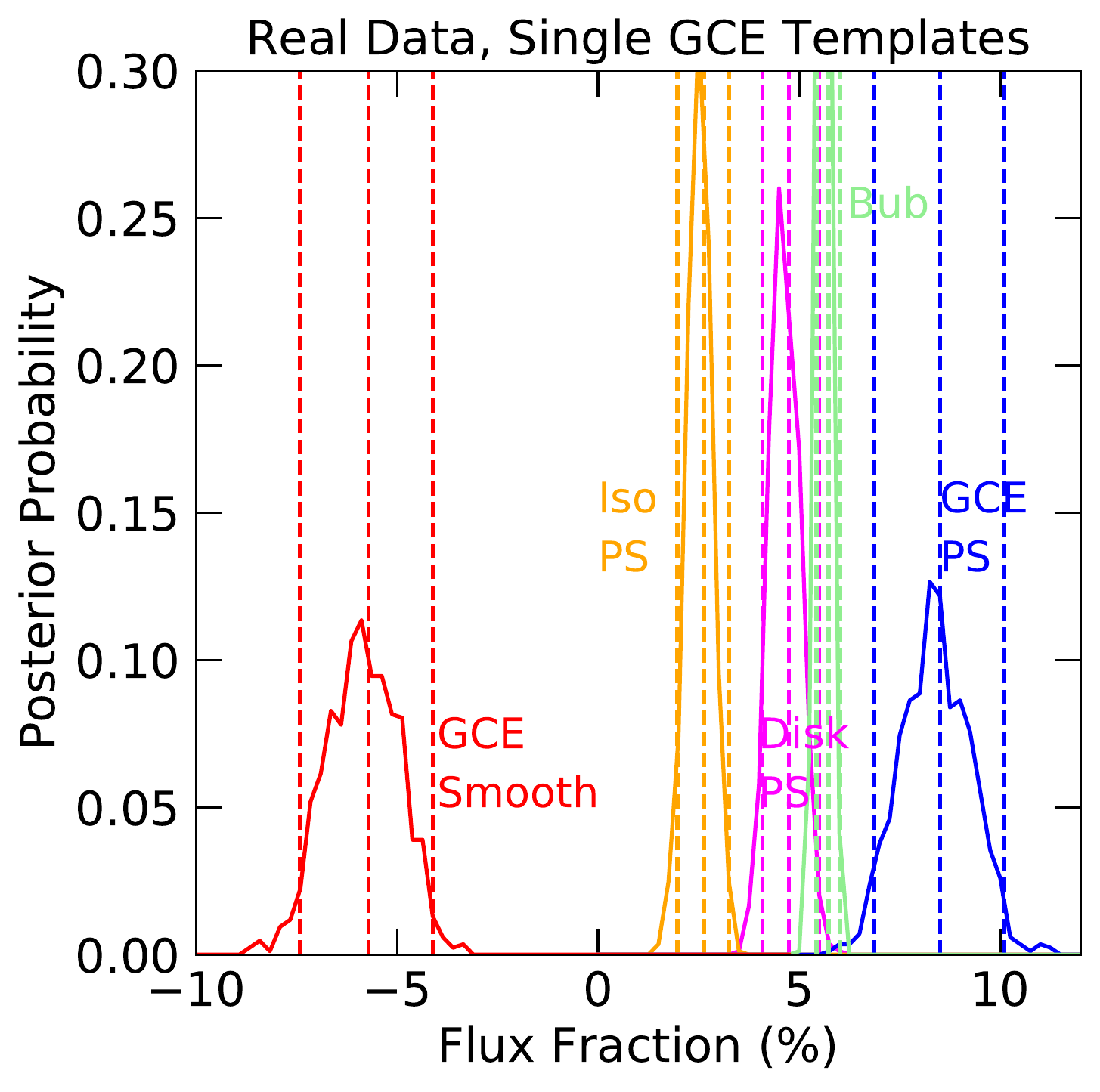}}
\subfigure{\includegraphics[width=0.32\textwidth]{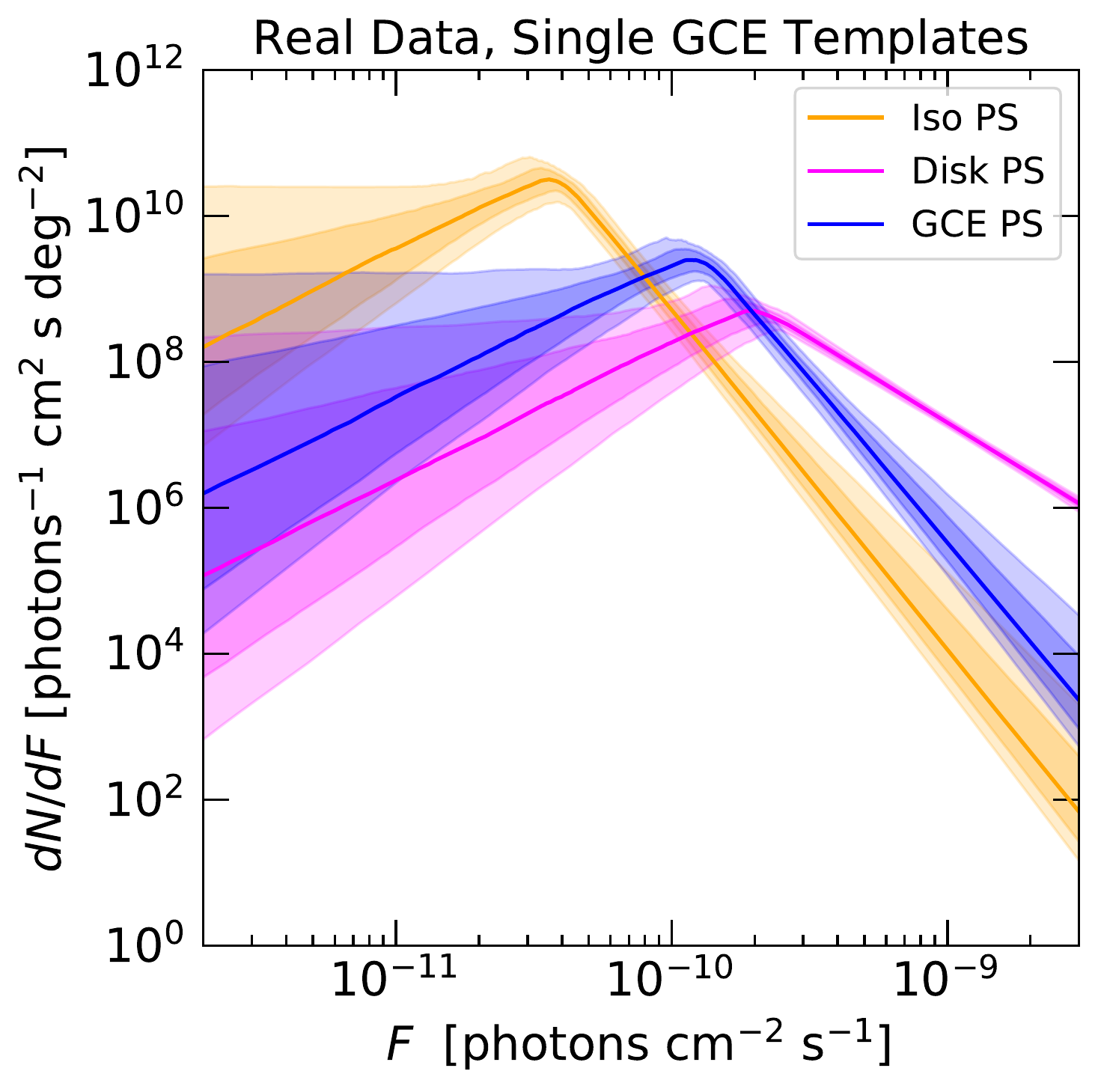}}\\
\subfigure{\includegraphics[width=0.32\textwidth]{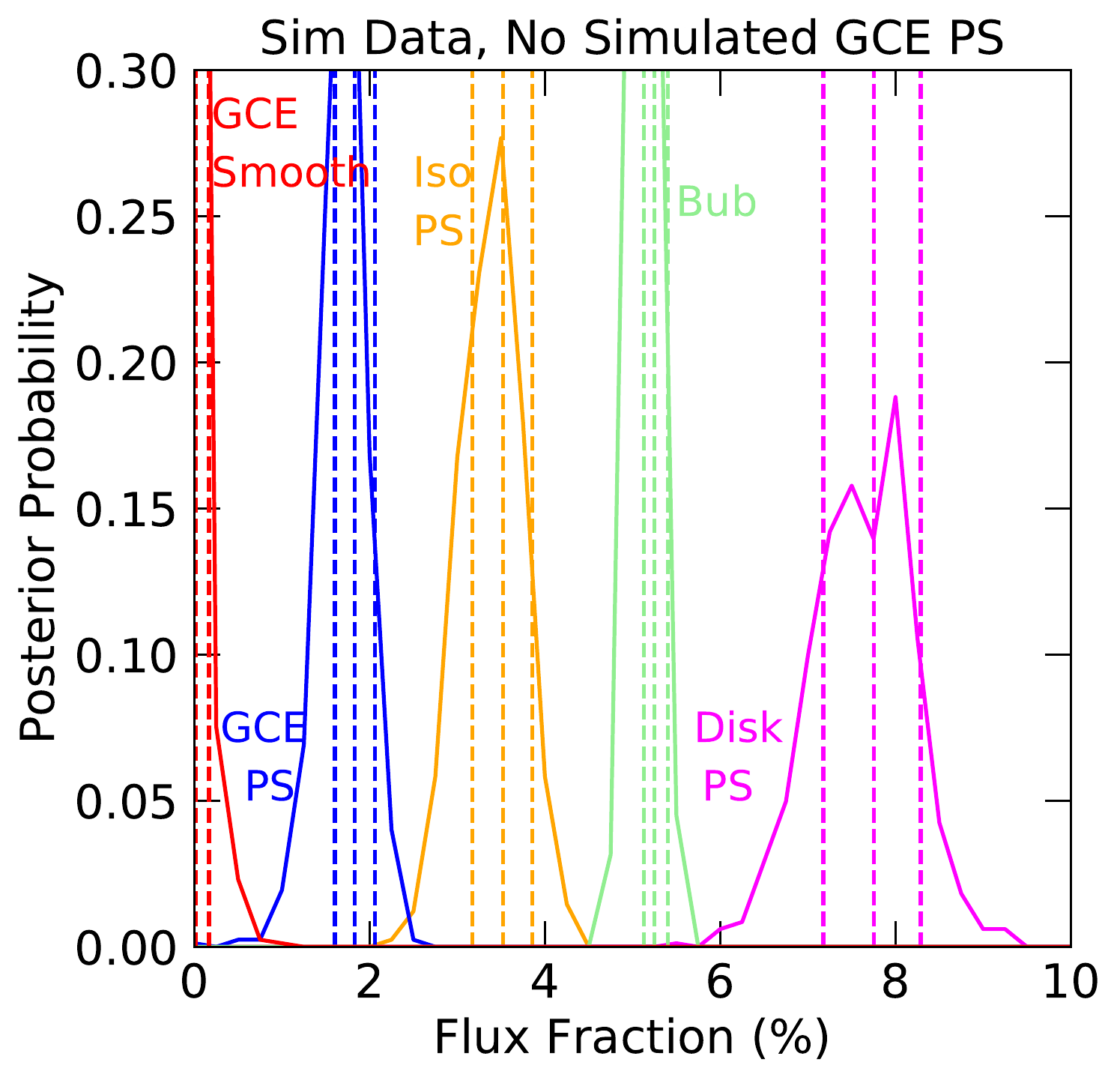}}
\subfigure{\includegraphics[width=0.32\textwidth]{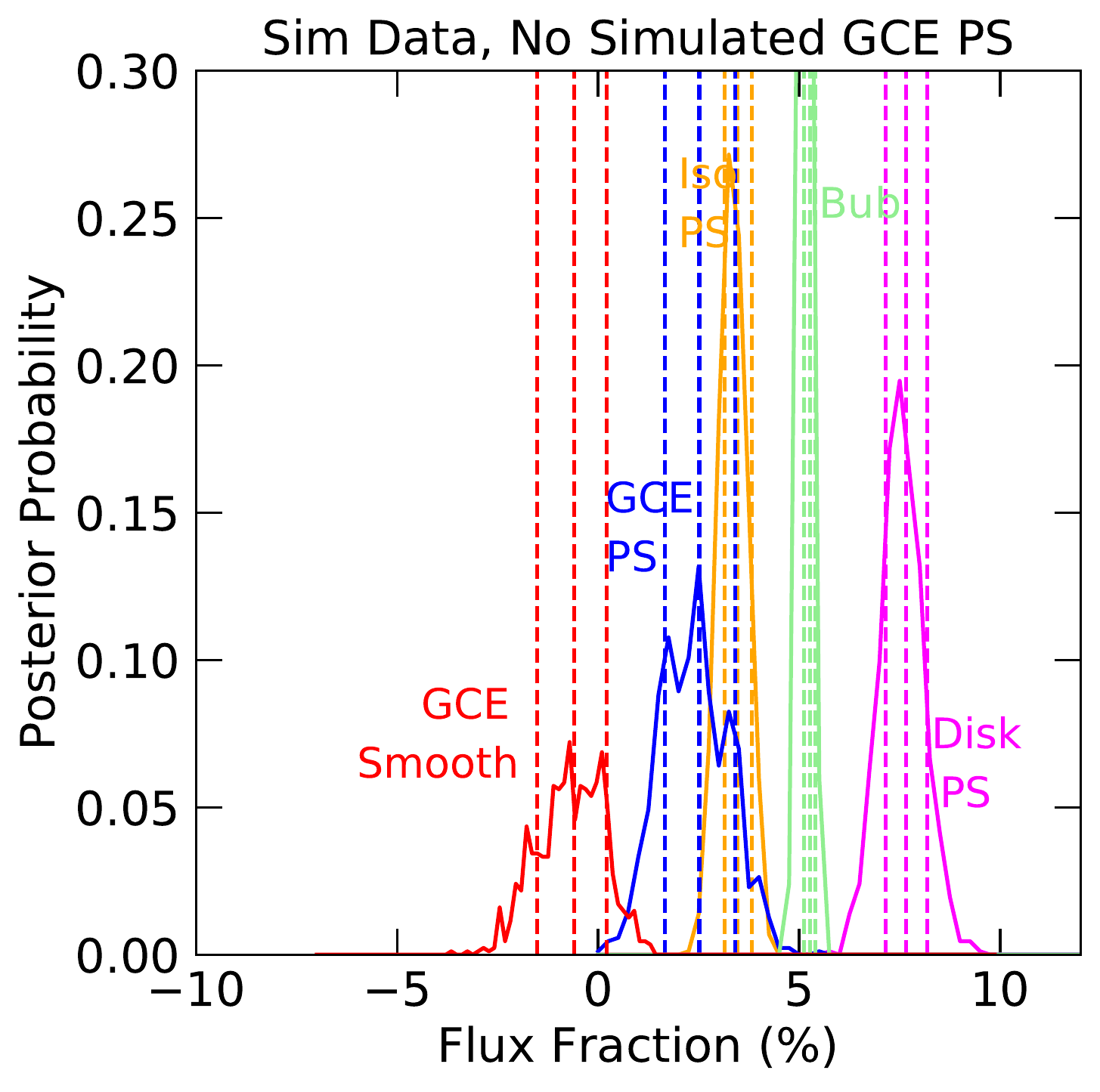}}
\subfigure{\includegraphics[width=0.32\textwidth]{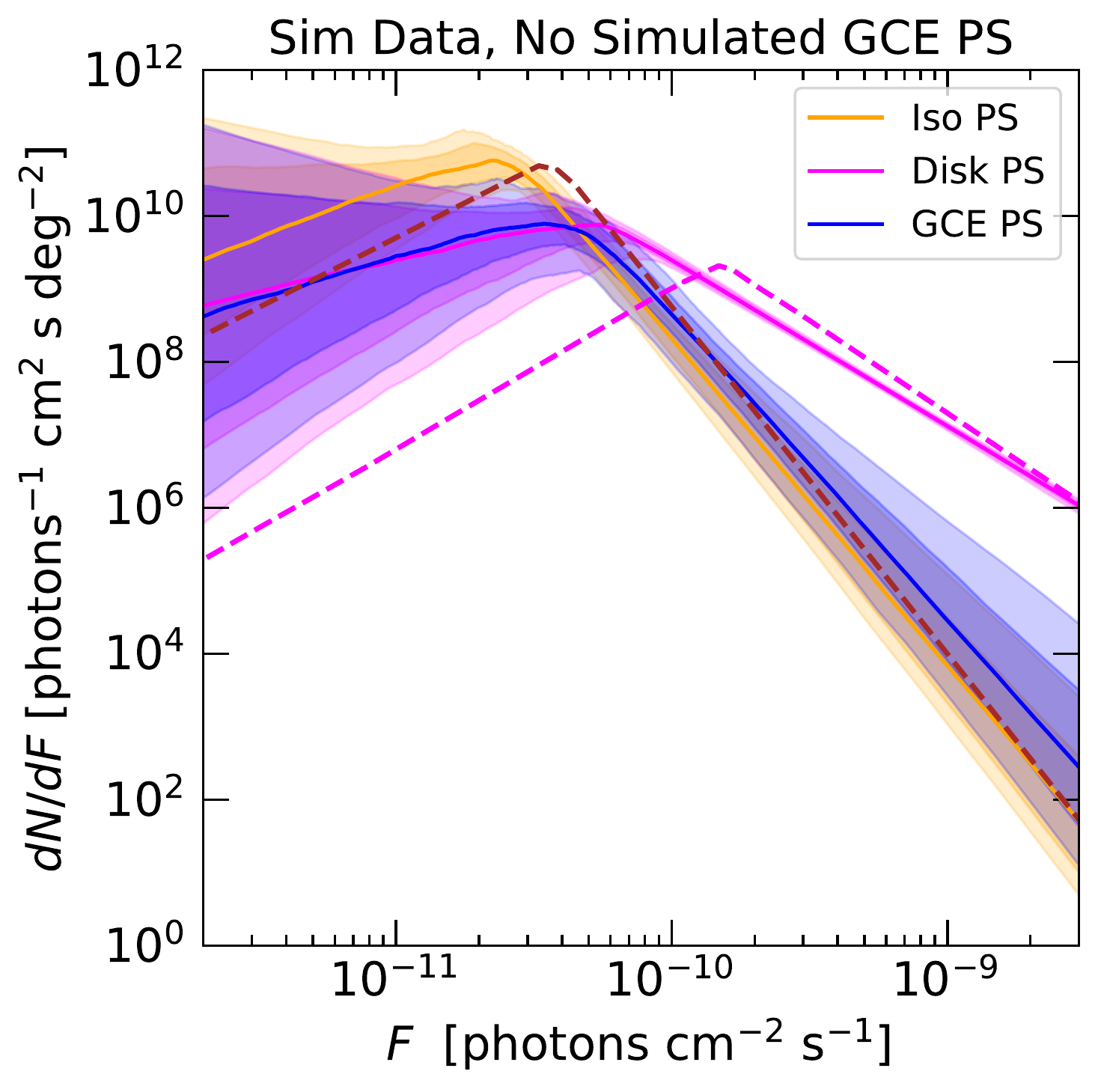}}\\
\caption{Comparison of real (\emph{top row}) and simulated (\emph{bottom row}) data in a 30$^\circ$ radius ROI; in all cases the analyses used single GCE templates (smooth and PS) over the whole ROI. The simulated dataset is based on the best fit model (fluxes shown in Fig.~\ref{fig:asym30}) using separate Poissonian templates for the northern and southern GCE; no GCE PSs were simulated.
\textbf{Left column:} Flux posteriors for the fit where the GCE Smooth component is constrained to have positive coefficient. \textbf{Middle column:} Flux posteriors for the fit where the GCE Smooth component is allowed to float to negative values. \textbf{Right column:} SCF corresponding to the left column. The dashed lines show the simulated disk SCF (pink) and simulated Iso PS SCF (brown).}
\label{fig:30deg}
\end{figure*}

We then perform 20 simulations based on the posterior median parameters from a fit to real data, where the GCE is assumed to be smooth but allowed to be asymmetric. 

Figure~\ref{fig:30deg} compares the real data to a selected realization (again, where no GCE PSs are simulated), but for the 30$^\circ$ radius ROI. As discussed above, we see that when the smooth GCE component is allowed to float negative, it achieves quite negative values in the real data for this ROI. This is particularly noticeable when the negativity of the smooth component is expressed as a fraction of the total GCE flux (in the fit where all GCE components are forced to be non-negative); the posterior median for the GCE Smooth flux peaks at around $-2\times$ the total GCE flux in this ROI, whereas in the $10^\circ$ ROI the preferred magnitude of the negative GCE Smooth component is smaller than the total flux in the GCE.

\begin{figure*}[t]
\leavevmode
\centering
\subfigure{\includegraphics[width=0.35\textwidth]{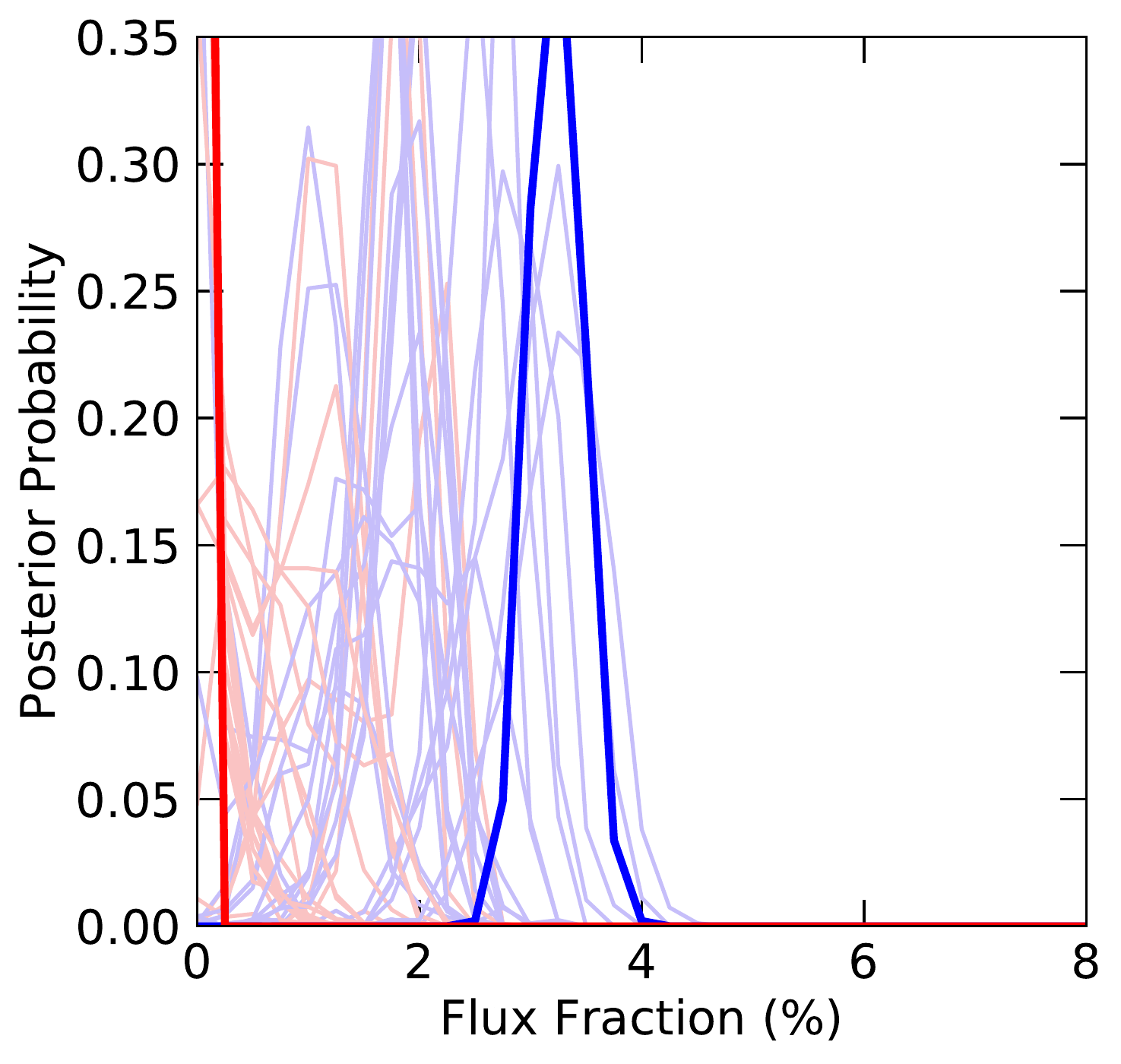}}
\subfigure{\includegraphics[width=0.34\textwidth]{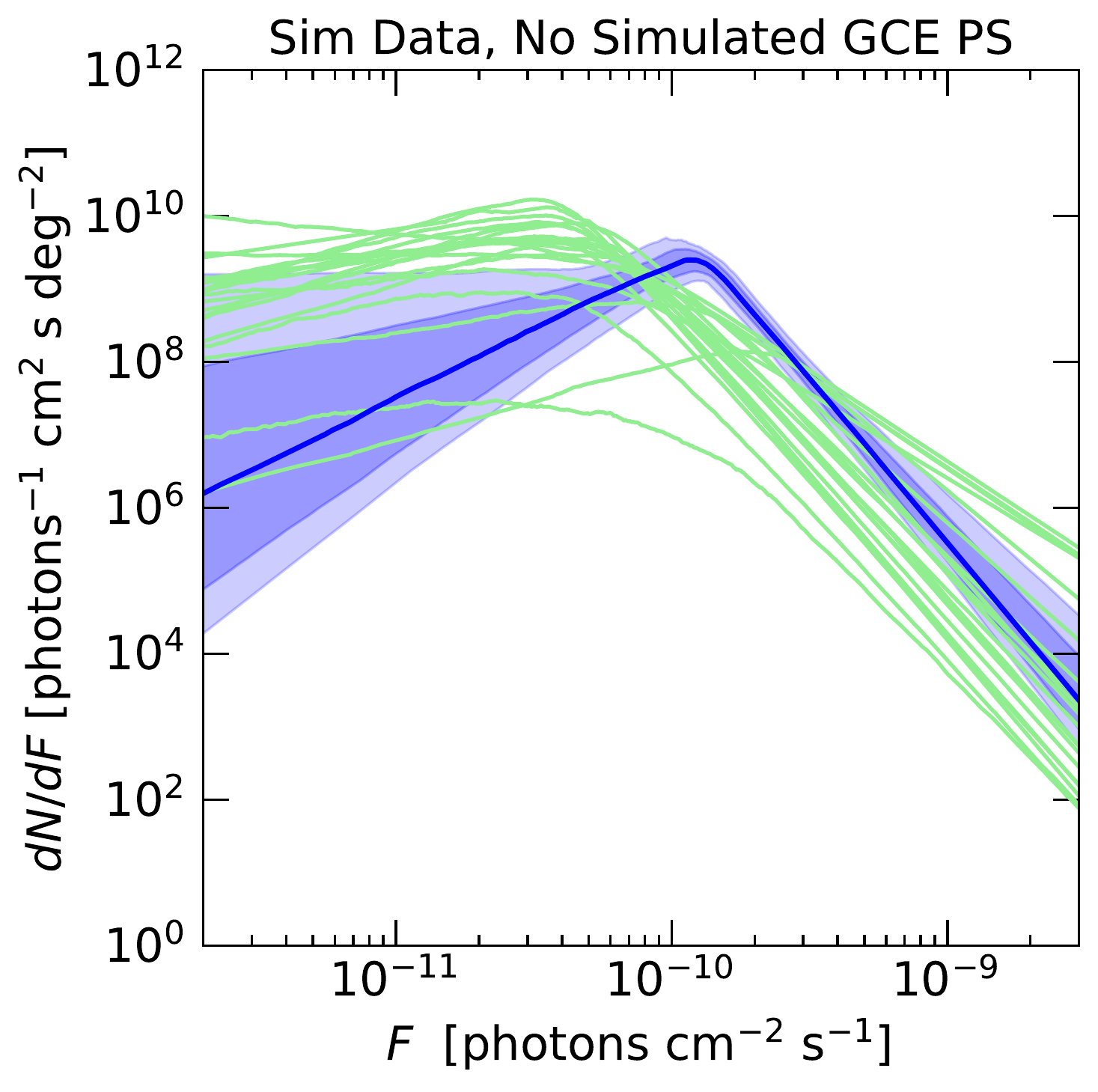}}
\caption{Spread of analysis results on 20 simulated-data realizations (for diffuse model \texttt{p6v11}) in the 30$^\circ$ radius ROI, extending Fig.~\ref{fig:30deg}. In all cases, the analyses use symmetric GCE templates (PS and Smooth). The simulated dataset is based on the best fit model (fluxes shown in left panel of Fig.~\ref{fig:asym30}) using separate GCE Smooth templates for the northern and southern GCE; no GCE PSs were simulated. \textbf{Left:} Flux fraction posteriors.  Fainter blue (red) lines correspond to the GCE PS (GCE Smooth) posteriors for simulated realizations, bold darker lines are the real data. \textbf{Right:} the SCF obtained in the real data using one symmetric GCE PS template is shown in blue, the posterior median values of the reconstructed SCFs for GCE PSs, in the simulations, are shown in green.}
\label{fig:30degspread}
\end{figure*}

We see in our simulated data that while unmodeled north-south asymmetry in the $10^\circ$ ROI could fully explain the observed degree of negativity in the GCE Smooth component, this is not true in the $30^\circ$ ROI. We do not observe the same preference for large negative normalizations of the smooth GCE when we simulate an asymmetric GCE and then fit it with symmetric GCE templates, at least within the realizations we have tested. The realization shown has a BF of $10^4$, while fitting the real data in this ROI yields a BF of $\sim10^{36}$ in favor of GCE PSs.

Figure~\ref{fig:30degspread} shows the spread of the results over the 20 realizations in the 30$^\circ$ ROI. The reconstructed flux posteriors overlap with the results in real data, albeit with a wide spread. Across the 20 realizations we observe a range of SCFs, all which contain sources well above the one-photon threshold; however, within this set of realizations, the peak of the SCF is consistently at a lower flux than observed in the real data. Note that across these realizations, the BF for PSs ranges from $\sim1-10^{14}$; just as in the 10$^\circ$ radius ROI, we see that failure to correctly model asymmetry of the GCE is expected to lead to a spurious preference for PSs. However, the larger BF for PSs in data, compared to simulations, is another indication that modeling the GCE as smooth and asymmetric does not fully explain the results in this larger ROI; this is not unexpected, as whatever effect is driving the smooth GCE deeply negative may well have the effect of producing a spuriously large preference for GCE PSs.

We can compare the SCFs for the GCE PS population found in real data, between the 10$^\circ$ and 30$^\circ$ radius ROIs. They are fairly similar but not identical; the best fit for the break is found to be $35_{-4.8}^{+5.3}$ photons/source in the 30$^\circ$ radius ROI, but $22.1_{-3.7}^{+4.2}$ photons/source in the 10$^\circ$ radius ROI (roughly $2\sigma$ discrepant). A significant discrepancy between the two would be further evidence in favor of a non-physical origin for the apparent source population, but is not apparent at this stage; it is also possible that the choice of SCF parameterization is obscuring differences or similarities between the two SCFs.

\subsection{Analyses with a 40$^\circ$ by 40$^\circ$ square, 2$^\circ$ plane mask ROI}

Previous studies of the morphology of the GCE have not detected the same north/south asymmetry that we find; in particular, Ref.~\cite{Calore:2014xka} explored the effects of subdividing the GCE into subregions, in the context of a Poissonian template analysis, and found the GCE appeared quite symmetric (depending only on Galactocentric distance). The authors of that work used a 40$^\circ$ by 40$^\circ$ square ROI centered on the Galactic center, masking the plane for $|b|<2^\circ$. They employed a range of Galactic diffuse emission models, finding comparable results across the board; thus we will focus on \texttt{Model A} and \texttt{Model F}. They also modeled known sources based on the 2FGL point source catalog, the \textit{Fermi} Bubbles, and isotropic emission.  They combined the gas-correlated bremsstrahlung and $\pi^0$ emission templates for the Galactic diffuse emission models into a single template (as we also do in our analyses), but gave additional freedom to the inverse Compton scattering (ICS) component, subdividing it into nine latitude strips with boundaries at $|b| =$ 2.0, 2.6, 3.3, 4.3, 5.6, 7.2, 9.3, 12.0, 15.5 and 20 degrees. The normalization of strips that are symmetric under $b \rightarrow -b$ are constrained to float to equal values.

Reproducing the analysis of Ref.~\cite{Calore:2014xka} (with the exception that we use a 3FGL PS model instead of 2FGL) and focusing on \texttt{Model F}, we find that for this ROI there is not any significant difference in the amount of flux attributed to north vs. south GCE templates, in agreement with the results of Ref.~\cite{Calore:2014xka} (we also confirm this result for \texttt{Model A}). We find that this conclusion holds even when the additional freedom in the ICS template is removed (i.e. we fit with a single ICS template rather than latitude strips), although there is a statistical preference for these additional degrees of freedom. Accordingly, the degree of asymmetry preferred in the GCE appears to depend on the combination of ROI and background modeling. (We have tested that other changes in the analysis, such as use of the 3FGL PS model rather than disk and isotropic PS templates, do not change the preference for asymmetry; it appears to be the change in ROI that drives the change in preference.)

Since \texttt{Model F} prefers a GCE asymmetry when the fit is performed in the $10^\circ$ radius ROI but not in this larger ROI, it is reasonable to ask whether perhaps this could reflect a preference that the \textit{core} of the GCE be north-south asymmetric, while its outskirts are more symmetric. However, the analysis of Ref.~\cite{Calore:2014xka} attempted to measure the GCE spectrum in symmetric regions north and south of the Galactic plane, with support only within $10^\circ$ of the GC, and found no indication of asymmetry. If this result persists in current data, and the discrepancy is significant (both questions which we leave for future studies), it would suggest that the small- and large-ROI analyses differ in their preference for the central region of the GCE to be asymmetric (despite using the same photons from that region), which in turn would imply that this difference is driven by the preferred normalizations of \textit{other} templates in the two fits. Accordingly, it seems very plausible that the apparent asymmetry of the GCE could be driven by cross-talk with the other templates. It is also, of course, possible that the GCE is truly asymmetric, and its apparent symmetry in earlier studies was a function of the same kind of cross-talk. However, it seems more coincidental to obtain an apparently symmetric signal via cross-talk with highly asymmetric templates, than to obtain an apparently asymmetric signal in the same way. Resolving this question will require a dedicated study which is beyond the scope of this work.

Since there is no preference for asymmetry in the large-ROI case for this background model, we have tested the preference for inclusion of GCE PSs, using a single (north-south-symmetric) template for both smooth and PS GCE components; in this case, we also replace the 3FGL PS model with templates for disk and isotropic source populations. We find that there is a preference for PSs in this case, with a modest BF of 200 when we use a single ICS template, and 400 when we subdivide the ICS template into strips. Thus this case serves as another existence proof (along with the 30$^\circ$ radius region) that a preference for PSs may persist once the possibility of north-south asymmetry of the GCE is taken into account, although of course similar unmodeled morphological errors may be present in this case, and even if taken at face value, the preference for PSs is not highly significant (comparable to a $3\sigma$ signal).

\subsection{Additional Template Variations}

We also test breaking the GCE template into four regions within the 10$^\circ$ radius ROI, consistent with the northern and southern inner ($<5^\circ$ from the Galactic Center) and outer ($5-10^\circ$ from the Galactic Center, $|b|>|l|$) regions in Ref.~\cite{Balaji:2018rwz}. We test floating the inner two regions as one template, and the outer two regions as one template (which effectively translates to allowing more freedom in the GCE profile slope), and find this is not preferred over the case with a single GCE template.
As such, in this work we focus on the effects of unmodeled north-south asymmetry in the signal template, rather than freedom in the profile slope (parameterized by separate inner and outer templates). However, we briefly consider other template variations.

\begin{figure*}[t]
\leavevmode
\centering
\subfigure{\includegraphics[width=0.31\textwidth]{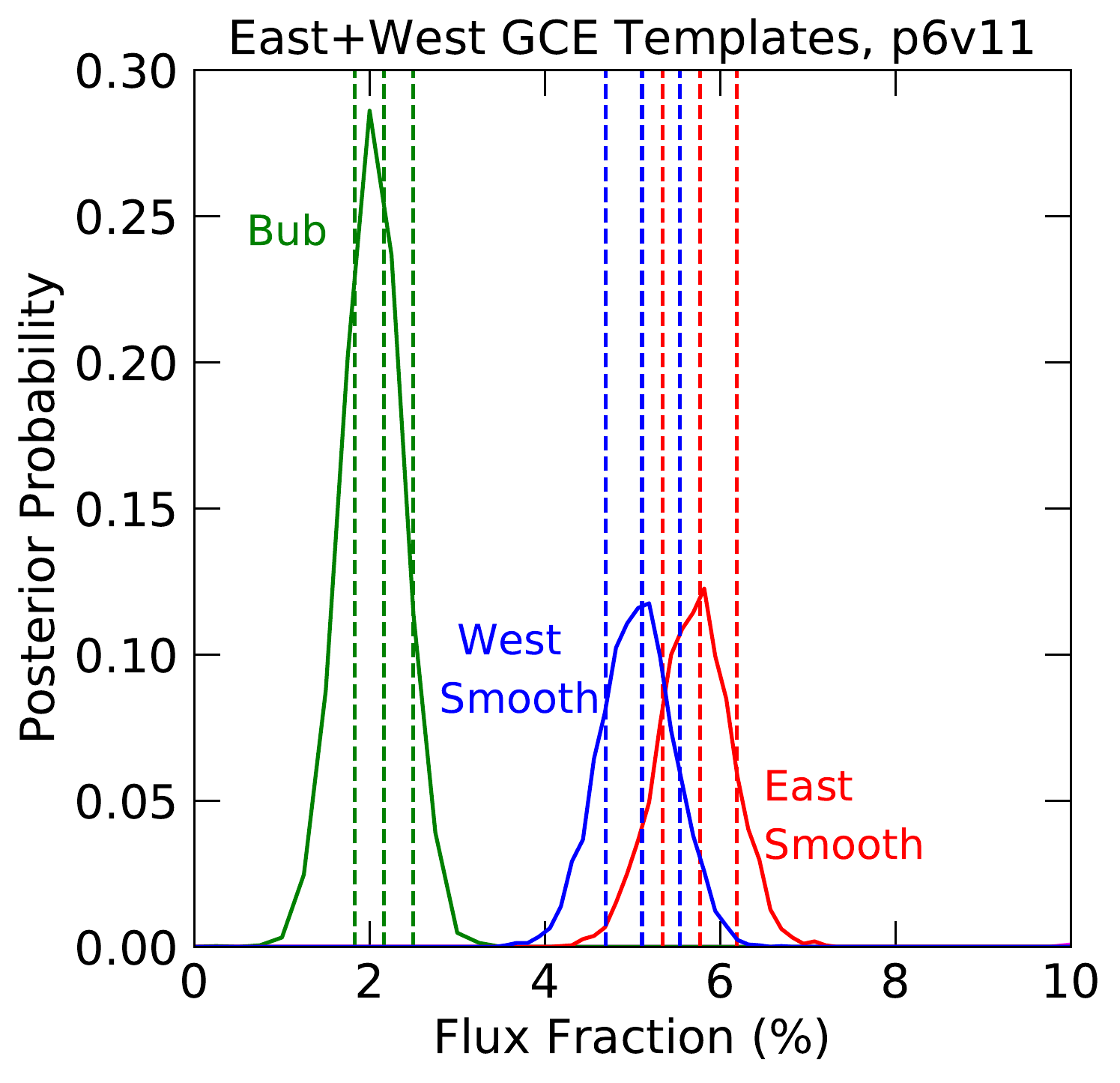}}
\hspace{1mm}
\subfigure{\includegraphics[width=0.31\textwidth]{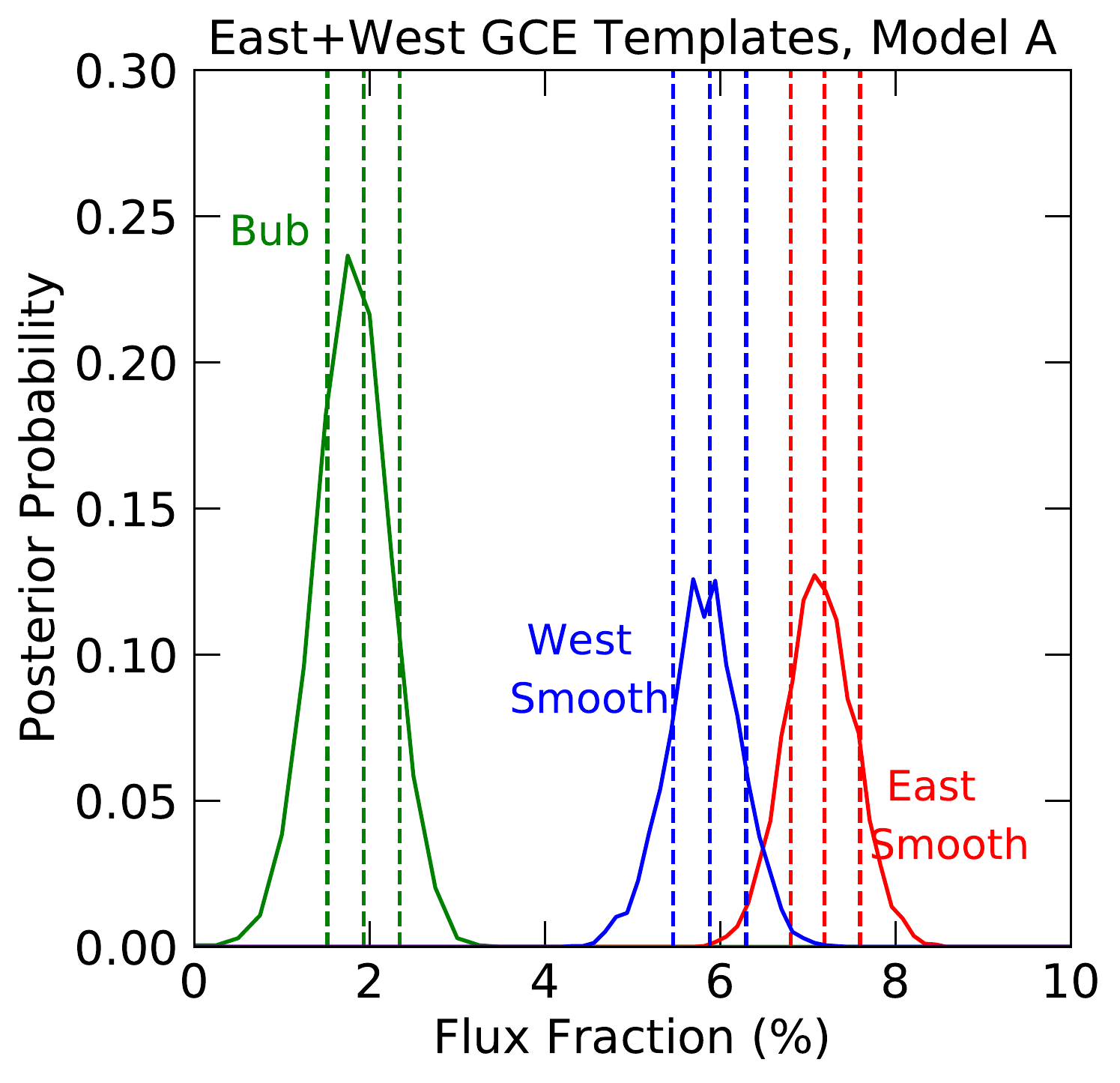}}
\hspace{1mm}
\subfigure{\includegraphics[width=0.31\textwidth]{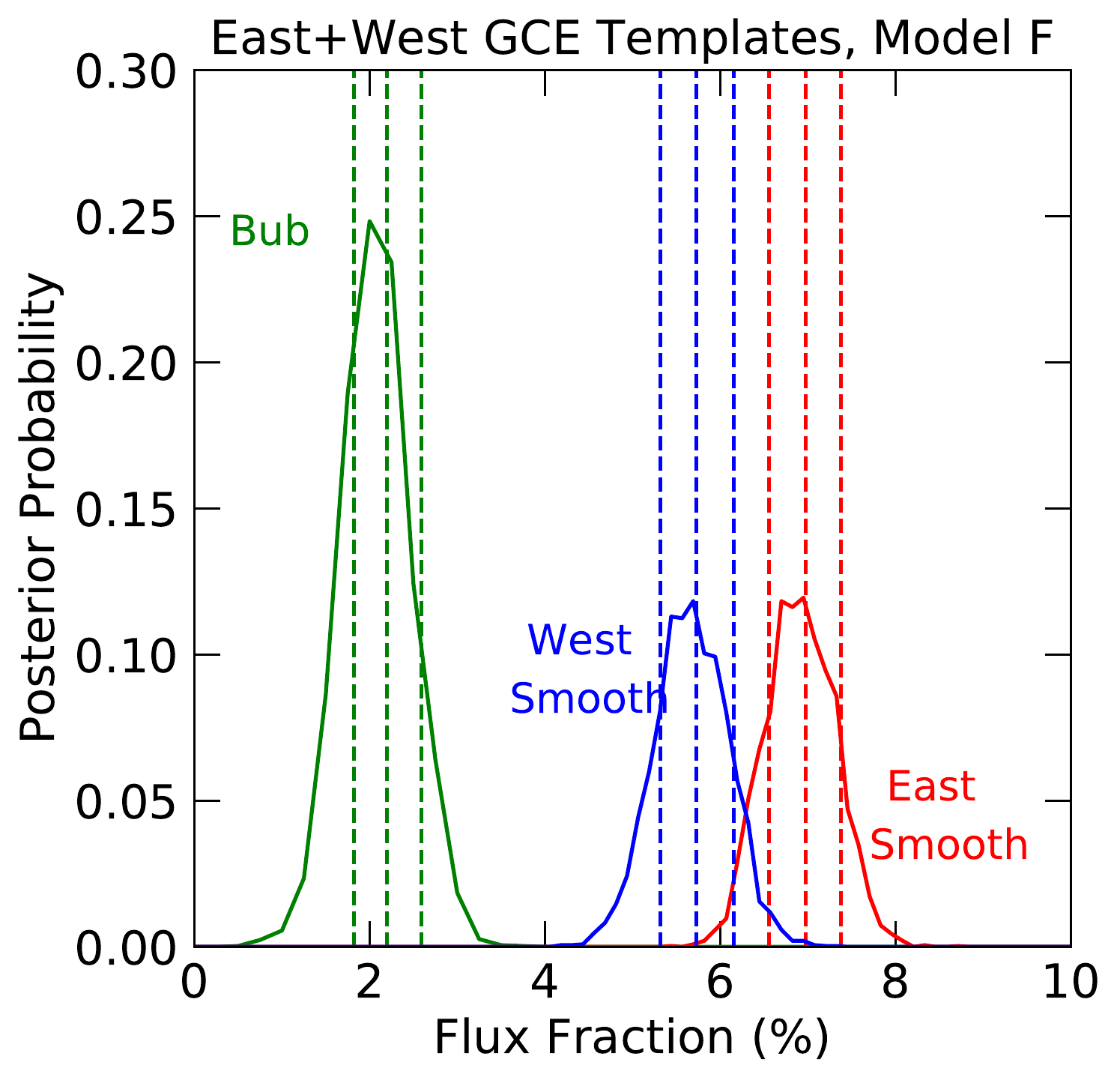}}
\caption{Flux posteriors when the GCE template is allowed to have separate East-West pieces, for diffuse models \texttt{p6v11} (\textit{left}), \texttt{Model A} (\textit{middle}), and \texttt{Model F} (\textit{right}).}
\label{fig:eastwest}
\end{figure*}

Figure~\ref{fig:eastwest} shows the impact of allowing east-west asymmetry in the signal template (within the 10$^\circ$ radius ROI). For \texttt{p6v11}, there is no preference for east-west asymmetry, and the BF for GCE PS when using east-west GCE pieces remains roughly consistent ($1\times10^{16}$) with one GCE template over the whole region (which yields $4\times10^{15}$). For \texttt{Model A} there is a preference for east-west asymmetry with BF of $\sim10^4$, and the BF for GCE PS when using east-west GCE pieces is $\sim10^3$.  For \texttt{Model F}, there is a preference for east-west asymmetry with BF of $\sim10^3$,  and the BF for GCE PS when using east-west GCE pieces is $\sim1$. Note for all diffuse models we test, the east-west GCE asymmetry is preferred less by the data than the north-south asymmetry.

It could also be interesting to test the effects of allowing further variations to the diffuse background model, and allowing asymmetry in the \textit{Fermi} Bubbles model. The modeling of the \textit{Fermi} Bubbles near the Galactic plane is highly uncertain; it would be an interesting direction for future studies to explore the impact of changing the Bubbles model and allowing it to have additional freedom. Certainly the north-south asymmetry explored in this work and Ref.~\cite{PhysRevLett.125.121105} does not exhaust the possibilities for systematic errors in the chosen templates to affect the PS preference.

\section{Implications}
\label{sec:discussion}

We have seen that allowing for one specific form of mismodeling (north-south asymmetry) removes the evidence for GCE PSs in some but not all NPTF analyses. In particular, while the behavior we have observed in the $10^\circ$ radius ROI appears quite stable over variations in the diffuse model, different prior choices, and different photon data selections, in larger ROIs a GCE PS preference can persist, and there is not always a significant north-south asymmetry. Since there is evidence for mismodeling in large ROIs that is not absorbed simply by allowing north-south asymmetry in the signal \cite{Leane:2019xiy}, there is some argument to simply dismiss these remaining PS preferences. The systematic mismodeling we have identified has proved capable of generating a spurious high-significance preference for PSs, with a SCF matching what is observed in real data, and consequently serves as a counterexample to a number of earlier arguments (summarized in Section~\ref{sec:previous}) for the robustness of NPTF-based GCE PS detections to mismodeling. 

However, one might still ask if it is likely that the GCE PSs nominally detected in the main analysis of Ref.~\cite{PhysRevLett.125.121105} (when GCE asymmetry is not allowed) are spurious, but also unrelated to the GCE PSs detected in other NPTF analyses. For example, a true GCE PS population could be fainter than the spurious sources we have identified, and beneath the sensitivity of our analysis in the $10^\circ$ ROI, but detectable when the background templates are informed by a larger ROI.
To explore the possibility that previous studies have found a true GCE PS population unrelated to the spurious PS signal identified in Ref.~\cite{PhysRevLett.125.121105}, let us examine the consistency of the SCFs obtained in the various analyses. 

\subsection{Source Count Function Consistency Across NPTF Analyses}

In all NPTF analyses of which we are aware, the preferred SCF implies a population of PSs where most of the photons originate from sources with fluxes very close to a break in the power-law index, $F_b$. The posterior SCF generally declines very steeply with increasing flux, above the break (often encountering the edge of prior ranges unless the prior is allowed to be very wide), but rises quite steeply in $F^2 dN/dF$ below the break as well -- for example, the original NPTF analysis in Ref.~\cite{Lee:2015fea} found a low-flux slope of $dN/dF \propto F^{0.66^{+0.90}_{-0.98}}$. For this reason, the SCF can largely be characterized by the location of $F_b$ (or the highest break, if multiple breaks are allowed), combined with an overall normalization.

\subsubsection{Source Count Functions in Past and Current NPTF Studies}

The original claim to detect a GCE-correlated population of PSs \cite{Lee:2015fea} found a flux break of $F_b = 1.76^{+0.44}_{-0.35} \times 10^{-10}$ photons/cm$^2$/s for the analysis where 3FGL sources were not masked, and a consistent but slightly lower value of $F_b = 1.62^{+0.45}_{-0.32} \times 10^{-10}$ photons/cm$^2$/s when the 3FGL sources were masked. Note that this analysis used a slightly different energy range than later studies using the public \texttt{NPTFit} package, $1.893-11.943$ GeV rather than $2-20$ GeV. However, using the broken-power-law spectrum for the GCE taken from Ref.~\cite{Calore:2014xka}, the difference in photon flux from GCE sources between these two bands is expected to be only $8\%$ (with the $2-20$ GeV band having slightly more photons), which is small compared to the statistical uncertainties on $F_b$.

The later analysis of Ref.~\cite{Leane:2019xiy}, using the top quartile by angular resolution of the \textit{Fermi} \texttt{Pass 8} data, and again a $30^\circ$ radius ROI, found a flux at the break of $F_b = 1.94^{0.34}_{-0.30} \times 10^{-10}$ photons/cm$^2$/s when the 3FGL sources were masked, and $F_b = 2.90^{+0.57}_{-0.46} \times 10^{-10}$ photons/cm$^2$/s when they were not masked. 

Both of these analyses allowed the slope of the SCF above the break to be very steep, yielding essentially a sharp cutoff. In our $10^\circ$ radius region (where we always unmask the 3FGL sources), using the top three quartiles in angular resolution, and allowing a sharp cutoff in the SCF, we obtain $F_b = 1.15^{+0.20}_{-0.17} \times 10^{-10}$ photons/cm$^2$/s. All of these results are quoted for the \texttt{p6v11} diffuse model. For \texttt{Model A}, the (unmasked) analysis of Ref.~\cite{Leane:2019xiy} yielded $F_b = 1.07^{+0.20}_{-0.16} \times 10^{-10}$ photons/cm$^2$/s .

We see that all of these results are broadly consistent with a break around $F_b=1-3 \times 10^{-10}$ photons/cm$^2$/s, but our smaller-ROI, three-quartile result, and the \texttt{Model A} result, prefer values on the low end of this range ($\sim 1 \times 10^{-10}$ photons/cm$^2$/s), while studies using \texttt{p6v11} in the $30^\circ$ radius ROI prefer higher values ($\sim 2-3 \times 10^{-10}$ photons/cm$^2$/s). This higher value may reflect whatever systematic mismodeling causes the smooth GCE component to prefer a very negative value in the $30^\circ$ radius ROI (although Ref.~\cite{Leane:2019xiy} found that \texttt{Model A} also prefers a negative value, albeit at not at the same level of significance).

\subsubsection{Sensitivity of the GCE SCF to Other Analysis Changes}

As discussed above, in the current work by default we have modified the prior on the slope above the break to reduce the sharpness of the cutoff, and this gives rise to a lower value of $F_b$ (as the contribution from sources with fluxes just above $F_b$ is substantially increased), $F_b = 0.79^{+0.15}_{-0.13}\times 10^{-10}$ photons/cm$^2$/s. Unlike the changes to the analysis discussed in the previous paragraphs, this does not represent a difference in the dataset or background models, just a change in the parameterization of the SCF. 

With this alternative prior choice, we can explore the effects on $F_b$ of other changes to the analysis. Using \texttt{Model A} instead of \texttt{p6v11} yields $F_b = 0.49^{+0.14}_{-0.11}\times 10^{-10}$ photons/cm$^2$/s; using a $30^\circ$ ROI instead of $10^\circ$ yields $1.25^{+0.19}_{-0.17}$; using a single angular resolution quartile instead of three quartiles yields (in the $10^\circ$ ROI) $F_b = 1.25^{+0.28}_{-0.23}\times 10^{-10}$ photons/cm$^2$/s for \texttt{p6v11} and $F_b = 1.02^{+0.29}_{-0.22}\times 10^{-10}$ photons/cm$^2$/s for \texttt{Model A}.  We see that with this prior, essentially all the results are consistent with our baseline analysis ($F_b = 0.79^{+0.15}_{-0.13}\times 10^{-10}$ photons/cm$^2$/s) within the uncertainties (at the $2\sigma$ level).

The most significant change is the effect of going to a larger ROI (while using the \texttt{p6v11} diffuse model); this variation creates an apparent preference for slightly brighter PSs that are not found by the fit in the smaller ROI. The $F_b$ values for \texttt{Model A} and \texttt{p6v11} in our current analysis are more consistent (less than $2\sigma$ discrepant, corresponding to a factor of 1.6 between median values of $F_b$), than in the ($30^\circ$ ROI) analysis of Ref.~\cite{Leane:2019xiy} (roughly 3.5$\sigma$ discrepant, corresponding to a factor of 2.7 between median values), and they are even more consistent in the case where only a single quartile is used ($\sim 20\%$ difference in median values, less than $1\sigma$ discrepant). Aside from the large-ROI analyses with \texttt{p6v11}, all the $F_b$ values from this and previous analyses seem broadly ($2\sigma$) consistent with each other, after accounting for the effect of the change in prior. 

\subsubsection{Implications of Source Count Function Consistency}

It would be a coincidence if the SCF of a real GCE PS population, undetectable in the $10^\circ$ radius ROI, happened to agree so closely with the spurious SCF induced by the asymmetry we have identified. Furthermore, the fact that the preferred SCF corresponds to \textit{brighter} sources in the larger ROIs is the opposite of one would expect if a true GCE PS population was present in both large and small ROIs, but being masked by a brighter spurious population (as identified in Ref.~\cite{PhysRevLett.125.121105}) in the smaller ROI. Of course, a real but fainter GCE PS population could still be present in any of these analyses, but it would not dominate the bright end of the extracted SCF.

 \subsection{Comment on the Source Count Function Shape}
 
We note that the tendency for the SCF to prefer a source population within a narrow flux range itself contrasts somewhat with typical estimates for the pulsar luminosity function (e.g. Ref.~\cite{Bartels:2015aea} assumed $dN/dF \propto F^{-1.5}$, based on Refs.~\cite{Petrovic:2014xra,Cholis:2014noa, Strong:2006hf,Venter:2014zea}). This in itself might be viewed as evidence that the inferred signal is not physical. However, the uncertainties are large, and Ref.~\cite{Chang:2019ars} has demonstrated that in simulations, the NPTF has a systematic tendency to reconstruct source count functions with less power at lower flux than the truth (albeit usually within the uncertainties). For these reasons, the very peaked shape of the SCF preferred by NPTF fits has not generally been thought to pose a problem in interpreting the results as evidence for a physical PS population. 

However, in this work we have demonstrated explicitly that just such a steeply peaked SCF is automatically preferred in the presence of an overly rigid signal template, in simulations where no GCE PSs exist. We thus find it very suggestive that all NPTF analyses to date appear to prefer this type of SCF.

 \subsection{Can We Exclude Hypotheses for the Origin of the GCE?}

Taking these results together, while we have not constructed explicit models that remove the GCE PS preference for all combinations of diffuse model and ROI, we observe that:
\begin{enumerate}
 \item All ROIs tested in previous analyses include our $10^\circ$ radius region.
 \item Previous analyses have found that the significance for the GCE dominantly comes from that subregion \cite{Daylan:2014rsa,Calore:2014xka}.
 \item As we have shown, the SCFs reconstructed in those analyses are similar both qualitatively and quantitatively to the inferred SCF for a population we have demonstrated to be \textit{spurious} in the $10^\circ$ radius subregion.
\end{enumerate}

For this reason, we do not currently consider the detection of GCE-correlated PSs in those analyses, or the attribution of the bulk of the GCE flux to those PSs, to be convincing. Of course, future analyses with the ability to identify and control for such systematic effects may change this conclusion.

We emphasize that we are not claiming to robustly exclude GCE-correlated PS populations with comparable SCFs to those found in previous works. That would be a much stronger claim than the one we make, which is that we do not think the hypothesis of a dominantly-smooth GCE can currently be discarded based on NPTF results. A robust exclusion of certain SCFs for a GCE PS population would require an analysis dedicated to that goal, with a careful study of systematic uncertainties, which we have not attempted to conduct here. While we have shown explicitly that systematic errors in the signal model can give rise to a spurious preference for PSs that mimics what is seen in the real data, it is also plausible that errors in various templates could conceal a true PS population. 

Similarly, we do not currently claim robust evidence that the GCE is intrinsically north-south asymmetric, given that this conclusion is not even uniform across the analysis variations we have tested (i.e., in larger ROIs it does not appear with all background models). It is plausible the asymmetry could arise from background mismodeling, given its dependence on the ROI and diffuse model. However, if a firm detection of north-south asymmetry in the GCE was established in future work, that would immediately provide significant constraints on its possible origin. In particular, it seems unlikely that a 2:1 asymmetry could be easily explained by simple DM annihilation models.

Finally, we should emphasize that none of the issues we have identified imply that the NPTF method or its \texttt{NPTFit} implementation is wrong in a mathematical/statistical sense; it has passed all tests of this type that we have conducted. It is in fact completely correct that if one has a strong prior that the underlying physical mechanism for the GCE is symmetric and there is no meaningful cross-talk with other templates that could induce an apparent asymmetry, then identifying a highly significant asymmetry serves as strong evidence for a GCE PS population -- it is much easier to generate an asymmetry with a higher-variance population. However, such a strong prior is hard to justify given the uncertainties in the background and signal models.

\section{Summary and Conclusions}
\label{sec:conclusion}

The physical origins of the GCE remain an enigma, in large part because of our lack of detailed knowledge of other contributions to the gamma-ray flux from the inner Galaxy. Novel analysis techniques, including NPTF, have aimed to shed light on this question, but run the risk of giving unreliable or deceptive results due to assumptions about the signal and backgrounds that are not borne out in reality. In this work and the companion paper~\cite{PhysRevLett.125.121105}, we have investigated the degree to which signal mismodeling can skew the conclusions of NPTF analyses.

In our companion paper~\cite{PhysRevLett.125.121105}, we demonstrated that in a 10$^\circ$ radius region around the Galactic Center, apparent NPTF evidence for GCE PSs is an artifact of unmodeled north-south asymmetry. We showed that expanding the signal model by allowing this asymmetry renders the preference for PSs insignificant, and the behavior of the real data is what would be expected from simulations which include an asymmetric but entirely smooth GCE.

In this work, we have built upon these results, aiming to understand why this behavior occurs, and to characterize its implications for NPTF analyses more broadly. We have developed an analytic approximation that allows for a simple intuitive understanding of the effect; in short, the presence of PSs can be thought of as a proxy for higher variance away from the signal expectation value (at least in the limit where the relevant likelihoods can be approximated as Gaussian). However, higher variance can be induced by mismodeling as well as PS populations, and so excess variance from a wide range of possible types of mismodeling -- e.g., in detector acceptance, energy / angular resolution, or template morphology -- could be interpreted instead as evidence for PSs. This approximation appears to work well to explain the observed preference for PSs, and the average brightness of those PSs, in the case of a very simplified simulation containing \textit{only} a smooth asymmetric GCE (with no backgrounds), when the NPTF analysis is performed assuming a symmetric GCE. A strong preference for PSs was also present when we simulated a smooth asymmetric GCE and a smooth background model only. These tests confirmed that higher variance from signal mismodeling could lead to a high-significance but spurious preference for a GCE PS population, even in the unrealistic case where there are no PSs present in the data at all.

We have explored the degree to which this behavior -- the preference for north-south asymmetry, and the disappearance of the preference for GCE PSs once it is included -- occurs consistently in the real gamma-ray data, under a number of variations to the baseline analysis presented in the companion paper. When testing alternate diffuse models in the 10$^\circ$ radius ROI, we found comparable behavior to Ref.~\cite{PhysRevLett.125.121105}: spurious GCE PSs can be created with the same significance as those found in the real data, by having an unmodeled smooth GCE asymmetry present. For the diffuse models we tested, the preference for a smooth GCE asymmetry persisted, as did the disappearance of any significant evidence for GCE PSs (although in one model there was no preference for GCE PSs to begin with). Restricting our photon dataset to the quartile of photons with the best angular resolution, we found very similar results, though including the top three quartiles led to higher expected sensitivity. Modifying the priors for the SCF for GCE PSs, in a way that changed its preferred shape, did not change the consistency between the results on real data and the results expected from simulations with an asymmetric smooth GCE.

In larger ROIs, we found that the GCE asymmetry persisted with \texttt{p6v11}, but that it could disappear with other diffuse models \texttt{Model A} and \texttt{Model F}. These results highlight why we do not claim the asymmetry is necessarily an intrinsic feature of the GCE. Instead, given the ROI and diffuse model dependence, it seems plausible that the asymmetry is itself an artifact of background mismodeling, where an unmodeled asymmetry in the backgrounds is transferred to the preferred signal model. However, if future analyses were to demonstrate that the central part of the GCE does indeed possess a robust and pronounced north-south asymmetry, that would strongly constrain possible GCE origins; for example, we expect it would be challenging to obtain a large asymmetry in a DM annihilation signal. 

We provided an example (\texttt{Model F} in a $40\times 40^\circ$ ROI) of a case where no north-south asymmetry is preferred and a modest preference for GCE PSs remains in the data (although we caution that other forms of mismodeling may be present in that case, and our analysis has demonstrated that mismodeling can have severe implications for the interpretation of apparent PSs). Likewise, when using the \texttt{p6v11} diffuse model in a $30^\circ$ radius region, we found that not all the behavior observed in the real data seems to be explained by accounting for the north-south GCE asymmetry; it is plausible that multiple mismodeling effects may contribute to the results in these larger ROIs.

We explored the consistency between the (apparently spurious) SCFs for GCE PSs obtained in analyses of our $10^\circ$ radius ROI with those obtained in larger ROIs, where a PS preference can persist even when asymmetry is allowed, and with previous NPTF analyses. We found that the SCFs were broadly consistent with a peak around $10^{-10}$ photons/cm$^2$/s, although studies in the smaller ROI tended to prefer slightly fainter PSs than the larger ROIs, when other analysis choices were held constant. This seems to disfavor the possibility that the studies of larger ROIs are detecting a true GCE PS population that is concealed by a brighter spurious population in the $10^\circ$ radius subregion. Instead, it suggests (albeit does not prove) that the SCFs detected in large ROIs may be as spurious as the very similar SCFs shown explicitly to be an artifact of signal mismodeling in the smaller ROI. Nonetheless, we cannot exclude the presence of GCE-correlated PSs; a GCE PS population dominated by sources much fainter than $10^{-10}$ photons/cm$^2$/s would almost certainly go undetected in our analysis, but even brighter GCE PSs could potentially be hidden by systematic errors.

Our principal conclusion is thus that NPTF studies do not currently provide a clear and convincing argument for (or against) a PS-dominated GCE. While systematic uncertainties have always been a concern in NPTF studies, this work and our accompanying paper have for the first time demonstrated that mismodeling can produce a convincing but spurious PS signal at high significance, and furthermore that this scenario is realized in the inner-Galaxy gamma-ray data. We have shown that the NPTF approach, of matching the photon statistics of the model to those of the real data, implies this potential for confusion; both PS populations and signal mismodeling lead to inflated variance relative to the expectation value, and consequently the level of mismodeling directly correlates with the brightness of the inferred PS population. Our simulations show that this type of mismodeling can yield an apparent preference for PSs comparable to what one would expect from a true PS population, and a SCF very consistent with results on the real data (both in our study and in previous work), which is then quite stable under a range of variations to the analysis. 

Despite what might seem to be a frustratingly inconclusive situation, our understanding of the GCE continues to actively improve. For example, recent work using wavelet methods \cite{Zhong:2019ycb} placed an upper limit on the contribution to the GCE from wavelet peaks previously attributed to GCE PSs \cite{Bartels:2015aea}. This suggests a large fraction of the GCE should originate from either relatively faint sources or smooth emission, as opposed to point sources that are bright enough to already appear as the faintest members of up-to-date source catalogs. As we have argued in this work, apparent NPTF evidence for a GCE dominated by relatively bright PSs may be due to signal mismodeling, in which case there is no inconsistency with an approximately smooth GCE. If this is the case, it remains to be understood whether the GCE originates from faint pulsars, cosmic ray interactions with gas or starlight, annihilating DM, or some other source.

\tocless \section{Acknowledgments}
We thank M. Buckley, M. Buschmann, L. Chang, G. Collin, R. Crocker, D. Curtin, D. Finkbeiner, P. Fox, D. Hooper, S. Horiuchi, T. Linden, M. Lisanti, O. Macias, S. McDermott, S. Mishra-Sharma, S. Murgia, K. Perez, N. Rodd, B. Safdi, T. Tait, and J. Thaler for helpful discussions. We thank the \textit{Fermi} Collaboration for the use of \textit{Fermi} public data. The work of RKL was performed in part at the Aspen Center for Physics, which is supported by NSF grant PHY-1607611.  RKL and TRS are supported by the Office of High Energy Physics of the U.S. Department of Energy under Grant No. DE-SC00012567 and DE-SC0013999, as well as the NASA Fermi Guest Investigator Program under Grant No. 80NSSC19K1515.
\\

\appendix
 
 \section{Comment on Chang et al (2019) (Ref.~\cite{Chang:2019ars})}
\label{sec:chang}

Motivated by the results of Ref.~\cite{Leane:2019xiy}, an independent group of authors~\cite{Chang:2019ars} explored the degree to which potential biases in the NPTF method could result in a misattribution of the GCE to PSs, exclusively using simulated data. That work demonstrated that even when the data are correctly modeled by the templates, there could be some misattribution of injected DM signals to PSs, and vice versa, with the misattribution effects being larger when the diffuse Galactic emission model used in the fit is different to that used in the simulation. It also demonstrated that the bright end of the PS SCF (i.e. the number of bright PSs) could, at least in some cases, be more robust to these particular sources of misattribution than the faint end (the number of faint PSs).

However, the degree of misattribution demonstrated in Ref.~\cite{Chang:2019ars} is not nearly as severe as what is observed in real data~\cite{Leane:2019xiy}. In the real data, even DM signals much larger than the GCE are mis-reconstructed as PSs. The analysis of Ref.~\cite{Chang:2019ars} focuses on injections of the same size as the GCE itself; in fact such injections were already found to be routinely somewhat mis-reconstructed in simulated data in Ref.~\cite{Leane:2019xiy}, with more striking qualitative differences between simulations and real data being observed for larger injections. Furthermore, the authors of Ref.~\cite{Chang:2019ars} attribute the behavior they observe to the degeneracy between a DM signal and a population of very faint sources. While this degeneracy can certainly be important, we do not find it plausible that it is the primary cause of the behavior observed in Ref.~\cite{Leane:2019xiy}. This is because:

\begin{enumerate}
 \item The best-fit parameters in Ref.~\cite{Leane:2019xiy} generically correspond to a zero (or negative, if allowed) total flux in the DM component + PSs below the degeneracy limit. This means that the fit does not prefer any contribution to the flux from photons for which this degeneracy is relevant. 
 \item Explicitly setting the SCF to zero below the one-photon threshold did not change the reconstructed DM coefficient, explicitly demonstrating point (1) above.
 \item Injecting an additional DM-like signal (and requiring physical priors on all components) was found to modify the bright end of the SCF, not solely the faint end.
\end{enumerate}

  Thus, we interpret the behavior observed in Ref.~\cite{Leane:2019xiy} as evidence that some mismodeling (of signal, background, or both) is leading to a preference for a bright PS contribution much larger than can be accommodated by the physical model, corresponding to a negative contribution from the degenerate DM + faint PS components. Injecting an extra DM signal while imposing physical priors (i.e. non-negative fluxes for all templates) provides more available flux that can be allocated to bright PSs, and hence changes the bright end of the reconstructed SCF. This is a separate problem to the degeneracy between the faint PSs and smooth DM component. \\

Finally, it is important to note that while Ref.~\cite{Chang:2019ars} considered a scenario where the DM signal was 100 percent DM, and found that the DM signal was reconstructed correctly in this case:

\begin{enumerate}
 \item As mentioned above, the effects discussed in Ref.~\cite{Chang:2019ars} do not explain the behavior of the real data, suggesting that there must be other systematic effects present in the real data that could modify reconstruction of a DM signal.
 \item That analysis was based on a simplified simulation where \emph{no} unresolved PSs at all were simulated in the region of interest (ROI) -- but indisputably there will be additional (non-GCE) PSs in the ROI in realistic \textit{Fermi} data, and Ref.~\cite{Leane:2019xiy} has explicitly demonstrated that their presence can affect DM signal recovery.
 \item Ref.~\cite{Chang:2019ars} also demonstrates (in an appendix) that when the diffuse background is mismodeled, a 100 percent DM signal is incorrectly reconstructed as originating from a PS population in a significant fraction of simulations, sometimes with high significance. 
\end{enumerate}

Thus we conclude from Ref.~\cite{Chang:2019ars} that while there are systematic effects that can lead to a degree of misattribution of injected DM-like signals to PSs even when the templates accurately describe the real data -- without leading to detection of a spurious PS population when the underlying signal is 100\% smooth -- those particular systematics do not adequately explain the behavior observed in existing NPTF analyses of the real data. In contrast, systematic errors in the signal and/or background templates have the potential to fully explain behavior observed in the real data without invoking a GCE-correlated PS population, as we have demonstrated in the current article and the accompanying paper \cite{PhysRevLett.125.121105}. Such errors can lead to spurious detection of even relatively bright PSs with high nominal significance, even when the underlying simulated signal is 100\% smooth.

\let\oldaddcontentsline\addcontentsline
\renewcommand{\addcontentsline}[3]{}
\bibliography{annulus}
\let\addcontentsline\oldaddcontentsline


\end{document}